%% file: lat2020.tex
\documentclass[epj]{svjour}

\usepackage{latexsym}
\usepackage{graphics}
\usepackage{amsmath,amssymb}
\usepackage[multiple]{footmisc}
\usepackage{cite}

\input{macros}

\begin{document}

\input{title.tex}

\input{sect1.tex}
\input{sect2.tex}
\input{sect3.tex}
\input{sect4.tex}
\input{sect5.tex}

\input{biblio.tex}

\end{document}

%% file: macros
\def\rmd{{\rm d}}

\def\proof{\noindent{\sl Proof:}\kern0.6em}

\def\frac#1#2{\hbox{$#1\over#2$}}
\def\dual{\mathstrut^*\kern-0.1em}

\def\lvec#1{\setbox0=\hbox{$#1$}
    \setbox1=\hbox{$\scriptstyle\leftarrow$}
    #1\kern-\wd0\smash{
    \raise\ht0\hbox{$\raise1pt\hbox{$\scriptstyle\leftarrow$}$}}
    \kern-\wd1\kern\wd0}
\def\rvec#1{\setbox0=\hbox{$#1$}
    \setbox1=\hbox{$\scriptstyle\rightarrow$}
    #1\kern-\wd0\smash{
    \raise\ht0\hbox{$\raise1pt\hbox{$\scriptstyle\rightarrow$}$}}
    \kern-\wd1\kern\wd0}
\def\slash#1{\setbox0=\hbox{$#1$}\setbox1=\hbox{$\kern1pt/$}
    #1\kern-\wd0\kern1pt/\kern-\wd1\kern\wd0}

\def\nabstar#1{{\nabla\kern0.5pt\smash{\raise 4.5pt\hbox{$\ast$}}
               \kern-5.5pt_{#1}}}

\def\drvstar#1{{\partial\kern0.5pt\smash{\raise 4.5pt\hbox{$\ast$}}
               \kern-6.0pt_{#1}}}

\def\ldrvstar#1{{\lvec{\,\partial}\kern-0.5pt\smash{\raise 4.5pt\hbox{$\ast$}}
               \kern-5.0pt_{#1}}}

\def\MeV{{\rm MeV}}
\def\GeV{{\rm GeV}}

\def\fm{{\rm fm}}
\def\MSbar{\overline{\rm MS\kern-0.5pt}\kern0.5pt}
\def\Nf{{N_{\rm f}}}
\def\Nh{{N_{\rm h}}}
\def\Nl{{N_{\ell}}}

\def\alphas{\alpha_{\rm s}}

\def\psibar{\overline{\psi}}

\def\zetabar{\bar{\zeta}}
\def\zetaprime{\zeta\kern1pt'}
\def\zetabarprime{\zetabar\kern1pt'}

\def\diracstar#1#2{
    \setbox0=\hbox{$\gamma$}\setbox1=\hbox{$\gamma_{#1}$}
    \gamma_{#1}\kern-\wd1\kern\wd0
    \smash{\raise4.5pt\hbox{$\scriptstyle#2$}}}

\def\tr{{\rm tr}}

\def\Obs{{\mathcal O}}

\def\Ds{D_{\rm s}}
\def\DsdagDs{\Ds{\Ds}^{\kern-1pt\dagger}}

\def\avg#1{{\kern1.0pt\overline{\kern-1.0pt#1\kern-1.0pt}\kern1.0pt}}

%% file: title.tex
% title.tex

\title{Past, present, and future of precision determinations of the QCD coupling from lattice QCD}
\author{Mattia Dalla Brida\inst{1,2,3,}\thanks{Affiliation with institutions 1 and 2 ended on 31/10/2020.}}%

%
%\offprints{}       
%
\institute{Dipartimento di Fisica, Universit\`a di Milano-Bicocca,
	       Piazza della Scienza 3, I-20126 Milano, Italy \and  
	       INFN, Sezione di Milano-Bicocca, Piazza della Scienza 3,
	       I-20126 Milano, Italy \and 
           Department  of  Physics,  University  of  Cyprus,  
           P.O.~Box  20537,  1678  Nicosia,  Cyprus}
       
\date{Received: date / Revised version: date}

\abstract{Non-perturbative scale-dependent renormalization 
		  problems are ubiquitous in lattice QCD as they enter 
		  many relevant phenomenological applications. They require
		  solving non-perturbatively the renormalization group 
		  equations for the QCD parameters and matrix elements 
		  of interest in order to relate their non-perturbative
		  determinations at low energy to their high-energy
		  counterparts needed for phenomenology. Bridging the
		  large energy separation between the hadronic and
		  perturbative regimes of QCD, however, is a notoriously
		  difficult task.  In this contribution we focus on the
		  case of the QCD coupling. We critically address 
		  the common challenges that state-of-the-art lattice
		  determinations have to face in order to be significantly
		  improved. In addition, we review a novel 
		  strategy that has been recently put forward in order to 
		  solve this non-perturbative renormalization problem and 
		  discuss its implications for future precision determinations. 
		  The new ideas exploit the decoupling of heavy quarks to 
		  match $\Nf$-flavor QCD and the pure Yang-Mills theory. 
		  Through this matching the computation of the non-perturbative
		  running of the coupling in QCD can be shifted to
		  the computationally much easier to solve pure-gauge
		  theory. We  shall  present results for the
		  determination of the $\Lambda$-parameter of $\Nf=3$-flavor QCD
		  where this strategy has been applied and proven successful.
		  The results demonstrate that these techniques have the
		  potential to unlock unprecedented precision determinations 
		  of the QCD coupling from the lattice. The ideas are moreover
		  quite general and can be considered to solve other
		  non-perturbative renormalization problems.
} 

\maketitle

% end 

%% file: sect1.tex
% sect1.tex

\section{Introduction}
\label{sec:intro}

Renormalization is a fundamental step in order to extract
(meaningful) phenomenologically relevant results from 
lattice QCD calculations. For the lattice theorist it is 
natural to renormalize the bare parameters of the 
lattice QCD Lagrangian and the composite operators of 
interest in terms of some hadronic renormalization schemes 
(cf.~refs.~\cite{Sommer:1997xw,Sommer:2015kza}). In order to 
make the determinations accessible to phenomenologists, however, 
it is often necessary to translate the results obtained 
in the chosen hadronic schemes to results in the (perturbative)
schemes and at the scales commonly considered in phenomenology. 
In practice, this requires the determination of the 
non-perturbative renormalization group (RG) running of the
renormalized QCD parameters and operators in some convenient 
intermediate scheme, from the hadronic scales 
where they were originally defined, up to some high-energy scale, 
where perturbation theory eventually applies and a matching
to phenomenological schemes can be performed.

Over the last decade or so, lattice QCD has entered a 
precision era for an increasingly large set of quantities
(cf.~ref.~\cite{Aoki:2019cca}). Renormalization is a relevant 
part of many of these computations where it can significantly 
impact the quality of the final results. Hence, as we are forced 
to become more aware of all possible sources of uncertainties in
the determination of the bare lattice quantities, the same care
must be reserved to their renormalization. In particular, as any
other lattice calculation, besides the statistical errors the
determinations of renormalized parameters and operators have 
their systematics to deal with, i.e. discretization effects, 
finite-volume effects, quark-mass effects, and, when a matching 
to phenomenological schemes is necessary, also perturbative
uncertainties. It is therefore important that the development in
strategies to compute (bare) lattice quantities is accompanied 
with new ideas to improve their renormalization, so to guarantee 
a precise and robust end result.

An extreme example of this situation, if we can call it this way, 
is the determination of the QCD parameters. In this case we can 
say that the problem is entirely a renormalization problem, which,
however, has very important phenomenological applications. On the
lattice, the QCD coupling and quark masses are renormalized in terms 
of hadronic masses and decay constants, while in phenomenology the 
QCD parameters are needed at energies of the order of a hundred 
$\GeV$ and above. One would thus think that lattice QCD is not the 
right tool for providing this information given the very high 
energies involved. It appears more natural indeed to obtain these 
parameters directly from high-energy quantities, rather than 
from the hadronic spectrum. As we shall recall later in this
contribution, this is actually not the case, as lattice 
techniques offer an ideal framework for these computations. 

For the last 10-15 years, lattice QCD has consistently delivered 
some of the most precise determinations for the QCD parameters, 
as in  particular for the QCD coupling $\alphas$
(see e.g.~refs.~\cite{Aoki:2019cca,dEnterria:2019its,Zyla:2020zbs}).%
\footnote{We here adopt the common notation 
		  $\alphas(\mu)\equiv\alpha_{\overline{\rm MS}}^{(5)}(\mu)$,
		  where $\alpha_{\overline{\rm MS}}^{(5)}(\mu)$
		  is the QCD coupling of the 5-flavor theory
		  renormalized in the $\overline{\rm MS}$-scheme 
		  (see e.g.~ref.~\cite{Zyla:2020zbs}). Note that for  
		  the ease of notation we often omit to write
		  explicitly the $\mu$-dependence of the coupling.}
The current world average for the QCD coupling evaluated for 
reference at the $Z$-boson pole mass $M_Z$ is
$\alphas(M_Z)=0.1179(10)$~\cite{Zyla:2020zbs}, and has a precision 
of about 0.8\%. The lattice determinations alone give
$\alphas(M_Z)=0.1182(8)$~\cite{Aoki:2019cca}, and are the most 
precise subcategory of those considered by the PDG. Besides the 
high precision of the individual state-of-the-art determinations, 
it is important to emphasize also their overall consistency. This 
is a rather non-trivial result considering the fact that even 
though all lattice determinations share some common systematics, 
these are probed quite differently by considering very different
strategies~\cite{Aoki:2019cca}. It is fair to say that such a
variety of approaches within a PDG subcategory is in fact 
unique~\cite{Zyla:2020zbs}.

Despite the tremendous efforts on and off the lattice, however, 
the current uncertainty on $\alphas$ is still large. It is one of 
the largest sources of uncertainty in several key processes, 
particularly so within the Higgs sector, and it is expected to be 
a limiting factor in many high-precision studies at future 
colliders (see e.g.~refs.~\cite{Salam:2017qdl,dEnterria:2019its}).
An uncertainty on $\alphas(M_Z)$ comfortably below the percent
level is desired for precision  applications. For these reasons,
there are  plans for future phenomenological determinations of
$\alphas(M_Z)$ aiming at reaching the extremely competitive
accuracy of 0.2\% using high-luminosity high-energy data 
(see 
e.g.~refs.~\cite{Gomez-Ceballos:2013zzn,Abada:2019lih,Agostini:2020fmq,dEnterria:2020cpv}). The lattice community needs to meet the challenge.

Reducing the current uncertainties on lattice determinations of 
$\alphas$ by such an important factor is not easy. 
Similarly to several phenomenological determinations most lattice
determinations of $\alphas$ are currently limited by systematic 
uncertainties related to the use of perturbation theory
at relatively low scales~\cite{Aoki:2019cca}. The issue is due to 
the fact that reaching high energy on the lattice requires small 
lattice spacings to be simulated and this is in general difficult 
without a dedicated strategy. 

A way around this has been known since a long time and it is based 
on the concepts of finite-volume renormalization schemes and 
finite-size scaling (or step-scaling)
techniques~\cite{Luscher:1991wu,Jansen:1995ck}. 
The methods have been recently applied for obtaining one of the 
most precise determinations of $\alphas$~\cite{Bruno:2017gxd}. 
The key feature of the approach is that it allows for reaching 
high energy with all systematics under control. This 
puts the lattice determinations in the privileged position of 
being able to reach in a clean and controlled way high energies
fully non-perturbatively. The systematics due to the application 
of perturbation theory, in particular, can be entirely avoided 
at the expenses of the statistical errors accumulated in running
from low up to high-enough energy. The net advantage of this
situation is that differently from systematic uncertainties,
statistical errors can be straightforwardly reduced. Nonetheless, 
a reduction of the current uncertainties on $\alphas$ by an
important factor is yet a computationally expensive task, 
even employing a step-scaling strategy 
(cf.~ref.~\cite{Bruno:2017gxd}).

In this contribution we want to review the recent proposal made 
in ref.~\cite{DallaBrida:2019mqg} which may allow for such error
reduction in a substantially cheaper way. The key feature of this
proposal is that one can replace the computation of the RG running
of the coupling in $\Nf$-flavor QCD with that in the pure-gauge
theory. It is clear that, regardless of the chosen strategy, this
allows for a substantial simplification of the problem.

In short, the idea is built on three main steps and exploits 
the decoupling of heavy quarks in a couple of ways.  In the first 
step, heavy-quark decoupling is used to connect a low-energy scale 
$\mu_{\rm dec}$ in $\Nf$-flavor QCD with the corresponding scale 
in the pure-gauge theory. This is achieved through the computation
of a massive renormalized coupling in an (unphysical) theory with 
$\Nf$ heavy quarks of mass $M\gg \mu_{\rm dec}$. In a second step,
by computing the non-perturbative RG running in the pure Yang-Mills
theory of a convenient coupling one obtains the pure-gauge
$\Lambda$-parameter in units of $\mu_{\rm dec}$, i.e. 
$\Lambda^{(\Nf=0)}_{\overline{\rm MS}}/\mu_{\rm dec}$.
Finally, perturbative decoupling relations are invoked at a 
scale $\mu\approx M$ to estimate the ratio of $\Lambda$-parameters 
in the $\Nf$-flavor and pure Yang-Mills theory, that is, 
$\Lambda^{(\Nf=0)}_{\overline{\rm MS}}/
\Lambda^{(\Nf)}_{\overline{\rm MS}}$.
Putting these steps together, one obtains
$\Lambda^{(\Nf)}_{\overline{\rm MS}}/\mu_{\rm dec}$,
and given the physical value of $\mu_{\rm dec}$ finds
$\Lambda^{(\Nf)}_{\overline{\rm MS}}$. Considering
$\Nf=3$ or 4, once $\Lambda^{(\Nf)}_{\overline{\rm MS}}$
is determined one proceeds as usual and 
applies perturbative decoupling relations at the
charm and/or bottom quark-mass scale to estimate
$\Lambda^{(\Nf=5)}_{\overline{\rm MS}}$
and from it $\alphas(M_Z)$. 

The strategy has already been proven successful in the 
determination of 
$\Lambda^{(\Nf=3)}_{\overline{\rm MS}}$~\cite{DallaBrida:2019mqg}.
The ideas presented in this reference are however general and  
may be applied to solve other non-perturbative scale-dependent
renormalization problems that face analogous challenges.

The outline of this contribution is the following. 

We begin in Sect.~\ref{sec:PrecisionDeterminations} by recalling 
the main challenges in solving scale-dependent renormalization problems
on the lattice. The emphasis will be on the
determination of the QCD coupling. Besides introducing important 
concepts for later sections, the presentation gives us the 
opportunity to discuss some recent interesting determinations. 
These clearly illustrate the difficulties that state-of-the-art
computations of the coupling have to face in order to be 
significantly improved.

In Sect.~\ref{sec:Decoupling}, we introduce the theory of 
heavy-quark decoupling and present the results of several recent
studies that systematically assessed the size of non-perturbative 
effects induced by heavy quarks. More precisely, the accuracy of 
using perturbative decoupling relations to match the
$\Lambda$-parameters of different $\Nf$-flavor theories is
investigated, as well as the corrections due to the heavy quarks in
low-energy quantities. These studies not only set the foundation 
for the renormalization strategy based on decoupling, but also
establish the precision at which $\alphas$ can be obtained from
results in $\Nf=3$ QCD. 

In Sect.~\ref{sec:RenByDec}, the application of heavy-quark
decoupling to the determination of the $\Nf$-flavor QCD coupling 
is described in detail and the results of
ref.~\cite{DallaBrida:2019mqg} for 
$\Lambda^{(\Nf=3)}_{\overline{\rm MS}}$ are presented. 
We conclude in Sect.~\ref{sec:Conclusions} with some
comments on the future prospects for $\alphas$ determinations 
in view of this new strategy.

We care to note that it is not the aim of the present contribution 
to discuss in detail the many different lattice approaches that are
currently considered to determine the QCD coupling. 
In particular, we do not provide a complete account of all recent
determinations. For such a discussion, we refer the interested reader
to the comprehensive work of FLAG~\cite{Aoki:2019cca} and to other
interesting recent reviews (see
e.g.~refs.~\cite{DelDebbio:2021ryq,Komijani:2020kst}). 

% end

%% file: sect2.tex
% sect2.tex

\section{Precision determinations: the case of $\alphas$}
\label{sec:PrecisionDeterminations}

Before presenting the renormalization ideas based on decoupling, 
we believe it is important to put these into context. The aim of
these strategies, in fact, is not simply that of providing
alternative ways to solve non-perturbative scale-dependent
renormalization problems. The goal is to develop a framework that
will allow us to improve significantly our control over the current
most relevant uncertainties. In this section, we thus want to recall
what the main challenges are in solving this class of problems 
and which are the common approaches that are used to tackle them. 
Many of the concepts and observations that will be presented 
are for the most part well known. However, these issues are now 
more current than ever given the high precision that lattice QCD
calculations have achieved, in particular in the determination of
the QCD parameters. For this reason, we think it is important to
address them also here. This gives us the opportunity to 
discuss some new insight that has been gathered from several recent
high-precision studies, as well as introducing key concepts for
later sections.

As anticipated, the discussion will focus on the case of the 
QCD coupling $\alphas$. This allow us to analyze in easier terms 
the main challenges that we need to face in high-precision
non-perturbative determinations of RG runnings while capturing 
all the relevant issues. Moreover, lattice determinations of 
the strong coupling are a distinct case of competitive
calculations which have the potential to deliver unprecedentedly
precise results for a very relevant and fundamental quantity.
Making a significant progress over the present state-of-the-art
determinations by mere brute force, however, is extremely demanding
from the computational point of view. It is therefore mandatory 
to develop new strategies with the clear scope of improving our
control on all sources of uncertainty.
 
\subsection{Determinations of $\alphas$ on and off the lattice}

As already mentioned, since more than a decade lattice QCD is 
providing the Particle Physics community with the most accurate 
determinations of $\alphas$ (see
refs.~\cite{Aoki:2019cca,dEnterria:2019its,Zyla:2020zbs}).
The reason behind  this is that, as we shall recall, lattice
determinations have some important advantages over their
phenomenological counterparts (see e.g. 
refs.~\cite{Sommer:2015kza,Salam:2017qdl,DelDebbio:2021ryq} for 
some reviews).

Any determination of $\alphas$, whether on the 
lattice or not, relies on the following basic strategy. 
One considers a short-distance observable $\Obs(q)$ 
which depends on a characteristic energy scale $q$. 
In the limit where $q\to\infty$, this observable is
compared with its theoretical prediction, $\Obs_{\rm th}(q)$, 
in terms of a perturbative expansion%
\footnote{Note that in general the coupling $\alphas$ 
		  to be considered here should be the QCD coupling of 
		  the relevant $\Nf$-flavor theory, i.e.
		  $\alphas(\mu)\equiv
		  \alpha_{\overline{\rm MS}}^{(\Nf)}(\mu)$, from
		  which $\alpha_{\overline{\rm MS}}^{(5)}(M_Z)$
		  can eventually be extracted
		  (cf.~e.g.~ref.~\cite{Aoki:2019cca} and
		   Sect.~\ref{subsec:EffectiveCouplings}). At this stage, 
		  however, we prefer to keep the discussion simple. 
		  We shall return to this point later in detail.}
\begin{equation}
	\label{eq:Oth}
	\Obs_{\rm th}(q)=\sum_{n=0}^N k_n(s)\alphas^n(\mu)
	+{\rm O}(\alphas^{N+1})+{\rm O}\bigg({\Lambda^p\over q^p}\bigg),
	\
	\mu={q\over s}\,.
\end{equation}
The functions $k_n(s)$ appearing in this equation are the 
coefficient functions defining the perturbative series. They 
are known up to some order $N$ and depend on the 
scale factor, $s>0$, that relates the renormalization scale 
$\mu$ at which $\alphas$ is extracted with the scale $q$.
The basic difference between phenomenological and lattice
determinations of $\alphas$ is the choice of the observable
$\Obs(q)$.
 
Requiring $\Obs_{\rm th}(q)=\Obs(q)$ for some finite $q$, clearly 
fixes the value of $\alphas(\mu)$ only up to some error. This 
error comes from several different sources, many of which
are common to all types of determinations. First of all, 
there is the precision $\delta \Obs(q)$ to which the observable
$\Obs(q)$ is known. This of course depends on the relevant
statistical and systematic uncertainties associated with the
determination of $\Obs(q)$. Secondly, there is the effect of 
the truncation of the perturbative series to a given order, 
i.e.~the size of the  ${\rm O}(\alphas^{N+1})$ terms in
eq.~(\ref{eq:Oth}). In addition to these there are contaminations
from ``non-perturbative contributions''. These are represented in
eq.~(\ref{eq:Oth}) by power corrections to the perturbative expansion
of ${\rm O}(\Lambda^p/q^p)$, where $p>0$ and $\Lambda$ is some
characteristic non-perturbative scale of QCD.%
\footnote{Our knowledge of the form of non-perturbative
	effects is in fact rather limited. In addition,
	strictly speaking, perturbative and non-perturbative
	contributions cannot really be separated due to 
	the asymptotic nature of the series (see
	e.g.~ref.~\cite{Martinelli:1996pk}). However, as the 
	discussion is at this point qualitative we simply adopt 
	the simplistic representation of non-perturbative effects 
	as power corrections to the perturbative expansion.} 
Thus, regardless of the chosen strategy, an accurate 
determination of $\alphas$ needs to have, \emph{at least}, 
these general sources of error under control. Note that for 
the most part these are systematic in nature.

Lattice determinations of $\alphas$ are in principle favored
in several ways in succeeding at this task. Firstly, on
the lattice the QCD parameters are first renormalized 
in terms of some precisely measured hadronic quantities
(e.g.~hadron masses, decay constants, etc.), for which experimental
uncertainties typically contribute only marginally to the end
result. Once these are fixed, one has lots of freedom in choosing 
an observable $\Obs(q)$ as the getaway to extract $\alphas$.
One can therefore \emph{devise} convenient observables 
which have small statistical and systematic uncertainties;
in particular there is no need for these quantities to be 
accessible in experiments. Phenomenological determinations
of $\alphas$, instead, rely on experimental data for the 
observable $\Obs(q)$. It is the typical situation that when 
$q$ becomes large, and therefore many sources of systematic
uncertainty in eq.~(\ref{eq:Oth}) become small, the experimental
errors $\delta\Obs(q)$ become large. It is thus difficult to find
in general a single experimental quantity $\Obs(q)$ that allows 
one to accurately follow its scale dependence over a wide range 
of  $q$-values. On the lattice, on the other hand, if
\emph{carefully} chosen, $\Obs(q)$ can be computed precisely from
low- up to very high-energy scales. This gives a unique handle on
controlling non-perturbative corrections and the contribution of
the missing perturbative orders in eq.~(\ref{eq:Oth}). 

Another advantage for the lattice theorist is that $\Obs$ is 
defined within QCD alone. Consequently, the theoretical 
description $\Obs_{\rm th}$ of eq.~(\ref{eq:Oth}) does 
not need to include contributions besides those from QCD.
In addition, no modeling of hadronization is needed when
comparing the observable $\Obs$ with its perturbative prediction
$\Obs_{\rm th}$. Different is again the situation for  
phenomenological determinations. In these cases, other Standard 
Model (SM) contributions may be needed in order to extract
$\alphas$ and some modeling of hadronization is
necessary. Depending on the process, these are known only up to
some accuracy and typically depend on the value of other SM 
parameters as well. The precision one can aim for $\alphas$ can 
therefore be limited by these factors.% 
\footnote{Experimental observables are in principle sensitive also 
	to any New Physics. How this affects the extracted value 
	of $\alphas$, however, is something hard to assess.}
Of course, lattice QCD determinations are not entirely exempted 
from this issue. In this case, however, the problem of disentangling
the QCD contributions from ``the rest'' is confined to the hadronic 
quantities entering the renormalization of the theory, rather
than to the observable $\Obs$ itself. As mentioned earlier, 
the uncertainties on the hadronic quantities have, at present,
limited impact on the results for $\alphas$. 

All current lattice QCD determinations of $\alphas$, 
on the other hand, have to deal with the fact that their
calculations are performed with an unphysical number of
quark flavors. The bottom and typically also the charm quark 
are in fact not included in the simulations. This brings up the 
issue of having to account for their missing contributions. 
We shall leave this very important discussion aside for the 
moment and come back to it in detail in the following 
(see Sect.~\ref{sec:Decoupling}).

\subsection{The challenge of reaching high energy}
\label{subsec:ChallengesLattice}

Having for the most part presented the pros that lattice
determinations of $\alphas$ in principle have, we now 
address what the main difficulties are in practice. 

As any lattice QCD observable, besides the statistical
uncertainties, $\Obs(q)$ is affected by several systematics 
that need to be controlled. These include general ones, i.e.
discretization errors, finite-volume effects, an unphysical 
number of quarks, and quark-mass effects, as well as others 
which depend on the specific choice of $\Obs$ and set-up that 
we make (e.g.~excited-state contaminations, finite-temperature 
effects, Gribov copies, topology freezing, etc.). Finite 
quark-mass effects are typically not a relevant issue in
determinations of the QCD coupling~\cite{Aoki:2019cca}. As
anticipated, we then leave the problem of having an unphysical
quark-content for later. We also ignore observable specific issues.
Here we focus instead on discretization and finite-volume effects.
The combination of having the two under control, in fact, can
severely restrict the accessible range of $q$-values, if the
renormalization strategy is not carefully chosen. 

In particular, if one is determined in resolving within the same
lattice simulation both the hadronic energy scales relevant for 
the renormalization of the lattice theory, 
and the energy scale $q$ at which $\alphas$ is extracted, 
then one is necessarily limited by the simultaneous constraints:
\begin{equation}
	\label{eq:Window}
	L^{-1}\ll \Lambda
	\quad
	\textrm{and}
	\quad
	\Lambda\ll q\ll a^{-1}\,.
\end{equation}
The first inequality expresses the fact that finite-volume
effects must be under control in hadronic quantities. 
The infrared cutoff set by the finite extent $L$ 
of the lattice must be smaller than the typical 
non-perturbative scales of QCD, denoted here by 
$\Lambda$. The second inequality, instead, encodes 
two separate requirements. On the one hand,  the scale $q$ must 
be much lower than the ultraviolet cutoff set by the lattice 
spacing $a$. In this way, discretization errors in $\Obs(q)$ 
are kept under control, and $\Obs(q)$ can be obtained in 
the continuum limit with controlled errors.%
\footnote{Here and in the following we assume that 
		  $\Obs(q)$ is a properly renormalized lattice quantity 
		  with a well-defined continuum limit and free from
		  infrared divergences. Alternative strategies 
		  extract $\alphas$ from bare lattice quantities by
		  taking $q\propto a^{-1}$ and expressing their 
		  expansion in lattice perturbation theory in terms of 
		  $\alphas$ (see e.g.~ref.~\cite{Mason:2005zx}). In these
		  cases, discretization errors and scale dependence of
		  $\Obs(q)$ are entangled. Addressing systematic
		  uncertainties becomes more subtle and requires a separate 
		  discussion (see ref.~\cite{Aoki:2019cca}).}
On the other hand, $q$ needs to be much larger than the 
scales $\Lambda$. Only in this situation perturbation theory
can reliably be applied to extract $\alphas$ by comparing
$\Obs(q)$ with eq.~(\ref{eq:Oth}).

The typical lattices that are simulated today have sizes
$L/a\lesssim100$. Taking for definiteness $m_\pi L=4$, with
$m_\pi=140\,\MeV$, the first condition in eq.~(\ref{eq:Window})
implies that for such ensembles $q \ll a^{-1}\approx 3.5\,\GeV$. 
Of course this is a crude estimate and somewhat higher 
energies might be achieved by compromising at different 
stages of the calculation (e.g.~ considering heavier pion masses 
or smaller volumes in order to reach smaller values of $a$, or 
taking $aq\lesssim 1$). Nonetheless, it is clear that although
convenient in practice, considering lattices devised for studying
hadronic physics to compute short-distance quantities necessarily
poses severe challenges on the feasibility of the approach, as the
accessible energies $q$ are quite limited. 

As a concrete example of this situation, we can point to the recent
determination of $\alphas$ of ref.~\cite{Cali:2020hrj}. 
For their computation the authors employ state-of-the-art
large-volume simulations by the CLS
initiative~\cite{Bruno:2014jqa,Bali:2016umi}.
The two-point functions of both axial and vector currents at
short-distances are used to extract $\alphas$. Given the range of
lattice spacings at their disposal, $a\approx0.04-0.08\,\fm$, the
accessible energies for which continuum limit extrapolations 
could be performed in a controlled way are limited to 
$q\lesssim 2\,\GeV$.

As we shall see below with explicit examples, 
a limited range of low $q$-values unavoidably implies a limited
attainable precision for $\alphas$ as perturbative truncation
errors become a major issue.

\subsubsection{The finite volume is your friend}
\label{subsubsec:FV}

In order to reach high precision we must tailor the 
lattice simulations to the problem at hand. Going back to 
eq.~(\ref{eq:Window}), there is in fact no reason to try 
to satisfy simultaneously the two conditions as these 
belong to separate problems. On the one hand, there is the
determination of the low-energy quantities used for the 
hadronic renormalization of the lattice theory, while, on 
the other hand, there is the determination of $\Obs(q)$ 
for large $q$. 

A more natural strategy is therefore to split 
the problem over several sets of lattice simulations, each
one covering a different range of energy scales. In this way,
systematic effects can be more easily kept under control, 
as the relevant conditions will be milder for each individual 
set of simulations. The way to effectively achieve this in
practice is to employ what are known as \emph{finite-volume
renormalization schemes}~\cite{Luscher:1991wu}. In this case, 
the scale $q$ at which the observable $\Obs(q)$ is evaluated 
is identified with the inverse linear extent of the finite volume,
i.e. $q=L^{-1}$. One may say that, in fact, the observable $\Obs$
considered is a \emph{finite-volume effect}. With this choice, 
one computes the non-perturbative RG running of $\Obs(L^{-1})$ 
by simulating lattices with different physical extent $L$. 
This strategy goes under the name of \emph{finite-size 
(or step-)scaling}~\cite{Luscher:1991wu,Jansen:1995ck}
(see ref.~\cite{Sommer:2015kza} for a recent account). 

More precisely, having fixed the bare QCD parameters through
some hadronic quantities, one computes $\Obs(L^{-1}_{{\rm had}})$ 
at a low-energy scale $q_{\rm had}\equiv L^{-1}_{{\rm had}}
\approx\Lambda$, and determines $L_{{\rm had}}$ in physical 
units. This is achieved by computing 
$\lim_{a\to0}(am_{\rm had})(L_{\rm had}/a)={\rm O}(1)$, where 
$m_{\rm had}$ is a known low-energy scale. No large scale
separations are involved in this step, and at common bare
parameters one can satisfy the conditions $L_{\rm had}/a\gg1$
and $am_{\rm had}\ll1$, as well as having finite-volume effects 
in $m_{\rm had}$ under control. 

Secondly, one computes in the continuum limit the change in
$\Obs(L^{-1})$ as $L$ is varied by a known factor, say,
$L\to L/2$. This step is repeated a number of times $n$,
going from each new $L$ to the next one. Once the energy scale
reached, $q_n=2^n/L_{\rm had}$, is large compared to the 
hadronic scales, perturbation theory can safely be applied 
to extract $\alphas(\mu_{\rm PT}=q_n/s)$ from the value of 
$\Obs(q_n)$ (cf.~eq.~(\ref{eq:Oth})).

It is important to emphasize that, if carefully chosen, 
the only source of systematic errors that affect the 
determination of $\Obs(L^{-1})$ are discretization effects. 
In particular, no matter what the scale $q=L^{-1}$ is, 
discretization effects are under control once $L^{-1}=q\ll a^{-1}$,
i.e. $L/a\gg1$. This approach elegantly exploits the freedom 
that we have in lattice QCD in choosing the observable $\Obs(q)$ 
to completely circumvent the issue of necessarily having a 
finite volume. In particular, within this strategy the 
computational power is entirely invested into controlling 
a single systematic uncertainty.

In principle there is quite some freedom in choosing the 
finite-volume observable $\Obs(q)$. For the strategy to be
successful in practice, however, this should have a number of
desirable properties (see e.g.~ref.~\cite{Sommer:2015kza}). 
First of all, it should be easily and precisely measurable in 
Monte Carlo simulations. It should be computable in perturbation 
theory to a sufficiently high-loop order in order to guarantee 
good precision in extracting $\alphas$ through eq.~(\ref{eq:Oth}).
It should preferably be gauge invariant, in order to avoid issues
with Gribov copies once studied non-perturbatively, and also 
be directly measurable for zero quark masses (see below). Finally, 
it should have, in general, small lattice artifacts. In fact,
it is not straightforward to find a single observable that 
has all these nice features for any range of $q$-values. This,
however, is not a real issue as different observables may 
be accurately combined in order to cover all the relevant range of
scales $q$. We shall see explicit examples of good complementary
observables in forthcoming sections.

\subsubsection{$\Lambda$-parameters, $\beta$-functions, and all that}
\label{subsubsec:LambdaParams}

As we leave the general discussion for entering a more 
quantitative analysis of the challenges of extracting  
$\alphas$, we find convenient to reformulate the problem 
in slightly different terms. First of all, it is useful 
to associate with the observable $\Obs$ a renormalized 
coupling $\bar{g}_\Obs$ through the relation:
\begin{equation}
	\label{eq:alphaO}
	\alpha_\Obs(\mu)\equiv{\bar{g}^2_\Obs(\mu)\over4\pi} \equiv 
	{\Obs(\mu) -k_0\over k_1} \overset{\mu\to\infty}{=} 
	\alphas(\mu) + {k_2\over k_1}\alphas^{2}(\mu)+\ldots\,,
\end{equation}
where the coefficients $k_i\equiv k_i(1)$ are those appearing 
in the perturbative expansion, eq.~(\ref{eq:Oth}). This simple
procedure defines a non-perturbative, regularization-independent, 
QCD coupling. In terms of these couplings, the extraction of
$\alphas$  is interpreted as the perturbative matching between
$\alpha_{\Obs}$ and $\alphas$, where different observables $\Obs$
define different renormalization schemes. The common normalization
allows us to compare the value of the couplings in different 
schemes as they approach the high-energy limit. This is useful 
in assessing the regime of applicability of the perturbative 
matching as the latter can be characterized by the value that
$\alpha_{\Obs}$ should reach.  
 
We recall at this point that the non-perturbative couplings
studied within lattice QCD are implicitly defined for a given
number of quark flavors, $\Nf$. A more proper notation to use 
for the couplings is therefore $\alpha_{\Obs}^{(\Nf)}(\mu)\equiv
[\bar{g}^{(\Nf)}_\Obs(\mu)]^2/(4\pi)$, which emphasizes the fact
that $\bar{g}^{(\Nf)}_\Obs(\mu)$ must be considered as a coupling
within the $\Nf$-flavor theory. For ease of notation, however, 
in this section we will often take the liberty of dropping the
superscripts $\Nf$ and leave these understood, unless they
are needed to avoid any confusion or for later reference. 

Having this noticed, a particularly convenient class of schemes 
to consider are \emph{mass-independent} (or simply massless)
renormalization schemes~\cite{Weinberg:1951ss}. These are 
defined in terms of observables $\Obs$ evaluated for vanishing 
quark masses. As a result, the RG running of these couplings 
is decoupled from that of the renormalized quark masses and
therefore simpler to solve. On the other hand, differently 
from the physical case of massive schemes, quarks 
do not decouple in the RG running of massless
schemes~\cite{Bernreuther:1981sg,Bernreuther:1983zp}.  
Hence, in the $\Nf$-flavor theory the latter is characterized by 
a fixed number of active flavors corresponding to $\Nf$.%
\footnote{We shall return on these points in more detail 
		  in Sect.~\ref{sec:Decoupling}.} 

To the coupling $\bar{g}^{(\Nf)}_\Obs$ in a given 
mass-independent scheme we can associate a quark-mass independent
$\Lambda$-parameter, $\Lambda^{(\Nf)}_\Obs$, defined as,
\begin{equation}
	\label{eq:LambdaParam}
	\begin{split}
	\Lambda^{(\Nf)}_\Obs&=\mu\,\varphi^{(\Nf)}_{\rm g,\Obs}
	(\bar{g}^{(\Nf)}_\Obs(\mu))\,, \\[1ex]
	\varphi^{(\Nf)}_{\rm g,\Obs}(\bar{g})&=
	(b_0(\Nf)\bar{g}^2)^{-{b_1(\Nf)\over 2b_0(\Nf)^2}}\,
	e^{-{1\over 2b_0(\Nf)\bar{g}^2}}\times\\[1ex]
	&\hspace*{-10.5mm}\times
	\exp\bigg\{
	-\int_{0}^{\bar{g}} 
	\rmd g \bigg[{1\over\beta^{(\Nf)}_\Obs(g) } + 
	{1\over b_0(\Nf) g^3} - {b_1(\Nf)\over b_0(\Nf)^2 g}\bigg]\bigg\}\,.
	\end{split}
\end{equation}
Through this relation, the value of the coupling
$\bar{g}^{(\Nf)}_\Obs(\mu)$ at any renormalization scale 
$\mu$ is in one-to-one correspondence with $\Lambda^{(\Nf)}_\Obs$,
provided that the $\beta$-function,
\begin{equation}
	\label{eq:BetaFunDef}
	\beta^{(\Nf)}_\Obs(\bar{g})
	\equiv\mu{\rmd\bar{g}^{(\Nf)}_\Obs(\mu)\over\rmd \mu}\bigg|_{\bar{g}}\,,
\end{equation}
is known. The $\beta$-function describes the dependence of the 
coupling on the renormalization scale. In perturbation theory,
it has an expansion which at the $N$-loop order reads:
\begin{equation}
	\label{eq:BetaFunctionPT}
	\beta_\Obs^{(\Nf)\,{\rm PT}}(\bar{g})\equiv
	-\bar{g}^3\sum_{k=0}^{N-1} b_k(\Nf)\,\bar{g}^{2k}\,,
\end{equation}
with
\begin{gather}
	\nonumber
	b_0(\Nf)={1\over(4\pi)^{2}}\bigg(11-{2\over 3}\Nf\bigg)\,,\\
	\label{eq:b0b1}
	b_1(\Nf)={1\over(4\pi)^{4}}\bigg(102-{38\over 3}\Nf\bigg)\,.
\end{gather}
The coefficients $b_0(\Nf),b_1(\Nf)$ are universal and shared 
by all mass-independent renormalization schemes. The scheme 
dependence only enters through the higher-order coefficients,
$b_i(\Nf)\equiv b^\Obs_i(\Nf)$, with $i\geq 2$. 

A first compelling property of $\Lambda$-parameters 
is that, differently from the case of couplings, 
their scheme dependence is in fact trivial. Leaving the 
$\Nf$-dependence implicit and taking the $\Lambda$-parameter 
in the $\overline{\rm MS}$-scheme, $\Lambda_{\overline{\rm MS}}$, 
as reference, we have that any other $\Lambda$-parameter 
$\Lambda_{\Obs}$ is \emph{exactly} related to
$\Lambda_{\overline{\rm MS}}$ by the relation:
\begin{equation}
	\label{eq:MStoOLambda}
	{\Lambda_{\overline{\rm MS}}\over\Lambda_{\Obs}}= 
	\exp\bigg\{{c_1(1)\over 2b_0}\bigg\}\,.
\end{equation}
In this equation $c_1(1)$ is the 1-loop coefficient of 
the perturbative matching relation between the 
corresponding couplings, which at $M$-loop order reads,
\begin{equation}
	\label{eq:MStoOCoupling}
	\bar{g}^2_{\overline{\rm MS}}(s\mu)=
	\bar{g}^{2}_{\Obs}(\mu)
	\Bigg(1+
	\sum_{k=1}^{M}
	c_k(s) \bar{g}^{2k}_{\Obs}(\mu)\Bigg)\,,
	\quad
	s>0\,.
\end{equation}
It is clear that the $\Lambda$-parameters are non-perturbatively
defined once the corresponding couplings and $\beta$-functions are.%
\footnote{Interestingly, eq.~(\ref{eq:MStoOLambda}) provides an
		  indirect non-perturbative definition of
		  $\Lambda_{\overline{\rm MS}}$ through the
		  $\Lambda$-parameter of any non-perturbative
  	  	  scheme.}
In addition, they are \emph{exact} solutions of the Callan-Symanzik
equations~\cite{Callan:1970yg,Symanzik:1970rt,Symanzik:1971vw}, 
and therefore RG invariants, i.e. $\rmd\Lambda_\Obs/\rmd\mu=0$. 
From a non-perturbative perspective this makes the
$\Lambda$-parameters natural quantities to compute, as they 
provide reference scales for both the low- and high-energy
regimes of QCD. We finally stress that, as their corresponding
couplings, the $\Lambda$-parameters are defined for a given 
number of quark flavors, $\Nf$ (cf.~eq.~(\ref{eq:LambdaParam})).
Each $\Nf$-flavor theory therefore has its own $\Lambda$-parameters.

\subsubsection{Systematic uncertainties in extracting
			  $\Lambda^{(\Nf)}_{\overline{\rm MS}}$}
\label{subsec:SystematicErrorsPT}
		  
Once the overall energy scale of the given $\Nf$-flavor 
theory has been set,%
\footnote{The issue of fixing the scale of a theory 
		  with an unphysical value of $\Nf$ will be addressed 
		  in Sect.~\ref{sec:Decoupling}.}
the non-perturbative value of the coupling, 
$\bar{g}_{\rm  PT}\equiv\bar{g}_\Obs(\mu_{\rm PT})$, 
in any scheme, at some high-energy scale $\mu_{\rm PT}$, 
allows for estimating $\Lambda_{\overline{\rm MS}}^{(\Nf)}$. 
Below we present two commonly employed strategies.

\paragraph{Strategy 1.} Using eq.~(\ref{eq:LambdaParam})
we first get an \emph{asymptotic} estimate for
$\Lambda_{\Obs}/\mu_{\rm {PT}}$ as,
\begin{equation}
	\label{eq:LambdaMuPT}
	{\Lambda_{\Obs}\over\mu_{\rm PT}}
	=
	\varphi_{\rm g,\Obs}(\bar{g}_{\rm PT})
	\overset{\bar{g}_{\rm PT}\to0}{\approx}
	\varphi^{\rm PT}_{\rm g,\Obs}
	(\bar{g}_{\rm PT})+	
	{\rm O}\big(\bar{g}_{\rm PT}^{2N-2}\big)\,.
\end{equation}
In this relation, $\varphi^{\rm PT}_{\rm g,\Obs}$ is defined
analogously to $\varphi_{\rm g,\Obs}$ of eq.~(\ref{eq:LambdaParam})
but with the replacement $\beta_\Obs\to\beta_\Obs^{\rm PT}$,
with $\beta_\Obs^{\rm PT}$ the perturbative $\beta$-function
of eq.~(\ref{eq:BetaFunctionPT}).  From the estimate of
${\Lambda_{\Obs}/\mu_{\rm PT}}$, 
${\Lambda_{\overline{\rm MS}}/\mu_{\rm PT}}$
is obtained, with no further approximation, using
eq.~(\ref{eq:MStoOLambda}). Finally, the knowledge of 
$\mu_{\rm {PT}}$ in physical units gives us
$\Lambda_{\overline{\rm MS}}$.

As anticipated by eq.~(\ref{eq:LambdaMuPT}), due to the 
truncation of $\beta_\Obs$ to $N$-loop order in the 
evaluation of ${\Lambda_{\Obs}/\mu_{\rm PT}}$, our estimate 
for ${\Lambda_{\overline{\rm MS}}/\mu_{\rm PT}}$, comes 
with a systematic uncertainty of 
${\rm O}\big(\bar{g}_{\rm PT}^{2N-2}\big)$. 
It is important to stress that eq.~(\ref{eq:LambdaMuPT})
is in fact only asymptotic. Similarly to what we discussed 
about eq.~(\ref{eq:Oth}), our estimate for 
${\Lambda_{\overline{\rm MS}}/\mu_{\rm PT}}$ is in principle 
also affected by ``non-perturbative corrections'', if the 
coupling $\bar{g}_{\rm PT}$ is not small enough for 
perturbation theory to be in the regime of applicability.
We shall come back to this issue shortly.

\paragraph{Strategy 2.} A second possibility to estimate
$\Lambda_{\overline{\rm MS}}$ is to first obtain from 
the non-perturbative value of $\bar{g}_{\rm PT}$ an estimate 
for $\bar{g}_{\overline{\rm MS}}(s\mu_{\rm {PT}})$ using 
the $M$-loop relation, eq.~(\ref{eq:MStoOCoupling}).
Given this, we can estimate 
${\Lambda_{\overline{\rm MS}}/(s\mu_{\rm PT}})$
using eq.~(\ref{eq:LambdaParam}) and the perturbative 
expression for the $\beta$-function, eq.~(\ref{eq:BetaFunctionPT}), 
in the $\overline{\rm MS}$-scheme. As before, the knowledge of
$\mu_{\rm {PT}}$ in physical units then gives us
$\Lambda_{\overline{\rm MS}}$.

In order to establish the perturbative uncertainties associated
with this second approach, we first recall that the 
$\beta$-function in the $\overline{\rm MS}$-scheme is currently
known to 5-loop 
order~\cite{Baikov:2016tgj,Luthe:2016ima,Herzog:2017ohr}. 
This introduces a systematic uncertainty in the determination of 
${\Lambda_{\overline{\rm MS}}/\mu_{\rm PT}}$ of 
${\rm O}\big(\bar{g}^8_{\overline{\rm MS}}(s\mu_{\rm {PT}})\big)
\approx{\rm O}\big(\bar{g}_{\rm PT}^{8}\big)$
(cf.~eqs.~(\ref{eq:LambdaMuPT}) and (\ref{eq:MStoOCoupling})). 
Secondly, using eq.~(\ref{eq:LambdaParam}) it is 
easy to show that the perturbative matching at $M$-loop 
order between $\bar{g}_{\rm PT}$ and 
$\bar{g}_{\overline{\rm MS}}(s\mu_{\rm {PT}})$
translates into a systematic uncertainty in 
${\Lambda_{\overline{\rm MS}}/\mu_{\rm PT}}$
of ${\rm O}\big(\bar{g}_{\rm {PT}}^{2M}\big)$. 

For all schemes $\bar{g}_{\Obs}$ used in lattice QCD 
determinations of $\Lambda_{\overline{\rm MS}}$ we have 
that $M\leq 3$ (see e.g.~ref.~\cite{Aoki:2019cca}). 
The systematic errors coming from the matching between 
couplings is therefore parametrically larger than the one 
from the truncation of $\beta_{\overline{\rm MS}}$.
Moreover, note that for all these schemes we have that 
$M=N-1$, where $N$ is the loop order at which the 
corresponding perturbative $\beta$-function, $\beta^{\rm PT}_\Obs$,
is known (cf.~eq.~(\ref{eq:BetaFunctionPT})).%
\footnote{This is the case because for all these schemes
		  the $N$-loop $\beta$-function, $\beta^{\rm PT}_\Obs$,
		  has been inferred from $\beta^{\rm PT}_{\overline{\rm MS}}$
		  at $N$-loops using the matching relation 
		  eq.~(\ref{eq:MStoOCoupling}) and 
		  $\beta^{\rm PT}_\Obs$ at $(N-1)$-loop order.}
This means that, for the schemes $\bar{g}_\Obs$ commonly used, 
\textbf{Strategy 1.} and \textbf{2.} result in the \emph{same}
parametric uncertainties of 
${\rm O}\big(\bar{g}_{\rm {PT}}^{2N-2}\big)$. 
Clearly, although parametrically the same, the actual 
size of the corrections might be different. 
In this second strategy, in particular, when matching the
couplings we have the freedom  to choose the parameter $s$
(cf.~eq.~(\ref{eq:MStoOCoupling})). Different choices can
result in different perturbative corrections to
${\Lambda_{\overline{\rm MS}}/\mu_{\rm PT}}$.\\

Devising different strategies like the ones above and 
comparing their outcome can help us assessing the systematic
uncertainties in $\Lambda_{\overline{\rm MS}}$ coming from 
the use of perturbation theory at $\mu_{\rm {PT}}$. A truly 
systematic study, however, requires to compare the determination 
of $\Lambda_{\overline{\rm MS}}/\mu_{\rm ref}$, where
$\mu_{\rm ref}$ is a common reference scale, for several 
different values of $\bar{g}_{\rm PT}$ as $\bar{g}_{\rm PT}\to0$. 
Only if agreement is found among all determinations, possibly 
including different strategies, one may be reassured that 
${\rm O}\big(\bar{g}_{\rm PT}^{2N-2}\big)$ terms, as well
as non-analytic terms in the coupling,
are negligible within the statistical uncertainties. 
In the case where, instead, the results for 
$\Lambda_{\overline{\rm MS}}/\mu_{\rm ref}$ show a clear 
dependence on $\bar{g}_{\rm PT}$, one should first confirm that
this is actually compatible with the expected 
${\rm O}\big(\bar{g}_{\rm PT}^{2N-2}\big)$
corrections. If this is the case, one may be confident 
that the asymptotic regime of the perturbative expansion 
is reached, no ``non-perturbative corrections'' are relevant, 
and one can therefore take as final estimate for 
$\Lambda_{\overline{\rm MS}}/\mu_{\rm ref}$ the extrapolated 
result for $\bar{g}_{\rm PT}\to0$.

Clearly, the program above is ambitious. The
running of the coupling $\bar{g}_{\rm PT}$ at high energies 
is only logarithmic in $\mu_{\rm PT}/\Lambda_\Obs$. Reducing 
the size of the perturbative truncation errors by a given factor 
hence requires an \emph{exponentially} larger change in the 
energy scale $\mu_{\rm {PT}}$. In order to accurately estimate 
the systematic uncertainties coming from the use of perturbation
theory one therefore needs to cover, non-perturbatively, a wide 
range of energies, reaching up to very high scales. 

If the chosen strategy to determine $\Lambda_{\overline{\rm MS}}$ 
does not allow for this and the accessible range of 
$\bar{g}_{\rm PT}$ is quite limited, one might be tempted 
to estimate the uncertainties due to the application of
perturbation theory in more simplistic ways. For example, 
one might opt for simply adding to the final result an uncertainty
$\delta\Lambda_{\overline{\rm MS}}/\Lambda_{\overline{\rm MS}}=
k\,\bar{g}_{\rm PT}^{2N-2}$, where $k$ is estimated
in some way from the available perturbative information.
Alternatively, one might estimate these uncertainties based on 
the spread of the results obtained at the smallest available
$\bar{g}_{\rm {PT}}$ from different strategies
(e.g. \textbf{Strategy 1.} vs \textbf{Strategy 2.}).
Given the asymptotic nature of the perturbative expansion, 
however, these practices cannot be considered reliable in
general. From the very definition of asymptotic series the 
only reliable way to assess its accuracy is to compare the 
series with the full function as $\bar{g}_{\rm PT}\to0$.%
\footnote{We recall that a series is said to be asymptotic 
		  to the function $f(\lambda)$, $\lambda\in\mathbb{R}$, 
		  if: $|f(\lambda)-\sum_{j=0}^N a_j\lambda^j\big| 
		  \overset{\lambda\to0}{\rightarrow}{\rm O}(\lambda^{N+1})$,
		  $\forall N$. Note in particular that at fixed $\lambda$,
		  larger $N$ does not necessarily imply a better
		  approximation of the series to the function.}
In order to do so, the coupling $\bar{g}_{\rm PT}$ must be varied 
by a sensible amount reaching down to small values. 

For the same reasons, it is not advisable to 
estimate the size of ``non-perturbative corrections'' using
some model assumption, or use some model to extrapolate
the results for $\Lambda_{\overline{\rm MS}}/\mu_{\rm ref}$  
to $\bar{g}_{\rm PT}\to0$. Our knowledge of the form of
non-perturbative effects is rather limited and the separation
between what is perturbative and non-perturbative is all but 
well defined. Hence, it is always debatable whether any model 
that tries to capture non-analytic terms in the coupling is 
really adequate to describe the data within the given accuracy.
Moreover, if the coupling $\bar{g}_{\rm PT}$ cannot be varied 
much, it is difficult to really distinguish, e.g. 
a power correction, from some higher-order term in 
$\bar{g}_{\rm PT}$, when statistical errors and other 
uncertainties are present. A more reliable practice is thus 
to avoid regions of large $\bar{g}_{\rm PT}$ where the 
${\rm O}\big(\bar{g}_{\rm PT}^{2N-2}\big)$ behavior has not 
clearly set in.

\subsection{The accuracy of perturbation theory at high energy}
\label{subsec:AccuracyPT}

In this section we want to review some recent
determinations of $\Lambda_{\overline{\rm MS}}$
which paid particular attention to the issue of 
the accuracy of perturbation theory in 
extracting $\Lambda_{\overline{\rm MS}}$~\cite{Brida:2016flw,DallaBrida:2018rfy,DallaBrida:2019wur,Husung:2020pxg,Nada:2020jay}. As we shall see,
the concerns exposed in the previous sections are
legitimate once the precision goals become competitive.
A robust analysis of perturbative truncation errors 
is essential to reach high accuracy.

\subsubsection{The high-energy regime of $\Nf=3$ QCD}
\label{subsubsec:AccuracyNf3}

\begin{figure}
	\centering
	%	\vspace*{5cm}      
	\resizebox{0.5\textwidth}{!}{%
	\includegraphics{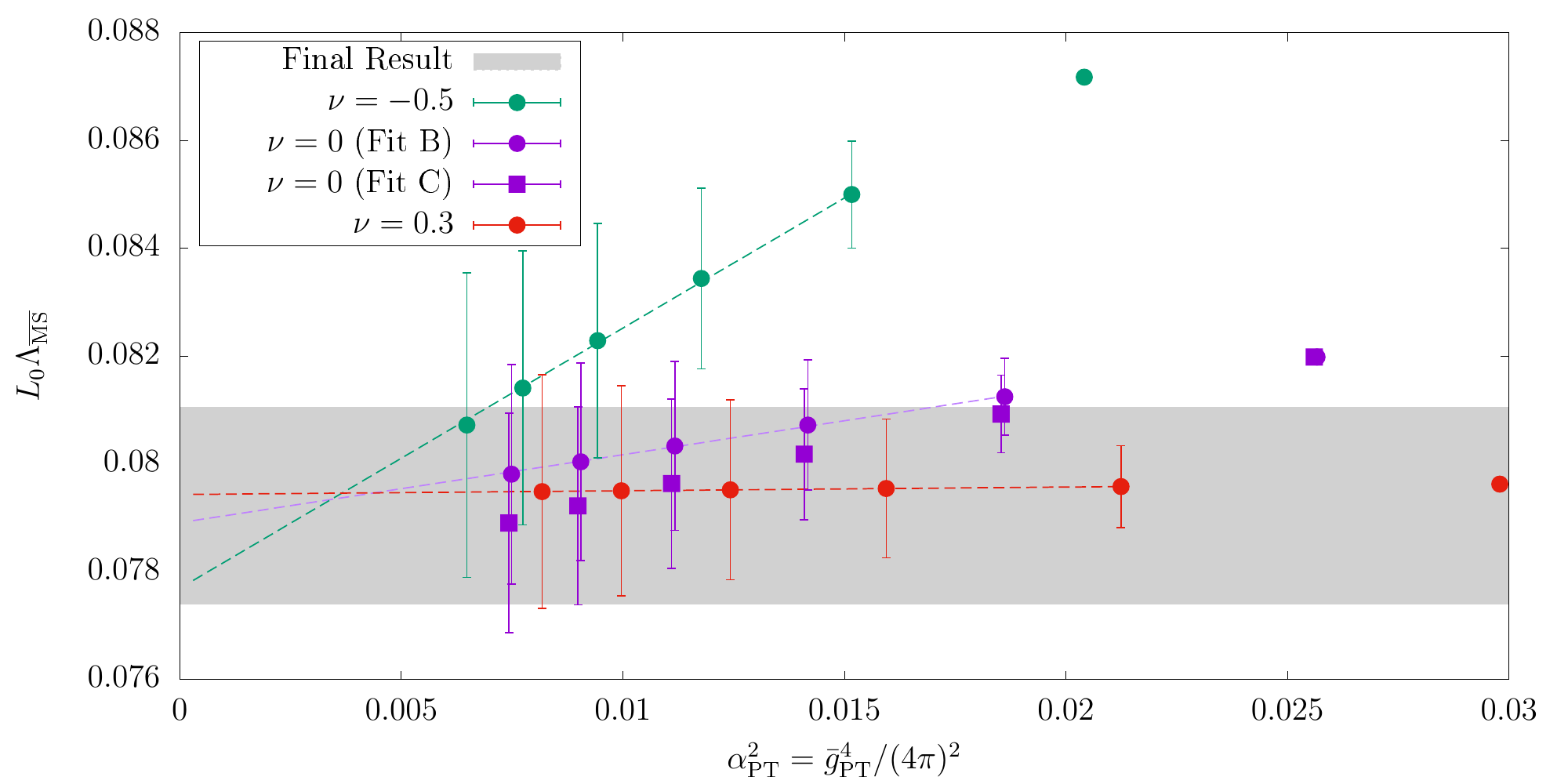}}
	\caption{Determination of $\Lambda_{\overline{\rm MS}}^{(3)}$ 
	in units of $\mu_{\rm ref}=L_0^{-1}$ for different values 
	of $\alpha_{\rm PT}$~\cite{DallaBrida:2018rfy}. 
	The	extraction in different ${\rm SF}_\nu$-schemes
	$(\nu=-0.5,0,0.3)$ is shown, as well as a comparison with 
	the final result $\Lambda^{(3)}_{\overline{\rm MS}}/
	\mu_{\rm ref}=0.0791(19)$~\cite{DallaBrida:2018rfy}.  As 
	the reader can see, when the extraction is performed at 
	high-enough energies $(\alpha_{\rm PT}\approx0.1)$, all 
	schemes nicely agree.}
	\label{fig:AccuracyNf3}       
\end{figure}

We begin with the high-energy studies of 
refs.~\cite{Brida:2016flw,DallaBrida:2018rfy} in $\Nf=3$ QCD. 
In Fig.~\ref{fig:AccuracyNf3}, we show the results from 
these references for $L_0\Lambda^{(\Nf=3)}_{\overline{\rm MS}}$ 
as a function of 
$\alpha^2_{\rm PT}\equiv\bar{g}^4_{\rm PT}/(4\pi)^2$. 
In this plot, $\mu_{\rm ref}\equiv L_0^{-1}\approx 4.3\,\GeV$,
is a convenient high-energy reference scale
and, as in previous sections, $\bar{g}_{\rm PT}\equiv\bar{g}_{\Obs}(\mu_{\rm {PT}})$
is the value of the coupling in the given scheme
at which perturbation theory is applied to extract
$\Lambda^{(3)}_{\overline{\rm MS}}/\mu_{\rm ref}$.
The ratio $\Lambda^{(3)}_{\overline{\rm MS}}/\mu_{\rm ref}$
is obtained following \textbf{Strategy 1.} of 
Sect.~\ref{subsec:SystematicErrorsPT}. 
The scales at which perturbation theory is used
correspond to $\mu_{\rm {PT}}=2^n\mu_{\rm ref}$, with
$n=0,\ldots,5$, and range from about $4\,\GeV$ to $140\,\GeV$. 
The couplings considered in this study,
$\bar{g}_\Obs(\mu)=\bar{g}_{{\rm SF}_\nu}(L^{-1})$,
belong to a family of finite-volume renormalization schemes
based on the QCD Schr\"odinger functional
(SF)~\cite{Luscher:1992an,Sint:1993un,Sint:1995rb}.
Different schemes within the family are identified by
different values of the parameter $\nu$. The precise 
definition of the schemes is not important and 
can be found in
refs.~\cite{Brida:2016flw,DallaBrida:2018rfy,Sint:2012ae}.%
\footnote{Traditionally only the $\nu=0$ scheme has been 
		  considered in applications, see
		  e.g.~refs.~\cite{Luscher:1993gh,Sint:1995ch,Bode:1998hd,Bode:1999sm,DellaMorte:2004bc,Aoki:2009tf} for some important examples.} 

In order to estimate $\Lambda^{(3)}_{\overline{\rm MS}}/\mu_{\rm ref}$ the 3-loop approximation to the relevant $\beta$-functions,
$\beta_{{\rm SF}_\nu}$, is used. The results for
$\Lambda^{(3)}_{\overline{\rm MS}}/\mu_{\rm ref}$
are therefore expected to show corrections of ${\rm O}(\alpha_{\rm PT}^2)$ as $\alpha_{\rm PT}\to0$. It is important to note at 
this point that the 3-loop coefficients of the $\beta$-functions 
in the different ${\rm SF}_\nu$-schemes are given for $\Nf=3$
by~\cite{Bode:1998hd,Bode:1999sm,DallaBrida:2018rfy}:%
\footnote{For comparison the 3-loop coefficient of
	      $\beta_{\overline{\rm MS}}$ for $\Nf=3$
	      is $(4\pi)^3 b_2^{\overline{\rm MS}}= 0.324$.}
\begin{equation}
	(4\pi)^3 b_{2}^{{\rm SF}_\nu}=
 	-(0.064(27) +\nu\times 1.259(10))\,.
\end{equation}
Hence, from the perturbative point of view all schemes 
with $|\nu|\lesssim 1$ appear to be on similar footing
and the perturbative expansion of their $\beta$-functions 
is well behaved. 

From this observation, one might naively expect that the  
${\rm O}(\alpha_{\rm PT}^2)$
corrections to $\Lambda^{(3)}_{\overline{\rm MS}}/
\mu_{\rm ref}$ obtained from different intermediate
schemes are similar too. Going back to Fig.~\ref{fig:AccuracyNf3},
we see that in all cases the results are well described 
by a ${\rm O}(\alpha_{\rm PT}^2)$ dependence over the whole
range of investigated couplings. This is compatible with the
expectation from the known leading non-analytic term in the
expansion which is expected to be quite small at these couplings, 
i.e. ${\rm O}(e^{-2.6/\alpha})$~\cite{DallaBrida:2018rfy}.
However, we clearly see a substantial difference in the 
size of the ${\rm O}(\alpha_{\rm PT}^2)$ corrections depending 
on the ${\rm SF}_\nu$-scheme that is considered. 
While one can find cases ($\nu=0.3$) where the 
${\rm O}(\alpha_{\rm PT}^2)$ corrections are insignificant 
within errors, other schemes ($\nu=-0.5$) show significant
corrections. The results for $\nu=-0.5$ at 
$\alpha_{\rm PT}\approx0.12$, for instance, show a 
$7-8\%$ deviation from the final result
$\Lambda^{(3)}_{\overline{\rm MS}}/\mu_{\rm ref}=0.0791(19)$  
quoted in ref.~\cite{DallaBrida:2018rfy}, which
corresponds to the gray band in the plot.

\begin{figure*}[!h]
	%	\vspace*{5cm}       % Give the correct figure height in cm
	\centering
	\resizebox{0.4\textwidth}{!}{%
	\includegraphics{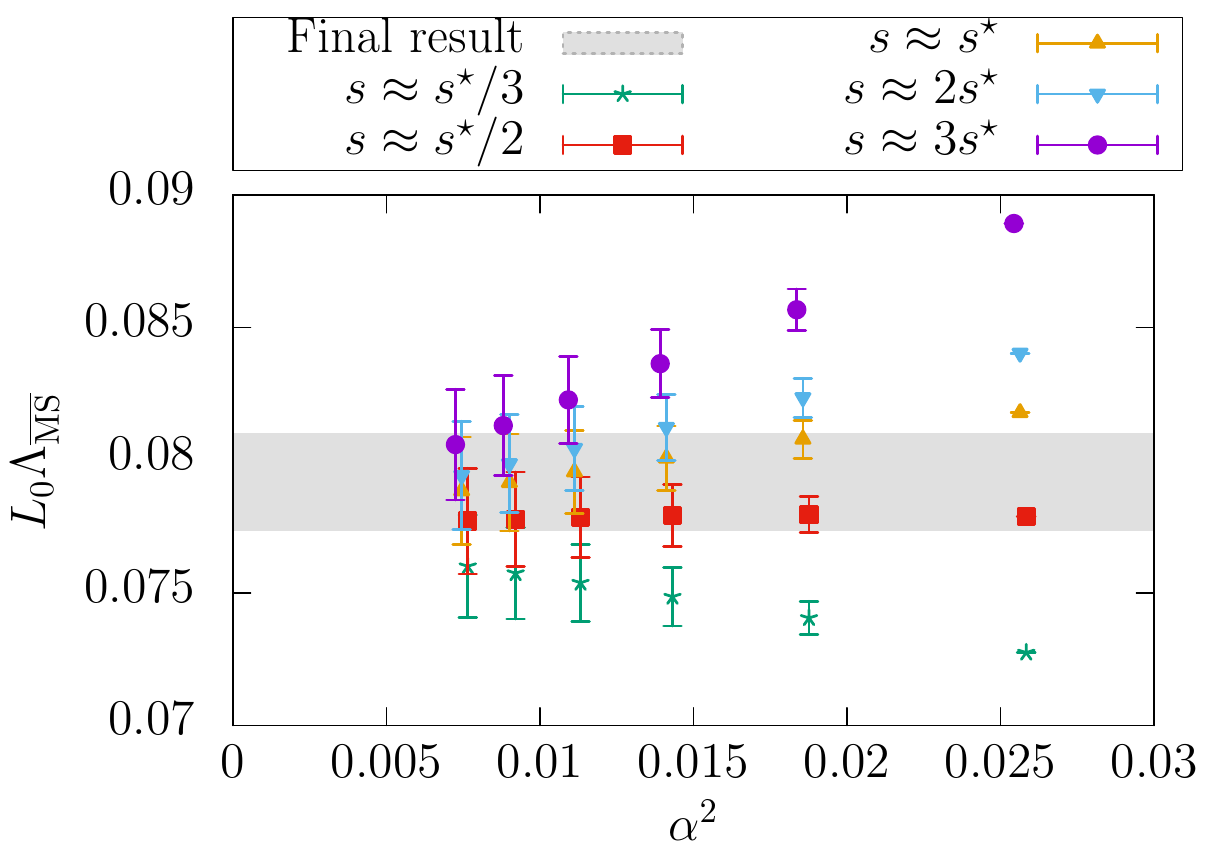}}
	\hspace*{10mm}
	\resizebox{0.4\textwidth}{!}{%
	\includegraphics{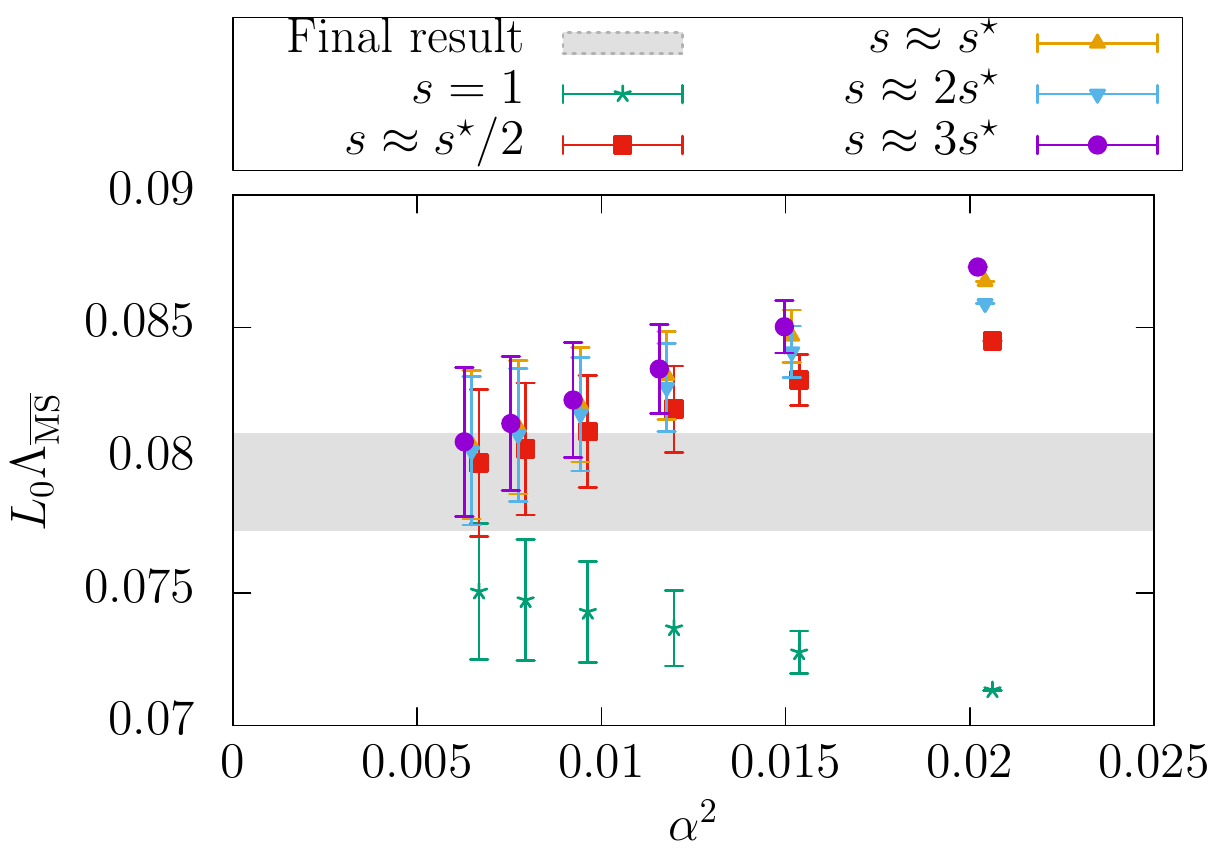}}
	\caption{Determination of $L_0\Lambda_{\overline{\rm MS}}^{(3)}=
	\Lambda_{\overline{\rm MS}}^{(3)}/\mu_{\rm ref}$
	at different values of $\alpha\equiv\alpha_{\rm PT}$, and using
	different renormalization scales (values of $s$) to match with
	the $\overline{\rm MS}$-scheme~\cite{DallaBrida:2018rfy}. 
	The left (right) panel uses the ${\rm SF}_\nu$-scheme with 
	$\nu= 0$ ($\nu=-0.5$), cf. text.}
	\label{fig:ScaleChangeNf3}
\end{figure*}

As the value of the coupling at which perturbation
theory is applied becomes smaller, any significant difference 
among the different determinations of $\Lambda^{(3)}_{\overline{\rm MS}}/\mu_{\rm ref}$ steadily fades away. In particular, 
once $\alpha_{\rm PT}\approx0.1$ is reached, any difference
is well below the statistical uncertainties on
$\Lambda^{(3)}_{\overline{\rm MS}}/\mu_{\rm ref}$,
which at these couplings are at the level of $2-3\%$.
A robust estimate for 
$\Lambda^{(3)}_{\overline{\rm MS}}/\mu_{\rm ref}$
can therefore be obtained by taking its value at 
$\alpha_{\rm PT}\approx0.1$ from one of the schemes 
that show milder perturbative corrections.
The result  $\Lambda^{(3)}_{\overline{\rm MS}}/\mu_{\rm ref}=0.0791(19)$ given above, for instance, corresponds 
to $\mu_{\rm PT}=2^4\mu_{\rm ref}\approx70\,\GeV$ 
from the $\nu=0$ scheme (cf.~$\nu=0$ (fit C) in
Fig.~\ref{fig:AccuracyNf3})  \cite{DallaBrida:2018rfy}.

The first important message from this study is that
it is in fact impossible to predict the actual size 
of perturbative truncation errors only from the 
available perturbative information. To reliably assess
these errors, perturbation theory must be tested against 
non-perturbative data over a wide range of energy scales. 
From the study we presented, in particular, we conclude that 
in order to be able to quote in full confidence the competitive 
precision of $2-3\%$ on $\Lambda^{(3)}_{\overline{\rm MS}}/
\mu_{\rm ref}$, one must reach non-perturbatively 
$\alpha_{\rm PT}\approx0.1$. At these couplings perturbative
truncation errors are fully under control and the error on
$\Lambda^{(3)}_{\overline{\rm MS}}/\mu_{\rm ref}$ is entirely
dominated by the statistical uncertainties coming from the
non-perturbative running of the coupling.

It is now instructive to look at the result of the analysis 
of the same data according to \textbf{Strategy 2} 
of Sect.~\ref{subsec:SystematicErrorsPT}.
The corresponding estimates for 
$\Lambda^{(3)}_{\overline{\rm MS}}/
\mu_{\rm ref}$ are shown in Fig.~\ref{fig:ScaleChangeNf3}
for the two cases, $\nu=0,-0.5$~\cite{DallaBrida:2018rfy}. 
The different determinations in each plot are obtained by 
varying the parameter $s$ entering the perturbative matching 
between the $\overline{\rm MS}$-coupling
and the ${\rm SF}_\nu$-couplings
(cf.~eq.~(\ref{eq:MStoOCoupling})). The values of $s$ considered
vary by about a factor $2-3$ around the value of 
\emph{fastest apparent convergence}, $s^*$.%
\footnote{The scale factor $s^*$ is defined by 		
		 (cf.~eq.~(\ref{eq:MStoOLambda})): 
		 \[
		 	s^*={\Lambda_{\overline{\rm MS}}\over\Lambda_{\Obs}}= 
		 	\exp\bigg\{{c_1(1)\over 2b_0}\bigg\}\,.
		 \]
		 where $b_0$ is the 1-loop coefficient of 
		 the $\beta$-function in eq.~(\ref{eq:b0b1}), and	 		 
		 $c_1(1)$ is the 1-loop coefficient of the matching
		 relation between the $\overline{\rm MS}$-coupling
		 $\bar{g}^2_{\overline{\rm MS}}(\mu)$ and the coupling
		 of interest $\bar{g}^2_\Obs(\mu)$ 
		 (cf.~eq.~(\ref{eq:MStoOCoupling})). With the choice $s=s^*$,
		 the $k=1$ term in eq.~(\ref{eq:MStoOCoupling}) vanishes.}
In phenomenological determinations of the QCD coupling the 
spread of the results obtained by varying $s$ around some 
``optimal'' value, typically by a factor $2$  or so, 
is commonly used to get an estimate of 
perturbative truncation errors (see e.g.~ref.~\cite{Zyla:2020zbs}).
Our intention is to test how this approach works in the
present case.

As one can see from Fig.~\ref{fig:ScaleChangeNf3}, for all choices 
of $s$ the data show the expected ${\rm O}(\alpha_{\rm PT}^2)$
scaling. The slope of the data, however, can vary 
significantly depending on the choice of the parameter $s$. 
As expected, the significance of these differences is reduced 
as $\alpha_{\rm PT}\to0$, and the different determinations come
together once $\alpha_{\rm PT}\lesssim0.1$.

What is clear from the results of
Fig.~\ref{fig:ScaleChangeNf3} is that the procedure of assigning 
a systematic error based on the spread of the results with $s$
at some fixed coupling is not always reliable. In the case of the
$\nu=0$ scheme (left panel), the spread in the results between,
say, $s^*/2$ and $2s^*$,  encloses the final estimate (gray band 
in the plot) for all coupling values in the range. If this
uncertainly was added to the statistical errors, it would give 
a conservative estimate for the total uncertainly. On the other
hand, in the case of the $\nu=-0.5$ scheme (right panel), the
procedure significantly underestimates the actual size of the 
${\rm O}(\alpha_{\rm PT}^2)$ corrections. Again
$\alpha_{\rm PT}\approx0.1$ has to be reached for the perturbative
uncertainties to be small compared to the statistical ones.

From this second analysis we reaffirm the conclusion that
it is very difficult to reliably estimate perturbative
truncation errors if the coupling $\alpha_{\rm PT}$ cannot be 
varied much, and if this is confined to values significantly 
larger than $\alpha_{\rm PT}\approx0.1$.

\subsubsection{The case of the pure Yang-Mills theory}
\label{subsubsec:YangMillsLambda}

\begin{figure*}
	\centering
	%	\vspace*{5cm}       % Give the correct figure height in cm
	\resizebox{0.875\textwidth}{!}{%
	\includegraphics{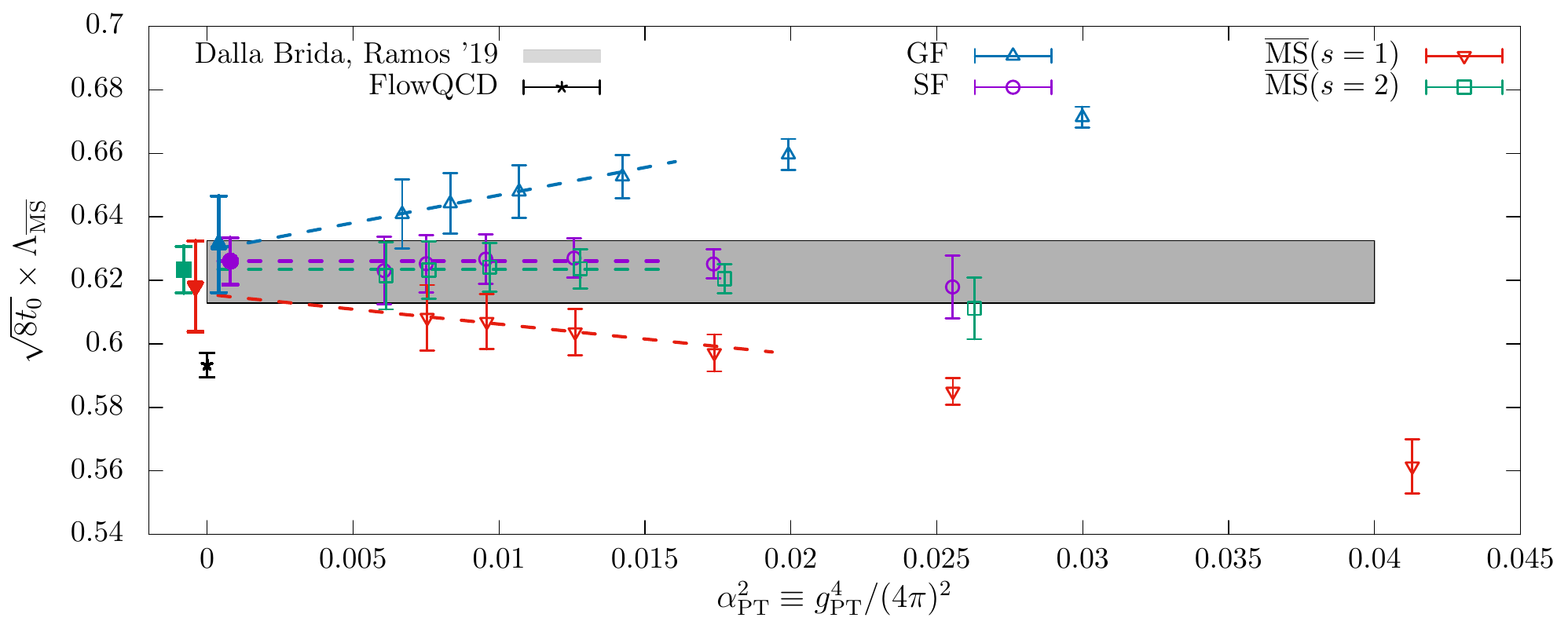}}
	\caption{The dimensionless product
	$\sqrt{8t_0}\Lambda_{\overline{\rm MS}}^{(0)}$ as a function
	of $\alpha_{\rm PT}$~\cite{Nada:2020jay}. The empty symbols
	represent the data at the given $\alpha_{\rm PT}$, while the
	filled symbols are extrapolations to $\alpha_{\rm PT}\to0$
	(shifted for better readability) of the	different approaches 
	to the perturbative matching (see text for more details). 
	The gray band is the result of ref.~\cite{DallaBrida:2019wur},
	while the data point labeled FlowQCD is the result of
	ref.~\cite{Kitazawa:2016dsl}.}
	\label{fig:PureGaugeLambda}       
\end{figure*}

\paragraph{Finite-volume schemes.} 

The second example that we consider is taken from the recent 
study of ref.~\cite{Nada:2020jay} in the pure Yang-Mills (YM)
theory. This work presents an independent analysis of the results 
from a previous study~\cite{DallaBrida:2019wur}, using 
novel techniques. Before entering the discussion, we care 
to stress that the case of the pure Yang-Mills theory is not 
just a curious example. As we shall see in the following section,
through the strategy of renormalization by decoupling precise
results for $\Lambda^{(5)}_{\overline{\rm MS}}$ can be obtained 
from the accurate knowledge of $\Lambda^{(0)}_{\overline{\rm MS}}$.
From this perspective, a robust determination of the 
$\Lambda$-parameter of the pure YM theory is very relevant.

In Fig.~\ref{fig:PureGaugeLambda} we show the results for
$\sqrt{8t_0}\Lambda^{(0)}_{\overline{\rm MS}}$ from
ref.~\cite{Nada:2020jay} as a function of $\alpha_{\rm PT}^2$.
The scale $\mu_{\rm had}=1/\sqrt{8t_0}$ is defined
in terms of the flow time $t_0$~\cite{Luscher:2010iy}, while 
$\alpha_{\rm PT}$ is once again the value of the relevant 
coupling at the renormalization scale $\mu_{\rm {PT}}$ where
perturbation theory is applied. Similarly to the case of  $\Nf=3$ QCD
discussed above, different schemes and strategies have been
considered in order to extract 
$\Lambda^{(0)}_{\overline{\rm MS}}/\mu_{\rm had}$
and study the perturbative truncation errors.
In all cases, the non-perturbative RG running from $\mu_{\rm had}$ 
up to $\mu_{\rm {PT}}$ is obtained using a finite-volume scheme
based on the YM gradient flow (GF)~\cite{Luscher:2010iy,Fodor:2012td,Fritzsch:2013je,DallaBrida:2016kgh}. The interested reader can find more details about the scheme 
in the main refs.~\cite{DallaBrida:2019wur,Nada:2020jay} 
(see also Sect.~\ref{subsec:CouplingFromDecoupling}). 

Once $\mu_{\rm {PT}}$ is reached, 
$\Lambda^{(0)}_{\overline{\rm MS}}/\mu_{\rm had}$ is 
estimated in a number of ways. For the case labeled as (GF) in 
the plot, \textbf{Strategy 1.} of 
Sect.~\ref{subsec:SystematicErrorsPT} is employed using the 
3-loop $\beta$-function in the GF-scheme of
choice~\cite{DallaBrida:2017tru}. In the other cases, the GF-scheme
is first non-perturbatively matched to the ${\rm SF}_{\nu=0}$-scheme
introduced in the previous section. Perturbation theory is then
applied either following \textbf{Strategy 1.} based on the 
${\rm SF}_{\nu=0}$-scheme (SF label in the plot), or by following
\textbf{Strategy 2.} and matching the ${\rm SF}_{\nu=0}$- and
$\overline{\rm MS}$-coupling ($\overline{\rm MS}(s=1,2)$ in the
figure). In the latter case, two values of the $s$-parameter,
$s=1,2$, are studied; note that $s^*\approx2$ in this case.
In all cases, the leading parametric
uncertainties in $\Lambda^{(0)}_{\overline{\rm MS}}/\mu_{\rm had}$ 
from the truncation of the perturbative expansion are of 
${\rm O}(\alpha_{\rm PT}^2)$. 

Going back to Fig.~\ref{fig:PureGaugeLambda} we see how 
two out of the four strategies ((SF) and $\overline{\rm MS}(s=2)$))
give results which are essentially independent on $\alpha_{\rm PT}$
over the whole range of couplings considered for the 
extraction of $\Lambda^{(0)}_{\overline{\rm MS}}/\mu_{\rm had}$. 
Note that in going from the largest to the smallest couplings
the energy scale varies by a factor $32$ while $\alpha_{\rm PT}$
changes by about a factor 2. On the other hand, the other two 
types of determinations ((GF) and $\overline{\rm MS}(s=1)$)) 
show a significant $\alpha_{\rm PT}$ dependence, roughly 
compatible with the expected ${\rm O}(\alpha_{\rm PT}^2)$ scaling. 
What is remarkable is that even considering values of 
$\alpha_{\rm PT}\approx0.08$ the different strategies give
estimates for $\Lambda^{(0)}_{\overline{\rm MS}}/\mu_{\rm had}$ 
which vary up to $\approx3\%$. This is about twice as large 
as the statistical errors on the points (cf.~Table 3 of 
ref.~\cite{Nada:2020jay}). In the case of the (GF) and
($\overline{\rm MS}(s=1)$) determinations, it is clear that a
trustworthy estimate for $\Lambda^{(0)}_{\overline{\rm MS}}/
\mu_{\rm had}$ can be quoted only by extrapolating the results 
for $\alpha_{\rm PT}\to0$. In general, perturbative truncation
errors are large also in the pure YM theory given the 
precision one can reach.

The results above show us once again the importance of 
an explicit non-perturbative calculation of the running of 
the coupling over a significant range of values, reaching down 
to small couplings, in order to assess the actual size of the
perturbative corrections. We join the authors of
ref.~\cite{Nada:2020jay}  and conclude that only by studying
non-perturbatively the limit $\alpha_{\rm PT}\to0$ one can 
avoid the dangerous game of estimating perturbative uncertainties
at some finite (potentially large) value of $\alpha_{\rm PT}$.
Without studying this limit, the determinations can 
easily be affected by perturbative truncation errors, even at
surprisingly small values of the coupling. 

\paragraph{Large-volume schemes.}

A precise determination of the $\Lambda$-parameter 
in the pure YM theory is certainly very much facilitated 
from the computational point of view with respect to
the case of QCD. However, as we have seen in the 
previous example, it is yet a non-trivial challenge to 
control perturbative truncation errors once a $1-2\%$
precision in $\Lambda_{\overline{\rm MS}}$ is reached. 

The disagreement among some recent determinations of
$\Lambda^{(0)}_{\overline{\rm MS}}$ is a clear signal 
that these difficulties should not be underestimated. 
The issue is well illustrated in Fig.~\ref{fig:PureGaugeLambda}, 
where the very precise results labeled (FlowQCD) from
ref.~\cite{Kitazawa:2016dsl} show a net tension with 
the determinations of refs.~\cite{DallaBrida:2019wur,Nada:2020jay}. 
We recall that the former result is based on extracting 
$\Lambda^{(0)}_{\overline{\rm MS}}/\mu_{\rm had}$ from 
the plaquette expectation value calculated in large-volume
simulations. Bare lattice perturbation theory at couplings 
$\alpha_{\rm PT}\approx0.095-0.12$ is used, with 
parametric uncertainties of O($\alpha_{\rm PT}^2$). We 
refer the reader to the given reference for the details. 
Here we just note that all the above determinations satisfy the 
most stringent criteria set by FLAG (cf.~ref.~\cite{Aoki:2019cca}).
Yet, one or more of these results have underestimated 
uncertainties.

\begin{figure}
	\centering
	%	\vspace*{5cm}       % Give the correct figure height in cm
	\resizebox{0.5075\textwidth}{!}{%
		\includegraphics{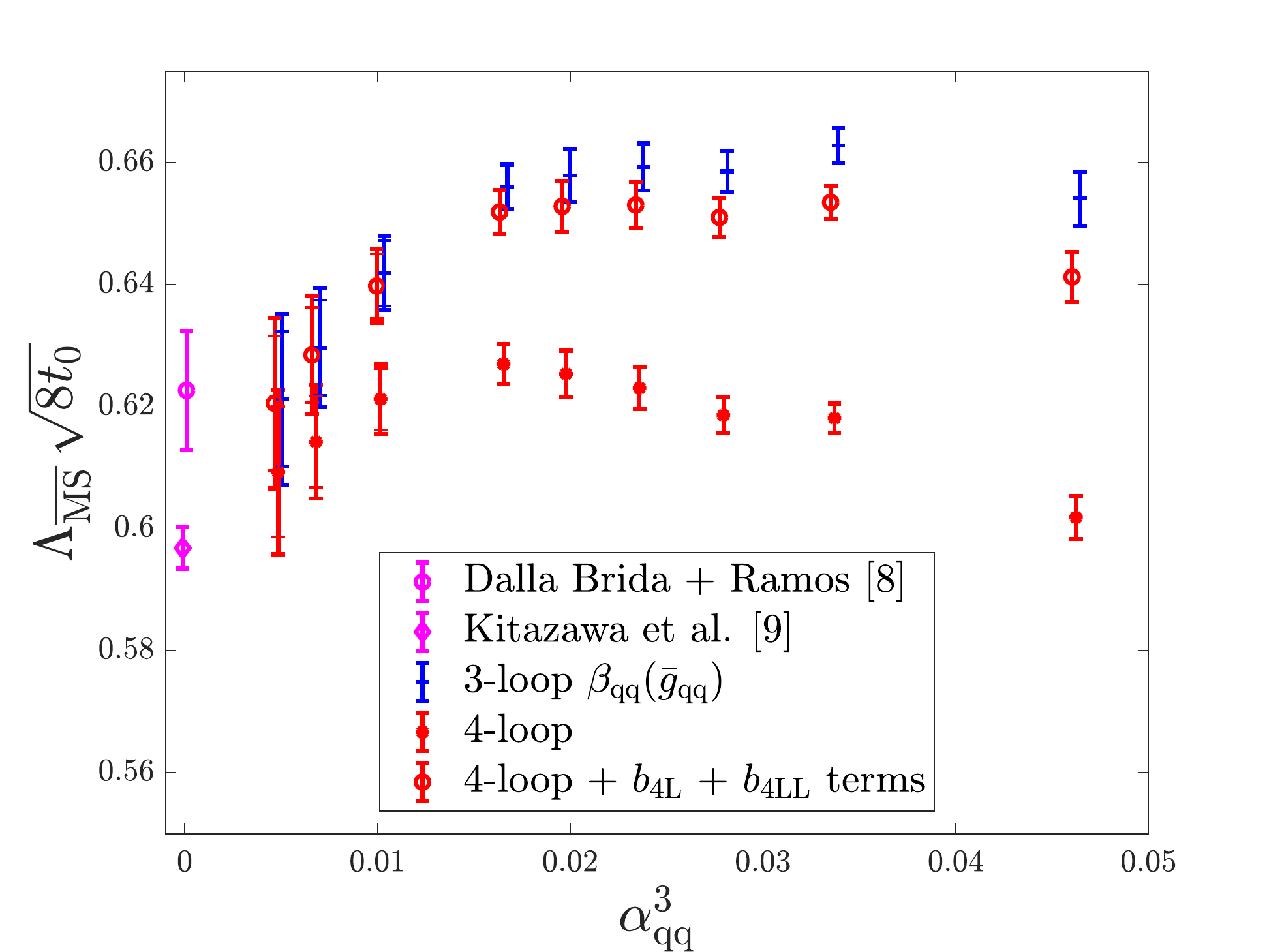}}
	\caption{$\Lambda_{\overline{\rm MS}}^{(0)}$ in units of 
		$\sqrt{8t_0}$ determined at various values of 
		$\alpha_{\rm qq}\equiv\alpha_{\rm PT}$~\cite{Husung:2020pxg}. 
		The results using different orders of perturbation
		theory for $\beta_{\rm qq}$ are shown,  as well
		as a comparison with the determinations of 
		refs.~\cite{DallaBrida:2019wur} and
		\cite{Kitazawa:2016dsl}. 
		(The reference numbers in the plot are those of
		ref.~\cite{Husung:2020pxg} from which the plot
		is taken.)}
	\label{fig:StaticPotentialC}       % Give a unique label
\end{figure}

Other groups have recently engaged in a precision determination 
of $\Lambda^{(0)}_{\overline{\rm MS}}$, also with the intent
of resolving the disagreement above. The recent results 
of ref.~\cite{Husung:2020pxg} based on the ${\rm qq}$-coupling,
$\alpha_{\rm qq}$, defined from the static
potential~\cite{Sommer:1993ce}, are particularly interesting in 
this respect.%
\footnote{A similar earlier study  on the challenges of 
		  extracting $\Lambda_{\overline{\rm MS}}$ in both 
		  $\Nf=2$ QCD and the pure-gauge theory using the 
		  ${\rm qq}$-coupling can be found in
		  ref.~\cite{Leder:2011pz}. For a recent application
	  	  of this scheme for the computation of $\alphas$
	  	  and a detailed account of the most recent results 
	  	  and developments see
	  	  refs.~\cite{Bazavov:2019qoo,Komijani:2020kst}.}
We report them in Fig.~\ref{fig:StaticPotentialC}. 
In the case of $\alpha_{\rm qq}$ the corresponding $\beta$-function,
$\beta_{\rm qq}$, is known up to 4-loop order, and some partial
information is available also at 5-loops 
(see e.g.~ref.~\cite{Tormo:2013tha}). Determinations of 
$\Lambda^{(0)}_{\overline{\rm MS}}/\mu_{\rm had}$
from $\alpha_{\rm qq}$ are hence expected to have 
asymptotically ${\rm O}(\alpha_{\rm qq}^3)$ corrections. 
In the plot, $\alpha_{\rm qq}$ refers to the coupling 
at which perturbation theory is used according to 
\textbf{Strategy 1.} of Sect.~\ref{subsec:SystematicErrorsPT}, 
i.e. it corresponds to $\alpha_{\rm PT}$ in our previous discussions.

Despite the accurate perturbative knowledge there
are a few challenges when using the ${\rm qq}$-scheme for 
precision determinations of 
$\Lambda_{\overline{\rm MS}}^{(0)}$~\cite{Husung:2020pxg}. 
The most relevant ones for our discussion are, first of all,
that the scheme is conventionally defined in an infinite 
space-time volume. In order to measure the coupling at small
lattice spacings one therefore needs large lattice sizes to
maintain the physical extent of the lattice large. In the
computation of ref.~\cite{Husung:2020pxg} lattice spacings 
down to $a\approx0.01\,\fm$ are reached while keeping the  
lattice extent $L\approx2\,\fm$. This means simulating 
lattices with up to $L/a\approx 200$. Secondly, the perturbative
expansion of $\alpha_{\rm qq}$ displays some infrared divergences
starting at 3-loop order in $\beta_{\rm qq}$. When resummed these
give rise to terms of the form, 
$\alpha_{\rm qq}^n\log(\alpha_{\rm qq})^m$, $n\geq3$, 
$1\leq m\leq n-2$, which are enhanced at small couplings 
(cf.~ref.~\cite{Husung:2020pxg}).

Figure~\ref{fig:StaticPotentialC} shows the results
for $\Lambda^{(0)}_{\overline{\rm MS}}/\mu_{\rm had}$ 
as a function of $\alpha_{\rm qq}^3$. The range of 
couplings covered by the data is
$\alpha_{\rm qq}\approx0.16-0.35$. As we can see from the 
plot, for couplings $\alpha_{\rm qq}\gtrsim0.21$ the
results for $\Lambda^{(0)}_{\overline{\rm MS}}/\mu_{\rm had}$
have good precision, but perturbative uncertainties
are large. This can be seen by looking at the difference between 
the 3-loop and 4-loop results (or analogously between the 
4-loop and 4-loop + 5-loop log-terms results). 
At these large couplings, the perturbative expansion seems to 
have reached its limit of applicability. This severely limits 
the precision one can aim at for 
$\Lambda^{(0)}_{\overline{\rm MS}}/\mu_{\rm had}$ if one is
restricted to this range of couplings. For couplings 
$\alpha_{\rm qq}\lesssim0.21$, the different orders of 
perturbation theory seem to start converging. On the other hand, 
the errors on the data become large. This is due to the 
difficulties in extrapolating the results to the continuum 
limit~\cite{Husung:2020pxg}. In fact, the errors are too large 
to make definite conclusions for the relevant limit 
$\alpha_{\rm qq}\to0$.

All in all, we see from this last example that a 
precise determination of  $\Lambda_{\overline{\rm MS}}^{(0)}$
is a challenge. Finite-volume renormalization schemes 
allow us to cover a wide range of couplings, reaching down
to rather small values. Yet, having control on perturbative
truncation errors requires care. When using large-volume schemes
the situation is further complicated by controlling
continuum limit extrapolations at the smallest (most relevant)
couplings. Small couplings require small lattice spacings, 
which require large lattice sizes in order to keep 
the physical volume large. As a result, even in the computationally
simpler case of the pure YM theory, one might have precise
data confined for the most part to a region of couplings 
too large to have perturbative uncertainties fully under control,
while at smaller couplings the data is not precise enough for a
competitive determination of $\Lambda$.

\subsection{The tricky business of continuum extrapolations}

Having discussed the difficulties of estimating perturbative
truncation errors in precision determinations of the 
$\Lambda$-parameters, we now want to touch on the issue of 
systematic uncertainties related to the continuum limit
extrapolations of the relevant couplings.

To give the reader a feeling of the pitfalls that these 
continuum extrapolations can conceal, we first consider 
the $\Nf=3$ results of ref.~\cite{DallaBrida:2016kgh}. 
The relevant quantity to look at in this case is the 
step-scaling function (SSF) of the finite-volume GF-coupling 
$\bar{g}^2_{\rm GF}(\mu)$ with SF boundary conditions
(see refs.~\cite{Fritzsch:2013je,DallaBrida:2016kgh} and
eq.~(\ref{eq:GFcoupling}) for the definition of this scheme). 
We recall that the SSF, $\sigma(u)$, encodes the change in the
coupling $u$ when the renormalization scale is varied by a 
factor of 2~\cite{Luscher:1991wu}. Specifically, having set
$\mu=L^{-1}$,
\begin{equation}
	\sigma(u)\equiv
	\bar{g}^2_{\rm GF}(\mu/2)|_{\bar{g}^2_{\rm GF}(\mu)=u},
	\quad
	\sigma(u)=\lim_{a/L\to 0} \Sigma(a/L,u)\,.
\end{equation}
It is clear from its definition that the SSF is a discrete 
version of the $\beta$-function. The latter can in fact be 
obtained once the SSF is known in a range of couplings 
(cf.~ref.~\cite{DallaBrida:2016kgh}).

On the lattice, the SSF is determined by
extrapolating to the continuum limit its discrete approximations,
$\Sigma(a/L,u)$. In order to compute the latter one must first
identify a set of lattice sizes $L/a$ and corresponding values 
of the bare coupling $g_0$ for which $\bar{g}^2_{\rm GF}(L^{-1})=u$,
with $u$ a specific value. The lattice SSFs $\Sigma(a/L,u)$ 
are then given by the couplings $\bar{g}^2_{\rm GF}((2L)^{-1})$
measured at the values of $g_0$ previously determined but on
lattices with sizes $2L/a$. 

\begin{figure}
	\centering
	%	\vspace*{5cm}       % Give the correct figure height in cm
	\resizebox{0.495\textwidth}{!}{%
		\includegraphics{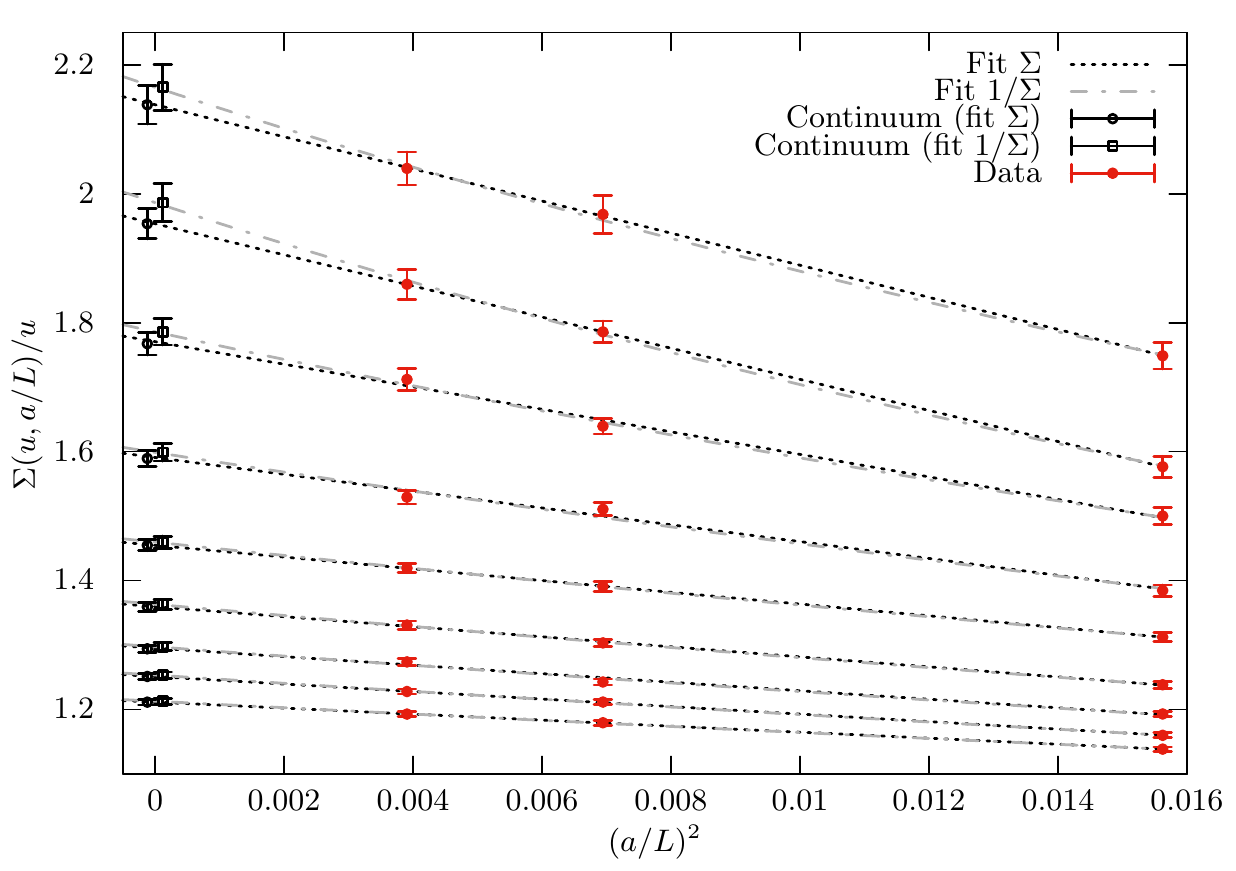}}
	\caption{Continuum extrapolations of $\Sigma(a/L,u)/u$
		for  9 different values of $u\in[2.1,6.5]$ and 3
		lattice sizes $L/a=8,12,16$.}
	\label{fig:SSFGFCoupling}       
\end{figure}

The results for the lattice SSFs of the GF-coupling of
ref.~\cite{DallaBrida:2016kgh} are shown in
Fig.~\ref{fig:SSFGFCoupling}. They correspond to 9 values 
of the coupling $u_i\in[2.1,6.5]$, $i=1,\ldots,9$. As one can see
from the figure, the lattice data are very precise. On the other
hand, discretization effects are in general large, particularly 
so at the largest couplings. The results for $\Sigma(u,a/L)$ vary
in fact by up to 20\% in the range of $L/a$ considered, which is
quite a significant change compared to the statistical errors on
the points.

Given the results in Fig.~\ref{fig:SSFGFCoupling}, we may
expect that a simple fit of the data linear in $(a/L)^2$ is all
that is needed to extrapolate these to the continuum limit.
In particular, we may consider individual continuum extrapolations
for each $u_i$ value using the functional form
\begin{equation}
	\label{eq:FitsGFA}
	\Sigma(u_i,a/L)=\sigma^{(A)}_i+{r}^{(A)}_i\times (a/L)^2
	\quad
	\text{(fit A)}\,,
\end{equation}
where $\sigma^{(A)}_i$, ${r}^{(A)}_i$ are fit parameters. 
Within the uncertainties, linearly in $(a/L)^2$ is in fact 
excellent and the above fits are very good 
($\chi^2/{\rm dof}\approx 0.7$). One is thus tempted to 
take the precise values for $\sigma^{(A)}_i$ as 
estimates for the continuum SSF. 

The continuum results so obtained are well described by 
the simple relation:
$\Delta\sigma_i^{(A)}\equiv1/\sigma^{(A)}_i-1/u_i\approx-0.082$.
Note that this is the functional form expected from the 
perturbative expansion of $\sigma(u)$ at 1-loop order, although 
the coefficient predicted by perturbation theory is slightly
different, i.e. $\approx-0.079$.%
\footnote{Perturbation theory predicts:
		$\sigma(u)\overset{u\to0}{\approx} u+s_0u^2+{\rm O}(u^3)$,
		with $s_0=2b_0(\Nf)\ln(2)$, where $b_0(\Nf)$ is the 
		universal 1-loop coefficient of the $\beta$-function,
		eq.~(\ref{eq:b0b1}) 
		(see e.g.~ref.~\cite{DallaBrida:2018rfy}).
		The close agreement between the non-perturbative data for 
		$\sigma(u)$ and 1-loop perturbation theory is quite 
		peculiar, considering the fact that it holds up to 
		$\sigma(u)={\rm O}(10)$. We refer the interested
		reader to ref.~\cite{DallaBrida:2016kgh} for a  
		detailed discussion about this point.}
This observation suggests us to perform alternative fits to 
the data in Fig.~\ref{fig:SSFGFCoupling} considering  
the functional form
\begin{equation}
	\label{eq:FitsGFB}
	1/\Sigma(u_i,a/L)=1/\sigma^{(B)}_i+r^{(B)}_i\times (a/L)^2
	\quad
	\text{(fit B)}\,,
\end{equation}
with $\sigma^{(B)}_i$, ${r}^{(B)}_i$ new fit parameters. 
The quality of these fits is as good as for the fits A
of eq.~(\ref{eq:FitsGFA}). Distinguishing between the 
two fit forms would require significantly higher statistical
precision than the present one. 

It is important to note at this point that any functional 
form that we consider for the continuum extrapolations, 
necessarily comes with assumptions. Both fits A and B above, 
for instance, assume discretization errors of O($a^n$), 
$n\geq 3$, to be negligible. Moreover, even focusing only
on the leading O($a^2$) effects, we know from Symanzik 
effective theory
(SymEFT)~\cite{Symanzik:1981hc,Symanzik:1983dc,Symanzik:1983gh}
that these are not simply given by a ``classical''
term $\propto (a/L)^2$. They are in fact a non-trivial 
combination of different terms which in the limit $a/L\to0$ 
are asymptotically $\propto(a/L)^2\ln(L/a)^{-\Gamma_i}$, or 
rather $\propto(a/L)^2 [\bar{g}^2(a^{-1})]^{\Gamma_i}$, 
where $\Gamma_i\in\mathbb{R}$, $i=1,2,\ldots,$ and
$\bar{g}^2(a^{-1})$ is the given renormalized coupling of
the effective theory evaluated at a scale $\mu=a^{-1}$
(cf.~refs.~\cite{Balog:2009yj,Balog:2009np,Husung:2019ytz,Husung:2020}).

If terms of higher order than $a^2$ as well as logarithmic
corrections to pure $a^2$ scaling were completely negligible 
in the data, the fit parameters $\sigma^{(A)}_i$ and 
$\sigma^{(B)}_i$ should perfectly agree. From the results in
Fig.~\ref{fig:SSFGFCoupling} we see that there is in fact
agreement within one standard deviation. However, the difference
between the results from the two fits is clearly systematic, 
with the results from fit B being always larger than those 
from fit A. 

The issue becomes more evident if one tries to obtain a 
smooth parameterization for the continuum SSF from the fitted
continuum values $\sigma_i$. As noticed earlier, a fit of
$\Delta\sigma_i=1/\sigma_i-1/u_i$ to a constant $\Delta\sigma$
provides a good description of the continuum data $\sigma_i$ 
in the whole range of $u\in[2.1,6.5]$; this is the case for both
$\Delta\sigma^{(A)}_i$ and $\Delta\sigma^{(B)}_i$
($\chi^2/{\rm dof}<1$). Given the 9 independent values of
$\Delta\sigma_i$ for each fit, the results for the 
corresponding $\Delta\sigma$ are 3 times more precise
than the individual $\Delta\sigma_i$. The systematic 
effect then becomes clearly noticeable as one finds:
$\Delta\sigma^{(A)}=-0.0823(4)$ and
$\Delta\sigma^{(B)}=-0.0832(4)$, for the constant fits to
$\Delta\sigma^{(A)}_i$ and $\Delta\sigma^{(B)}_i$, respectively.

The previous considerations show that the description
of discretization errors as pure $(a/L)^2$ effects is in 
this case not accurate enough for the level of precision 
claimed in the continuum limit. Even though the different 
functional forms in eqs.~(\ref{eq:FitsGFA}) and (\ref{eq:FitsGFB})
fit the data well and perfectly agree with each other at finite
$L/a$, the corresponding extrapolations for $a/L\to0$ 
are clearly affected by some systematics. 
In ref.~\cite{DallaBrida:2016kgh}, more conservative
error estimates and robust central values for the continuum 
results are eventually obtained by carefully accounting as
systematic uncertainties in the data
the not entirely negligible effects of the higher-order terms
neglected in eqs.~(\ref{eq:FitsGFA}) and (\ref{eq:FitsGFB})  
(cf.~ref.~\cite{DallaBrida:2016kgh} for the details).  

The example above might not seem too pessimistic. 
However, it should come as a warning for the more general
situation. Estimating properly the systematic uncertainties 
related to continuum limit extrapolations of high-precision data 
can easily become a challenge, particularly so if
discretization errors are not small.

As recalled earlier, the leading asymptotic dependence of
renormalized lattice quantities on the lattice spacing as 
$a\to0$ is given by a combination of terms 
$\propto a^n [\bar{g}^2(a^{-1})]^{\Gamma_i}$, where the 
number and values of the $\Gamma_i$, as well as $n$, depend 
on the chosen discretization and set-up. The $\Gamma_i$ are 
in fact inferred from the anomalous dimensions of the fields
defining the O($a^n$) counterterms in the SymEFT, and the order
of perturbative improvement that has been possibly implemented
(cf.~ref.~\cite{Husung:2019ytz}). Hence, when one considers a 
pure $a^{n}$ dependence for the discretization errors, one is
implicitly assuming that all $\Gamma_i\approx 0$. This, however,
cannot be taken for granted.

For most cases of interest, the leading discretization effects 
have $n=2$, i.e.~they are of O($a^2$).%
\footnote{A relevant exception is the case of the SF, for which
		  the leading discretization errors are parametrically
		  of O($a$) (cf.~Sect.~\ref{subsec:FiniteVolumeDecoupling}).
		  In applications, however, the O($a$) effects
		  are subdominant with respect to the O($a^2$) effects,
		  and often also compared to the statistical errors. 
		  The precision studies of
		  refs.~\cite{DallaBrida:2016kgh,DallaBrida:2018rfy,DallaBrida:2019wur} 
		  thus opt for treating the O($a$) effects
		  as (small) systematic uncertainties in the data, and
		  perform continuum extrapolations assuming leading
		  O($a^2$) effects. In this respect, we note that in
		  refs.~\cite{Husung:2019ytz,Husung:2020} the $\Gamma_i$
		  relevant for the O($a$) effects in the (pure-gauge) SF
		  have been computed. The results support the
		  treatment of O($a$) effects pursued in
		  refs.~\cite{DallaBrida:2016kgh,DallaBrida:2018rfy,DallaBrida:2019wur} 
		  (cf.~the given references for the details).} 
The results of refs.~\cite{Husung:2019ytz,Husung:2020} then 
show that in the case of QCD we have O(10) different terms 
that contribute in general, and $\Gamma_i\approx [-0.1,3]$ 
for several common discretizations and values of $\Nf=2-4$.%
\footnote{The results refer to the contributions to 
		  discretization effects coming from the lattice 
		  action, considering several popular options
		  (cf.~ref.~\cite{Husung:2020}). If the relevant 
		  observable is not a spectral quantity, additional 
		  effects originating from the lattice fields 
		  that define it are present. These depend on the 
		  specific observable and choice of discretization 
		  (see, e.g.~refs.~\cite{Husung:2019ytz,Husung:2020}).}%
\footnote{In the case of the pure-gauge theory only
		  two terms from the lattice action contribute to
		  the O($a^2$) effects. The $\Gamma_i$ for different 
		  options can be found in ref.~\cite{Husung:2019ytz}. 
		  In all cases, $\Gamma_i\gtrsim 0.6$.}	  
Having all $\Gamma_i\gtrsim 0$ is certainly positive. 
In particular, the contributions relevant in the massless
theory all have $\Gamma_i>0$, which implies a faster approach to 
the continuum limit with respect to pure $a^2$ terms. However, the
large number of terms contributing makes for a complicated pattern 
of discretization errors in the general case, with no clear
contribution(s) dominating. As a result, it may be difficult 
in practice to identify the terms that are actually relevant.
Moreover, terms of the form $a^2[\bar{g}^2(a^{-1})]^{\Gamma_i}$
with $\Gamma_i\approx 2-3$ can be hard to distinguish from 
$a^3$ or $a^4$ terms in a limited range of lattice spacings
when statistical uncertainties are present.%
\footnote{It is clear that even though the SymEFT can predict
		  the form of the leading asymptotic discretization	
		  errors, it cannot predict the region where these 
		  dominate over formally suppressed contributions. 
		  In practice, it may thus be difficult to establish 
		  the regime of applicability of the results from 
		  SymEFT.}
The continuum estimates obtained by including different
contributions, on the other hand, may vary appreciably. 
In this situation, precise and robust final estimates are 
not easily achieved.

We stress that it is particularly important to take 
these considerations into account when aiming for precise 
determinations of short-distance quantities like the couplings. 
As discussed in previous sections, in the most interesting 
region of high energy, $\mu\gg\Lambda$, $a\mu$ may not be so 
small. Continuum extrapolations are thus likely to be difficult
and require special attention. Following the lines of
refs.~\cite{Husung:2019ytz,Husung:2020} one should take
the non-trivial $a$-dependence predicted by SymEFT into account,
provided the information is available. If this is not the case, 
one should try at least to estimate the uncertainties associated
with neglecting logarithmic corrections to classical scaling,
e.g.~by considering terms 
$\propto a^2[\bar{g}^2(a^{-1})]^{\Gamma_i}$, with
$\Gamma_i\approx1-3$, in the fit ans\"atze. Ideally, one 
would like to be in the situation where within the uncertainties
the continuum estimates do not sensibly depend on whether these
terms are considered or not. 

Given the observations above, we want to bring the reader's
attention to a recent study where the non-trivial 
$a$-dependence of discretization effects was found to be a
relevant issue. Specifically, we consider the computations 
of refs.~\cite{DallaBrida:2016dai,DallaBrida:2017tru}
of the GF-coupling in the pure Yang-Mills theory 
using Numerical Stochastic Perturbation Theory.%
\footnote{A recently expanded discussion in ref.~\cite{Nada:2020jay} 
		  provides another clear illustration of the difficulties 
		  of continuum limit extrapolations of precise coupling 
		  data using results from the pure-gauge theory 
		  (cf.~Figs. 6 and 7 of this reference and related discussion). 
		  We strongly recommend the interest reader to consult this reference. 
		  We moreover refer to the important pioneering studies of 
		  refs.~\cite{Balog:2009yj,Balog:2009np} in the non-linear $\sigma$-model 
		  in two dimensions.}  
In this framework, the lattice theory is numerically solved
through a Monte Carlo simulation up to a finite order in the 
bare coupling $g_0$~\cite{DiRenzo:1994sy,DiRenzo:2004hhl}. 
From expectation values in this ``truncated theory'' one  can
obtain the perturbative coefficients of the expansion of 
lattice quantities in $g_0$.

\begin{figure*}
	\centering
	%	\vspace*{5cm}       % Give the correct figure height in cm
	\resizebox{0.495\textwidth}{!}{%
		\includegraphics{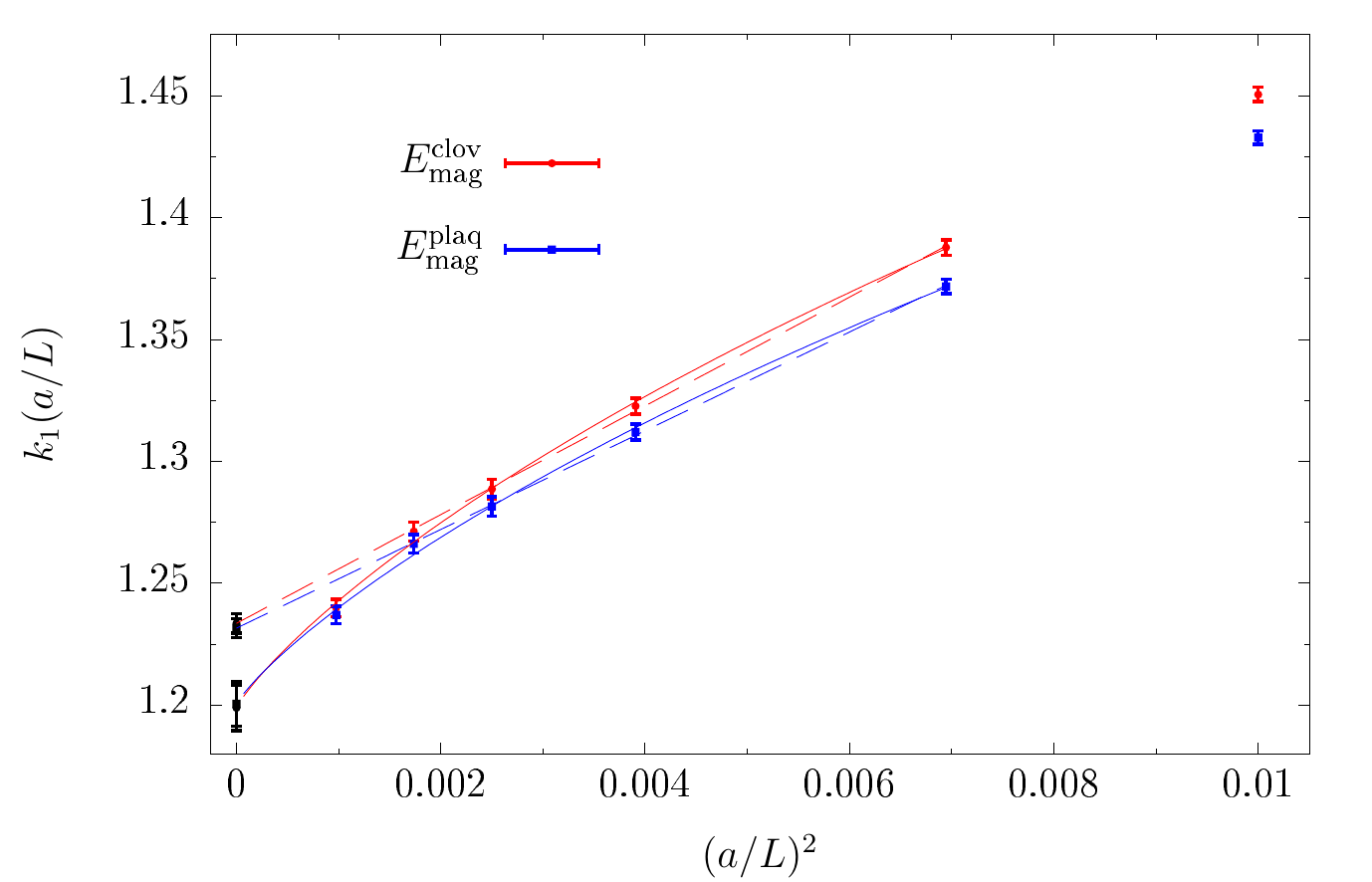}}
	\resizebox{0.495\textwidth}{!}{%
		\includegraphics{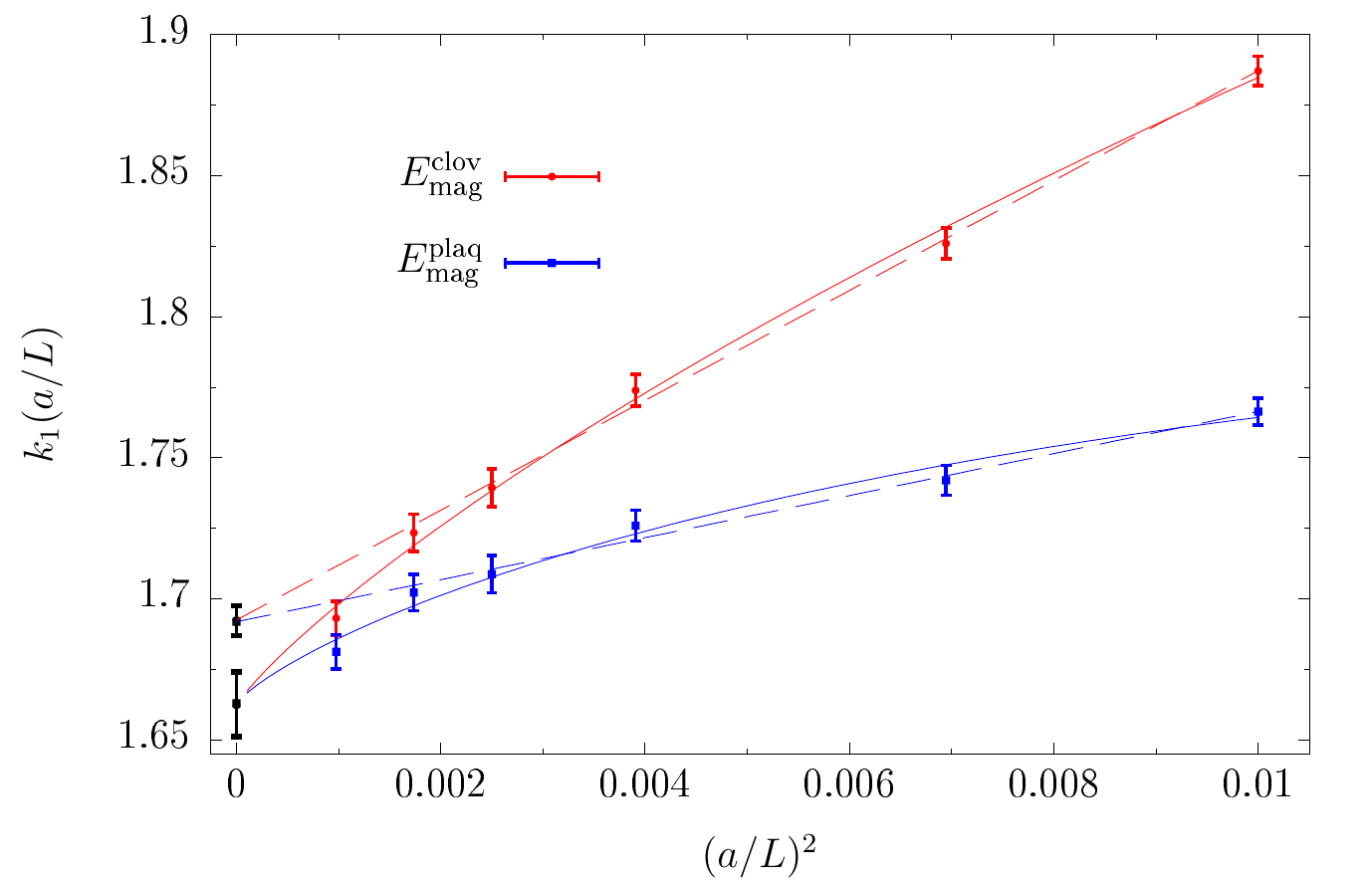}}
	\caption{Continuum extrapolations for $k_1(a/L)$. 
		Results for $c=0.3$ (left panel) and $c=0.4$ (right panel)
		are shown, for two different discretizations of the relevant
		observable. Different fits to the data are considered, 
		cf. text.}
	\label{fig:ContinuumLimitsNSPT}       
\end{figure*}

In refs.~\cite{DallaBrida:2016dai,DallaBrida:2017tru},  
the GF-coupling with SF boundary conditions 
$\bar{g}^2_{\rm GF}(\mu)$ has been computed up to two-loop 
order in $g_0^2$. Using the relation between 
$\alpha_0\equiv g_0^2/(4\pi)$ and
$\alpha_{\overline{\rm MS}}\equiv\alpha^{(\Nf=0)}_{\overline{\rm MS}}$~\cite{Luscher:1995nr,Luscher:1995np}, one can thus infer 
the relation
\begin{equation}
	\alpha_{\rm GF}(\mu)
	=\alpha_{\overline{\rm MS}}(\mu)
	+ k_1(a/L)\alpha^2_{\overline{\rm MS}}(\mu)
	+ k_2(a/L)\alpha^3_{\overline{\rm MS}}(\mu)+\ldots.
\end{equation}
The coefficients $k_1(a/L)$, $k_2(a/L)$ are functions of 
the resolution $a/L$ considered for the lattice.  
In order to obtain the matching relation between the couplings
in the continuum limit the coefficients must  be extrapolated 
for $a/L\to0$. Focusing on the 1-loop coefficient, $k_1(a/L)$, 
from SymEFT we expect that 
(see refs.~\cite{DallaBrida:2016dai,DallaBrida:2017tru})
\begin{equation}
	\label{eq:k1}
	k_1(a/L)\overset{a/L\to0}{\sim}
	k_1(0)+ \sum_{m=2}^\infty 
	\sum_{n=0}^1 r_{mn} (a/L)^m \ln(L/a)^n\,,
\end{equation}
with $r_{mn}$ some constants. Note that the coefficient $r_{21}$ 
of the leading term $\propto\ln(L/a)$ implicitly depends on 
the $\Gamma_i$ predicted by SymEFT (cf.~Sect.~5.2. of
ref.~\cite{Husung:2019ytz} and also ref.~\cite{Husung:2020}). 
Compared to the case of the full theory, the results from the
truncated theory have a simpler (yet non-trivial) cutoff
dependence. Given the high precision reached in these
calculations, this allows for a clean illustration of the
difficulties in continuum extrapolations.

In Fig.~\ref{fig:ContinuumLimitsNSPT} we show the results 
from ref.~\cite{DallaBrida:2016dai} for $k_1(a/L)$ for two
different values of the parameter $c$, $0.3$ and $0.4$, that 
specifies the GF-scheme
(cf.~refs.~\cite{DallaBrida:2016dai,DallaBrida:2017tru} 
and eq.~(\ref{eq:GFcoupling})). Two different discretizations 
of the observable defining the coupling ($E^{\rm clov}_{\rm mag}$,
$E^{\rm plaq}_{\rm mag}$) are also considered. The simulated
lattices have sizes $L/a=10-32$. 

Starting from the results for $c=0.3$ (left panel of
Fig.~\ref{fig:ContinuumLimitsNSPT}), we see how the data 
is very precise but discretization effects are sizable. 
In the plot we then show two types of extrapolations to the 
continuum limit. For the first type (solid lines), lattices 
with $L/a=12-32$ are fitted to the asymptotic form,
eq.~(\ref{eq:k1}), considering the leading terms $m=2,n=0,1$. 
The fits are good, $\chi^2/{\rm dof}\sim1$, and the
extrapolated results for the two discretizations agree well. 
The $m=2,n=1$, term is in fact crucial to obtain good fits. 
For the second set of extrapolations (dashed lines), we 
consider instead lattices with $L/a=12-24$. In this case 
the data can be very well described by a pure $(a/L)^2$ term 
($m=2, n=0$) over the whole range of lattice sizes. The continuum
extrapolated values obtained from these fits have significantly
smaller statistical errors than the ones from the previous fits,
and yet there is perfect agreement between the two discretizations. 
On the other hand, the results deviate from the previous
estimates by several of their standard deviations. 

The results for $c=0.4$ exhibit qualitatively the same features,
although the statistical errors on $k_1(a/L)$ are about a 
factor 2 larger and the two discretizations now show rather 
different lattice artifacts. On the other hand, cutoff effects 
are generally smaller than for $c=0.3$, and we thus 
include $L/a=10$ in the fits. It is clear that in both
cases, $c=0.3,0.4$, a reliable continuum extrapolation for
$k_1(a/L)$ is challenging due to the non-trivial $a$-dependence 
of the data. In particular, larger lattices than the ones
considered here are clearly needed in order to obtain accurate
continuum results (cf.~ref.~\cite{DallaBrida:2017tru} for the 
final determination).

In conclusion, through these examples we saw how 
assessing the systematics related to the continuum limit
extrapolations of couplings can be challenging. This is
especially true when one wants to maintain the high precision
reached on the lattice data also in the continuum limit, but
discretization errors are large. It then becomes hard to 
avoid systematic biases in the final determinations. To this 
end, it is crucial to test all the assumptions that 
enter the functional forms chosen for the extrapolations. 
In particular, we must keep in mind that good fits do 
not necessarily mean good results for parameters, especially
for extrapolations outside the range covered by the data.

% end

%% file: sect3.tex
% sect3.tex

\section{Heavy-quark decoupling}
\label{sec:Decoupling}

So far we focused on the main challenges 
that stand on the way of a precise determination of
$\Lambda^{(\Nf)}_{\overline{\rm MS}}$ and discussed  
in detail the cases of $\Nf=0,3$. The interesting 
quantity for phenomenology, however, is
$\Lambda^{(5)}_{\overline{\rm MS}}$. At present, lattice estimates 
of $\Lambda^{(5)}_{\overline{\rm MS}}$ are for the most part 
based on determinations of $\Lambda^{(3)}_{\overline{\rm MS}}$,
while just a handful are obtained from 
$\Lambda^{(4)}_{\overline{\rm MS}}$
(cf.~ref.~\cite{Aoki:2019cca}). As we shall recall in the 
next subsection, the most common strategy to obtain
$\Lambda^{(5)}_{\overline{\rm MS}}$  is in fact to 
non-perturbatively compute $\Lambda^{(3)}_{\overline{\rm MS}}$
through simulations of the $\Nf=3$ theory and then rely on 
perturbative decoupling relations for the heavy quarks to 
estimate the ratios $\Lambda^{(4)}_{\overline{\rm MS}}/
\Lambda^{(3)}_{\overline{\rm MS}}$
and $\Lambda^{(5)}_{\overline{\rm MS}}/
\Lambda^{(4)}_{\overline{\rm MS}}$ 
(see e.g.~ref.~\cite{Aoki:2019cca}).

The main reason for this is because, as is well-known, simulating 
the charm quark dynamically is at present challenging, let 
alone the case of the bottom quark. While the inclusion of 
the charm quark in the computation of the running of the QCD
coupling may be only moderately challenging with a suitable strategy 
(see e.g. ref.~\cite{Tekin:2010mm}), it does pose 
important difficulties in large-volume hadronic simulations. 
Besides the increased computational cost in simulating an 
additional quark with respect to $\Nf=2+1$ simulations, and the
more complicated tuning of the bare QCD parameters necessary to
define proper lines of constant physics, discretization effects
are a serious source of concern. Given the currently most 
accessible lattice spacings in hadronic simulations, say 
$a\gtrsim 0.05\,\fm$, we  have that $am_c\gtrsim 0.3$ and
$am_b\gtrsim 1$, where for definiteness we took
$m_c\approx1.27\,\GeV$ and $m_b\approx4.2\,\GeV$. In the
hadronic regime it is therefore a real challenge to control 
the discretization effects induced by including the charm quark 
in simulations, and unrealistic for the case of the bottom
quark. This is particularly true for the case of Wilson quarks 
where the charm quark can potentially introduce large O($am_c$)
effects, unless a complete Symanzik O($a$) improvement programme 
is carried out, which is certainly no simple task
(see e.g.~ref.~\cite{Fritzsch:2018kjg}). 

In this situation, it is mandatory to assess the 
reliability of the strategy presented above for the 
determination of $\Lambda^{(5)}_{\overline{\rm MS}}$.
To this end, in the following we shall recall the general theory 
of decoupling of heavy quarks and critically address  
its application in lattice determinations of 
$\Lambda^{(5)}_{\overline{\rm MS}}$. This includes
both the usage of perturbation theory for the inclusion
of heavy-quark loops in the running of the QCD coupling,
that is to estimate the ratios 
$\Lambda_{\overline{\rm MS}}^{(4)}/\Lambda_{\overline{\rm MS}}^{(3)}$, $\Lambda_{\overline{\rm MS}}^{(5)}/\Lambda_{\overline{\rm MS}}^{(4)}$, 
as well as the determination of the 
physical units of $\Lambda^{(5)}_{\overline{\rm MS}}$ 
from scale setting in the $\Nf=2+1$ theory. As we shall 
see, given the current precision on 
$\Lambda^{(3)}_{\overline{\rm MS}}$, accounting for 
heavy-quark effects by means of perturbation theory in the 
running of the QCD coupling is remarkably accurate, even 
for the  case of the charm quark. In addition, charm-quark 
effects in (dimensionless) low-energy quantities are found
to be quite small, supporting the fact that $\Nf=2+1$ QCD 
is accurate enough for establishing the physical scale. 
As a result, competitive determinations of
$\Lambda^{(5)}_{\overline{\rm MS}}$ are possible 
from results in the $\Nf=3$ flavor theory.

\subsection{The effective theory for heavy-quark decoupling
	 		and the QCD couplings}

In this subsection we introduce the effective theory
of heavy quarks and recall how this is conventionally applied 
in the determination of $\Lambda_{\overline{\rm MS}}^{(5)}$.
We refer the reader to refs.~\cite{Bruno:2014ufa,Athenodorou:2018wpk}
for a more detailed presentation.

\subsubsection{The effective theory for heavy-quark decoupling}
\label{subsec:HQET}

We begin by considering QCD with $\Nf$ flavors of quarks, 
which in short we denote ${\rm QCD}_{\Nf}$. Of these, 
$\Nl$ are considered to be light, while the other
$\Nh\equiv\Nf-\Nl$ are heavy. For simplicity, we assume that 
the light quarks are degenerate with mass $m$, while the
heavy quarks are also degenerate but with a mass $M\gg \Lambda$.
The effective theory associated with the decoupling of the
heavy quarks is formally obtained by integrating out in the
functional integral the fields associated with the 
heavy quarks~\cite{Weinberg:1980wa}. The field theory that results 
is characterized by having an infinite number of non-renormalizable
interactions, which are suppressed at low energies by negative
powers of the heavy-quark masses $M$. The couplings of the effective
theory can be fixed order by order in $M^{-1}$ by requiring that, 
at each given order, a finite number of observables is equal to 
the corresponding ones in the fundamental theory. Once the 
couplings are fixed up to a certain order $M^{-n}$, 
the effective theory is said to be matched to the fundamental 
one at this order, and can be used to describe the effects of 
the heavy quarks at low energies up to corrections
of O($M^{-n-1}$). In this sense, we say that as $M\to\infty$ the
heavy quarks decouple from low-energy physics as their effects
eventually fade away~\cite{Appelquist:1974tg}.

In formulas, the Lagrangian of the effective theory 
is of the general form (see e.g.~ref.~\cite{Athenodorou:2018wpk})
\begin{equation}
	\label{eq:EffectiveLagrangian}
	\mathcal{L}_{\rm dec}=\mathcal{L}_0 + 
	{1\over M}\mathcal{L}_1 + {1\over M^2}\mathcal{L}_2+\ldots\,,
\end{equation}
where the leading order corresponds to the Lagrangian of 
${\rm QCD}$ with $\Nl$ light quarks, i.e.
$\mathcal{L}_0=\mathcal{L}_{\rm QCD_\Nl}$, while the corrections
$\mathcal{L}_k$, $k\geq 1$, consist of linear combinations of 
local fields of mass dimension $4+k$, i.e.
\begin{equation}
	\label{eq:CountertermsLagrangians}
	\mathcal{L}_k= \sum_i \omega^{(k)}_i \Phi^{(k)}_i\,,
	\qquad
	[\Phi^{(k)}_i]=4+k\,,
\end{equation}
with $\omega^{(k)}_i$ dimensionless couplings. The fields
$\Phi^{(k)}_i$ are built from the light-quark and gluon 
fields, and include possible powers of the light-quark masses. 
They must respect the symmetries of the fundamental 
${\rm QCD}_{\Nf}$ theory, as in particular gauge invariance,
Euclidean (or Lorentz) symmetry, and chiral symmetry.

In the case where the light quarks are massless, the leading-order
effective theory, ${\rm QCD}_\Nl$, has a single parameter:
the gauge coupling $\bar{g}^{(\Nl)}(\mu)$. The effective 
and fundamental theory are therefore matched at leading order in
$M^{-1}$ once $\bar{g}^{(\Nl)}(\mu)$ is matched. This requires
that $\bar{g}^{(\Nl)}(\mu)$ is properly prescribed at a given
renormalization scale in a given renormalization scheme in terms 
of the coupling $\bar{g}^{(\Nf)}(\mu)$ of the fundamental theory
and the heavy-quark masses $M$.%
\footnote{For ease of notation we do not use any symbol to 
		  indicate the generic scheme of renormalized
		  couplings. We however assume that, unless otherwise
		  stated, the couplings are defined in a 
		  mass-independent scheme.}	  
In addition, provided that the fundamental 
theory is defined on a manifold without boundaries,%
\footnote{This means that the theory lives in infinite space-time 
		  or in a finite volume with (some variant of) 
		  periodic boundary conditions. The special 
		  but relevant case of a finite volume with 
		  boundaries will be considered in
		  Sect.~\ref{subsec:FiniteVolumeDecoupling}.}
it is possible to show that $\mathcal{L}_1=0$ and 
therefore O($M^{-1}$) corrections are
absent~\cite{Athenodorou:2018wpk}.%
\footnote{Note that for the sake of argument we
		 exclude the uninteresting case of $\Nl=1$, for which 
		 a proof of this result is to our knowledge missing.}
In this situation, the leading-order corrections 
induced by the heavy quarks are suppressed as $M^{-2}$
at low energy. 

In the case where the light quarks have a non-vanishing mass, 
the only mass-dimension 5 fields allowed in $\mathcal{L}_1$ 
are given by the fields of the leading-order Lagrangian
$\mathcal{L}_{\rm QCD_\Nl}$ multiplied by the light-quark 
masses $m$~\cite{Athenodorou:2018wpk}. Their effect can be
reabsorbed into a redefinition of the gauge coupling and 
light-quark masses of O($m/M$). Of course, in the case of 
massive light quarks the matching of the effective and 
fundamental theory at leading order also requires the matching 
of the light-quark masses. In addition, the couplings
$\omega_i^{(k)}$ of the corrections $\mathcal{L}_{k\geq 2}$ 
will depend in general on the light-quark masses, too.

\subsubsection{The effective theories and their couplings}
\label{subsec:EffectiveCouplings}

The application of the effective theory for heavy quarks 
in the determination of the QCD couplings was first advocated 
by Weinberg in his seminal paper on effective field
theories~\cite{Weinberg:1980wa}. The idea is based 
on the observation that for mass-independent renormalization
schemes the RG equations of the renormalizable couplings 
of the effective theory completely decouple from the others. 
This means that the couplings of the non-renormalizable 
interactions can be completely  ignored when determining 
the variation of the running coupling $\bar{g}^{(\Nl)}(\mu)$ 
of the effective theory with the energy scale $\mu$. 
The heavy quarks affect the value of the coupling of the 
effective theory only through the matching
with the coupling of the fundamental theory, $\bar{g}^{(\Nf)}(\mu)$.

The matching relation between $\bar{g}^{(\Nl)}(\mu)$ and 
$\bar{g}^{(\Nf)}(\mu)$ can in principle be established in
perturbation theory. This is best done at a scale 
$\mu_{\rm match}
\approx M$~\cite{Weinberg:1980wa,Bernreuther:1981sg}. 
Assuming the validity of perturbation theory at this scale, 
if the running coupling $\bar{g}^{(\Nl)}(\mu_{\rm match})$ 
in the effective theory is known, one can turn 
tables and obtain $\bar{g}^{(\Nf)}(\mu_{\rm match})$ from 
inverting the matching conditions. 

In phenomenological applications of this strategy 
(see e.g.~ref.~\cite{Zyla:2020zbs}), the value of
$\bar{g}^{(\Nl)}(\mu_{\rm match})$ is extracted from 
the value of the coupling $\bar{g}^{(\Nl)}(\mu_{\rm low})$  
at some lower energy scale, $\mu_{\rm low}\ll M$. The 
latter is obtained by comparing the perturbative expansion 
for some process $\Obs(q)$ with characteristic energy scale
$q\approx\mu_{\rm low}$, with its experimental
results. The effects of the heavy quarks in $\Obs(q)$ are 
expected to be suppressed as O($(q/M)^2$)
(cf.~Sect.~\ref{subsec:HQET}). Hence, assuming that these
effects can be neglected, the perturbative expansion of $\Obs(q)$
can be considered in the $\Nl$-flavor theory, which allows 
the coupling $\bar{g}^{(\Nl)}(\mu_{\rm low})$ to be extracted.
As observed above, the determination of the running of the 
effective coupling in the $\Nl$-flavor theory does not require
any input from the fundamental theory: one can thus readily 
obtain $\bar{g}^{(\Nl)}(\mu_{\rm match})$ from 
$\bar{g}^{(\Nl)}(\mu_{\rm low})$. Clearly, for this strategy 
to work in practice the energy scale $\mu_{\rm low}\ll M$ must 
be yet sufficiently high for perturbation theory to apply. 

In non-perturbative applications on the lattice, the strategy
presented above is realized in the following 
way (see e.g.~ref.~\cite{Aoki:2019cca}). Firstly, as discussed in
Sect.~\ref{subsec:SystematicErrorsPT}, through the study of the
non-perturbative running of a given massless coupling within 
the effective $\Nl$-flavor theory, one determines the ratio 
$\Lambda_{\overline{\rm MS}}^{(\Nl)}/\mu_{\rm had}$,
where $\mu_{\rm had}$ is a convenient (not necessarily physical)
low-energy scale. Assuming that the effects of the heavy quarks
can be neglected in the ratio of low-energy scales
$\mu_{\rm had}/\mu_{\rm phys}$, where $\mu_{\rm phys}$ is an
experimentally accessible hadronic quantity, this can also be
computed within $\Nl$-flavor QCD. The physical units of 
$\mu_{\rm had}$ and therefore $\Lambda_{\overline{\rm MS}}^{(\Nl)}$
can then be established by taking $\mu_{\rm phys}$ from experiments.
As a second step, as we shall see in detail in the following
subsection, the matching relation between the couplings of 
the effective and fundamental theory is expressed as a 
relation between their $\Lambda$-parameters.  
The ratio $\Lambda_{\overline{\rm MS}}^{(\Nf)}/
\Lambda^{(\Nl)}_{\overline{\rm MS}}$ is thus  
estimated using perturbation theory at a scale $\mu\approx M$.
From this estimate and the results for 
$\Lambda_{\overline{\rm MS}}^{(\Nl)}$, 
$\Lambda_{\overline{\rm MS}}^{(\Nf)}$ is obtained.

It is important to note that in extracting
$\Lambda_{\overline{\rm MS}}^{(\Nl)}$ from the running 
of the chosen non-perturbative scheme, perturbation theory 
can be applied at arbitrarily large energy scales. The perturbative 
matching between the effective and fundamental theory, instead, 
is best performed at a scale $\mu\approx M$, where $M$ is set 
by the mass of the given heavy quarks that decouple. Whether 
perturbation theory is accurate in this step hence depends on 
how heavy these quarks are. As we shall see, in the 
$\overline{\rm MS}$-scheme the perturbative matching is 
remarkably accurate already for masses $M$ close to that of 
the charm quark.

In the next subsection we describe in detail the perturbative
matching of the QCD couplings. We follow the lines of
the presentation in ref.~\cite{Athenodorou:2018wpk}, and start 
by reformulating this matching in terms of $\Lambda$-parameters~\cite{Bernreuther:1981sg}.
After doing so, we investigate the accuracy of a perturbative
matching within perturbation theory itself, and later present 
a discussion on the typical size of non-perturbative corrections 
one can expect in these relations.
 
\subsection{Perturbative decoupling}

\subsubsection{Definitions}
\label{subsec:DecouplingPT}

As we are interested only in the QCD coupling we 
assume for simplicity that the relevant effective theory 
is given by massless ${\rm QCD}_{\Nl}$. As discussed above, 
at leading order in $M^{-1}$ the only parameter of the 
effective theory that needs to be fixed is therefore the 
running coupling $\bar{g}^{(\Nl)}(\mu)$. The effective theory 
hence predicts any observable once the coupling in the chosen 
scheme is specified at some scale. 

The general form of the relation between the  couplings
of the leading-order effective theory $\bar{g}^{(\Nl)}$ 
and of the fundamental theory $\bar{g}^{(\Nf)}$, 
reads~\cite{Weinberg:1980wa,Bernreuther:1981sg,Bernreuther:1983zp}
\begin{equation}
	[\bar{g}^{(\Nl)}(\mu/\Lambda^{(\Nl)})]^2= 
	F_\Obs\big([\bar{g}^{(\Nf)}(\mu/\Lambda^{(\Nf)})]^2,M/\mu\big)\,,
\end{equation}
where for later convenience we explicitly wrote the dependence 
of the couplings on their corresponding $\Lambda$-parameter. 
The function $F_\Obs$ depends in principle on the specific
observable $\Obs$ that is used to establish the matching between 
the two theories. The dependence on the observable, however, 
is suppressed by powers of $M^{-1}$. In perturbation theory, 
these power corrections can be uniquely isolated from the
logarithmic terms in $M$ and can therefore be dropped. This is
consistent with matching the theories at leading order 
in $M^{-1}$. In doing so, the relation between the couplings 
becomes universal, i.e independent on the specific matching
condition.  It only depends  on the renormalization schemes 
chosen for the couplings.

In the $\overline{\rm MS}$-scheme the matching relation
(also referred to as decoupling relation)
is known up to 4-loop order. Below, we consider this 
only for the convenient choice of matching scale $\mu=m_*$, 
where $m_*$ is implicitly defined by the equation
$\overline{m}_{\overline{\rm MS}}(m_*)=m_*$,
where $\overline{m}_{\overline{\rm MS}}(\mu)\equiv
\overline{m}^{(\Nf)}_{\overline{\rm MS}}
(\mu/\Lambda_{\overline{\rm MS}}^{(\Nf)})$ are
the running masses of the heavy quarks in the 
fundamental theory in the $\overline{\rm MS}$-scheme.
Given this choice, the matching relation reads:
\begin{gather}
	\nonumber
	\big[\bar{g}^{(\Nl)}_{\overline{\rm MS}}(m_*/\Lambda^{(\Nl)}_{\overline{\rm MS}})\big]^2=
	g_*^2 C(g_*),\,
	\quad
	g_*\equiv \bar{g}^{(\Nf)}_{\overline{\rm MS}}(m_*/\Lambda^{(\Nf)}_{\overline{\rm MS}})\,,\\[1.5ex]
	\label{eq:CouplingMatchingPT}
	C(g_*)=1+h_2 g_*^4 + h_3 g_*^6 + h_4 g_*^8 + \ldots\,,
\end{gather}
where the $h_i$ are pure numbers that depend on both
$\Nh$ and $\Nl$~\cite{Grozin:2011nk,Chetyrkin:2005ia,Schroder:2005hy,Kniehl:2006bg,Gerlach:2018hen}. As explained in these references,
the particular choice of matching scale makes all
contributions
$\propto\log(\overline{m}_{\overline{\rm MS}}(\mu)/\mu)$ 
appearing in the matching relation vanish, and it is
considered to be optimal. This implies 
in particular that the $g_*^2$ term in $C(g_*)$ is absent.
In the general case, the scales 
$\overline{m}_{\overline{\rm MS}}(\mu)$ and $\mu$ should
anyway not be chosen too separated. Large coefficients 
otherwise appear in the perturbative matching relation
which can compromise its applicability.

As anticipated, from the perspective of lattice applications 
it is compelling  to recast the matching relation,
eq.~(\ref{eq:CouplingMatchingPT}), in terms of RG-invariant
(RGI) quantities. Specifically, this means the
$\Lambda$-parameter of the effective theory,
$\Lambda^{(\Nl)}_{\overline{\rm MS}}$, that of the fundamental
theory $\Lambda^{(\Nf)}_{\overline{\rm MS}}$, and the RGI 
quark-mass $M$ of the heavy quarks. The latter can be defined as
\begin{align}
	\nonumber
	M&\equiv\overline{m}^{(\Nf)}_{\overline{\rm MS}}(\mu)
	\varphi^{(\Nf)}_{\rm m, \overline{\rm MS}}(\bar{g}^{(\Nf)}_{\overline{\rm MS}}(\mu))\,,\\
	\nonumber
	\varphi^{(\Nf)}_{\rm m,\overline{\rm MS}}(\bar{g})&\equiv(2b_0(\Nf)\bar{g}^2)^{-{d_0\over 2b_0(\Nf)}}\times\\
	\label{eq:MassRGI}
	&\times\exp \bigg\{-\int_0^{\bar{g}}\rmd g
	\bigg[{\tau_{\overline{\rm MS}}^{(\Nf)}({g})\over 
	\beta_{\overline{\rm MS}}^{(\Nf)}({g})} - {d_0\over b_0(\Nf)g} \bigg\}\,,
\end{align}
where $b_0(\Nf)$ is the 1-loop coefficient of the $\beta$-function,
eq.~(\ref{eq:b0b1}), and the function
\begin{equation}
	\tau^{(\Nf)}_{\overline{\rm MS}}(\bar{g})\equiv
	{\mu \over \overline{m}^{(\Nf)}_{\overline{\rm MS}}(\mu)}
	{\rmd \overline{m}^{(\Nf)}_{\overline{\rm MS}}(\mu)\over\rmd \mu}
	\bigg|_{\bar{g}}
\end{equation}
encodes the scale-dependence of the quark masses in the
$\Nf$-flavor theory (in the $\overline{\rm MS}$-scheme). 
It has a perturbative expansion
\begin{equation}
	\label{eq:tauMSbar}
	\tau^{(\Nf)}_{\overline{\rm MS}}(\bar{g})\overset{\bar{g}\to0}{\approx}
	-\bar{g}^2\{d_0+\bar{g}^2 d_1+\ldots\}\,,
	\quad
	d_0=8/(4\pi)^2\,,
\end{equation}
known up to
5-loops~\cite{Larin:1993tq,Chetyrkin:1997dh,Vermaseren:1997fq,Baikov:2014qja}, where the actual scheme and $\Nf$-dependence start 
from $d_i$, $i\geq 1$. Note that  
even though in eq.~(\ref{eq:MassRGI}) we conveniently 
defined the RGI mass $M$ through the $\overline{\rm MS}$-scheme, 
its value is in fact scheme independent, as long as
mass-independent schemes are considered for the quark masses.
This means in particular that $M$ can be non-perturbatively
defined through any non-perturbative (massless) renormalization
scheme.

Using the above definitions, together with the definitions
in eq.~(\ref{eq:LambdaParam}), and the matching relation
eq.~(\ref{eq:CouplingMatchingPT}), it is immediate to conclude
that~\cite{Bruno:2014ufa,Athenodorou:2018wpk}
\begin{equation}
	\label{eq:Plf}
	P_{\ell,\rm f}(M/\Lambda_{\overline{\rm MS}}^{(\Nf)})\equiv
	{\Lambda_{\overline{\rm MS}}^{(\Nl)}\over
	 \Lambda_{\overline{\rm MS}}^{(\Nf)}}\Bigg|_{\rm matched}=
 	{\varphi_{\rm g,\overline{\rm MS}}^{(\Nl)}(g_*\sqrt{C(g_*)})
 	 \over 	\varphi_{\rm g,\overline{\rm MS}}^{(\Nf)}(g_*)}\,.
\end{equation}
As anticipated by our notation,  $P_{\ell,\rm f}$ can be considered
as a function of $M/\Lambda_{\overline{\rm MS}}^{(\Nf)}$, since 
the value of the coupling $g_*$ can be expressed in terms of the 
RGI parameters through the relation
\begin{equation}
	\label{eq:gstar}
	{M\over\Lambda_{\overline{\rm MS}}^{(\Nf)}}=
	{\varphi_{\rm m,\overline{\rm MS}}^{(\Nf)}(g_*)
	\over
	\varphi_{\rm g,\overline{\rm MS}}^{(\Nf)}(g_*)}\,,
	\qquad
	g_*=\bar{g}^{(\Nf)}_{\overline{\rm MS}}(	{M/\Lambda_{\overline{\rm MS}}^{(\Nf)}}).
\end{equation}
In the limit where $M/\Lambda^{(\Nf)}_{\overline{\rm MS}}\to\infty$,
the coupling $g_*$ goes to zero and the function $P_{\ell,\rm f}$
admits an asymptotic perturbative expansion in terms of $g_*^2$. 

\subsubsection{Perturbative uncertainties}
\label{subsubsec:DecouplingPT}

We now want to study the behavior of the  
perturbative expansion of $P_{\ell,\rm f}(M/\Lambda)$. 
We shall focus our attention on the functions 
$P_{\rm 3,4}$ and $P_{\rm 4,5}$, which are the 
relevant ones for estimating 
$\Lambda^{(5)}_{\overline{\rm MS}}$ given
$\Lambda^{(3)}_{\overline{\rm MS}}$. 

\begin{figure*}[h]
	\centering
	%	\vspace*{5cm}       % Give the correct figure height in cm
	\resizebox{0.375\textwidth}{!}{%
	\includegraphics{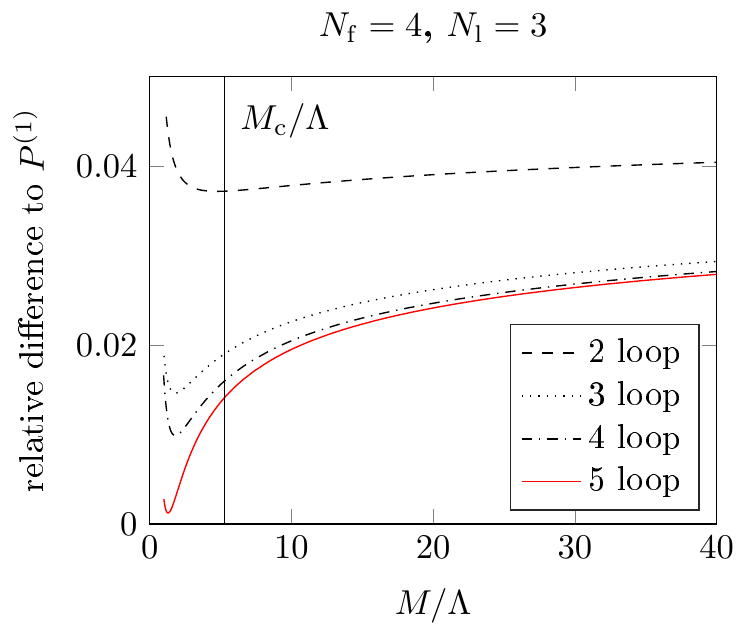}}
	\hspace*{10mm}
	\resizebox{0.375\textwidth}{!}{%
	\includegraphics{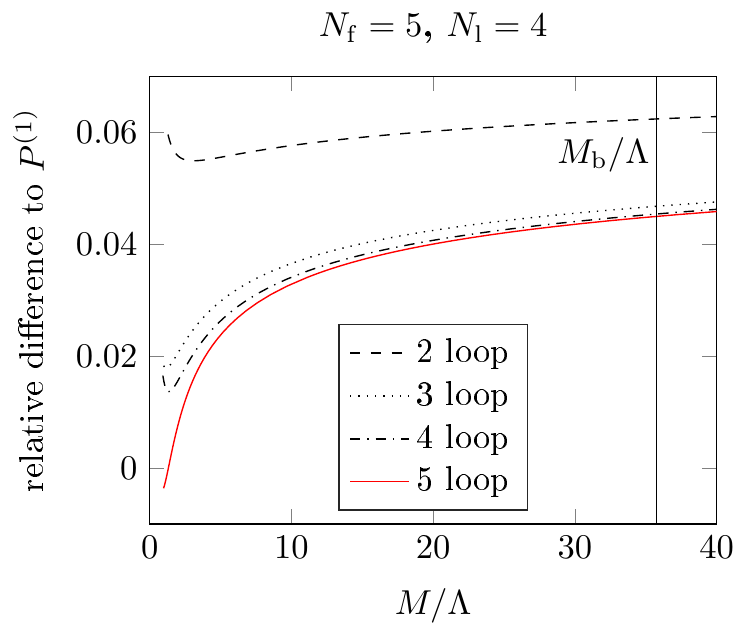}}
	\caption{Relative differences from the 
	``1-loop approximation'' 
	$P^{(1)}_{\ell, \rm f}(M/\Lambda)=(M/\Lambda)^{\eta_0}$, 
	$\eta_0=2\Nh/(33-2\Nl)$, for different orders of the
	perturbative expansion of $P_{\ell, \rm f}(M/\Lambda)$
	as a function of $M/\Lambda$~\cite{Athenodorou:2018wpk}. 
	The results for $\Nl=3$, $\Nf=4$
	($\Nl=4$, $\Nf=5$) are given in the left (right) panel.
	The values for the RGI charm ($M_c$) and bottom ($M_b$) quark
	masses in units of the proper $\Lambda$-parameters are 
	marked by vertical lines.}
	\label{fig:PerturbativePlf}       
\end{figure*}

In fig.~\ref{fig:PerturbativePlf} we present the results 
taken from ref.~\cite{Athenodorou:2018wpk} for the relative
deviation 
\begin{equation}
	\label{eq:P1Rel}
	{(P_{\ell,\rm f}-P_{\ell,\rm f}^{(1)})\over P^{(1)}_{\ell,\rm f}} 
\end{equation}
of $P_{\rm 3,4}$ (left plot) and $P_{\rm 4,5}$ (right plot), 
from their \emph{unsystematic} 1-loop approximation
$P^{(1)}_{\ell, \rm f}(M/\Lambda)=(M/\Lambda)^{\eta_0}$, 
where $\eta_0=2\Nh/(33-2\Nl)$~\cite{Athenodorou:2018wpk}.%
\footnote{The unsystematic approximation 
		  $P^{(1)}_{\ell, \rm f}$ obtained in
		  ref.~\cite{Athenodorou:2018wpk} happens to
		  be quite close to the different orders of 
		  the perturbative expansion of $P_{\ell,\rm f}$
		  for the values of $\Nf$ and $\Nl$ considered
		  and $M/\Lambda\lesssim 30$. For this 
		  reason it is used here to get the overall 
		  magnitude of the function $P_{\ell, \rm f}$. 
		  On the other hand, the different orders of 
		  perturbation theory are not expected to converge 
		  to $P^{(1)}_{\ell, \rm f}$ as
		  $M/\Lambda\to\infty$ since this approximation 
		  does not have the correct asymptotic
		  limit~\cite{Athenodorou:2018wpk}.}
The results are shown 
as a function of $M/\Lambda$ and for different orders
of perturbation theory. The order in perturbation theory
refers to the order at which the $\beta$-functions 
entering eqs.~(\ref{eq:Plf})-(\ref{eq:gstar}) are
considered. The $\tau$-function, eq.~(\ref{eq:tauMSbar}),
as well as the matching function $C(g_*)$ of
eq.~(\ref{eq:CouplingMatchingPT}) are considered to 
a consistent order in the expansion (see
ref.~\cite{Athenodorou:2018wpk} for the details). The values 
of the RGI masses of the charm and bottom quarks ($M_c$ and 
$M_b$, respectively) in units of the relevant $\Lambda$-parameter
are also shown. These are inferred from the results of the
PDG~\cite{Zyla:2020zbs}.

Starting from the case of $P_{4,5}$, we see how
the difference between the 3-loop and 2-loop results 
is around 2\% at the $b$-quark mass. The 4- and 5-loop 
results then differ only by very tiny corrections
from the 3-loop approximation. 
Looking at the behavior of the perturbative series alone, 
the perturbative prediction for the decoupling of the 
$b$-quark appears to be very reliable and accurate. 
Similar conclusions can be drawn for the decoupling of the
charm quark. Also in the case of $P_{3,4}$, the difference 
between the 3- and 2-loop results is about 2\% at the 
charm-quark mass, and higher-order corrections are all much
smaller. 

In conclusion, the perturbative results suggest 
that a perturbative treatment of the matching between the 
relevant effective and fundamental theories introduces only 
errors at the sub-percent level in the functions $P_{\ell,\rm f}$, 
and therefore in the connection of their $\Lambda$-parameters.

\subsection{Non-perturbative decoupling}

\subsubsection{Definitions}
\label{subsubsec:NonperturbativeDecDef}

Judging from perturbation theory alone, the perturbative
description of the decoupling of heavy quarks seems to 
work remarkably well. The rapid convergence of the different 
orders of the expansion of $P_{\ell,\rm f}$ at the values 
of both the charm- and bottom-quark masses, seems to suggest 
that the series is well within its regime of applicability 
and higher-order corrections are small. On the other hand, 
the perturbative expansion cannot tell us anything about 
the size of non-perturbative corrections to the decoupling
relations and whether perturbation theory actually applies at all.
Whether non-perturbative effects are significant within the 
target precision at the relevant quark masses can only be
established through a non-perturbative investigation. 

In order to set the grounds for estimating the error that 
one makes when using a perturbative approximation for 
$P_{\ell,\rm f}$ to extract $\Lambda_{\overline{\rm MS}}^{(\Nf)}$ 
from $\Lambda_{\overline{\rm MS}}^{(\Nl)}$, let us begin by
recasting the matching of the effective and fundamental theory 
in more non-perturbative terms.%
\footnote{For simplicity we consider also here the case where 
		  the light quarks are massless.}
	  
The leading-order effective theory describes the fundamental
theory at low energy only when the corresponding $\Lambda$-parameter
$\Lambda^{(\Nl)}$ is a properly chosen function of the scale 
$\Lambda^{(\Nf)}$ of the fundamental theory and of the 
RGI mass $M$ of the heavy quarks. To be more precise,
consider a low-energy mass-scale $\mathcal{S}\ll M$. This could be,
for instance, a hadronic mass or any of the popular
technical scales,
$t^{-1/2}_0,r^{-1}_0,w^{-1}_0$~\cite{Luscher:2010iy,Sommer:1993ce,Borsanyi:2012zs}.
Matching then means fixing the scale $\Lambda^{(\Nl)}$ through the
condition~\cite{Athenodorou:2018wpk}
\begin{equation}
	\label{eq:MatchingNPT}
	{\Lambda^{(\Nl)}\over \mathcal{S}^{(\Nl)}}= 
	P_{\ell,\rm f}^{\mathcal{S}}(M/\Lambda^{(\Nf)})
	{\Lambda^{(\Nf)}\over \mathcal{S}^{(\Nf)}(M)}\,.
\end{equation} 
In these equations $\mathcal{S}^{(\Nl)}$ and
$\mathcal{S}^{(\Nf)}(M)$ refer to the given low-energy scale
computed in the effective and fundamental theory, respectively. 
Note in particular that the matching function 
$P_{\ell,\rm f}^{\mathcal{S}}(M/\Lambda^{(\Nf)})$ depends
on the scale $\mathcal{S}$ that is considered 
in the matching relation. We also stress again that while 
the value of $\Lambda^{(\Nf)}$ does not depend on $M$
the one of $\Lambda^{(\Nl)}$ does depend through the 
matching condition. 

Once $\Lambda^{(\Nl)}$ is properly fixed through 
eq.~(\ref{eq:MatchingNPT}) in terms of $\mathcal{S}$, 
for any other low-energy quantity $\mathcal{S}'$ we expect that
\begin{equation}
	\mathcal{S}'^{(\Nl)}=\mathcal{S}'^{(\Nf)}(M)
	+{\rm O}\big(\big({\Lambda^{(\Nf)}/ M}\big)^2\big)\,.
\end{equation}
Note that ratios of low-energy scales, instead, do not depend
on the value of the $\Lambda$-parameters and are therefore 
insensitive to their matching. For these, it readily holds
that
\begin{equation}
	\label{eq:RatioS'S}
	{\mathcal{S}^{(\Nl)}\over \mathcal{S}'^{(\Nl)}}=
	{\mathcal{S}^{(\Nf)}(M)\over \mathcal{S}'^{(\Nf)}(M)}
	+{\rm O}\big(\big({\Lambda^{(\Nf)}/ M}\big)^2\big)\,,
\end{equation}
with $\mathcal{S}$ and $\mathcal{S}'$ any two low-energy scales.
Given this observation, from eq.~(\ref{eq:MatchingNPT}) we
conclude that 
\begin{equation}
	P_{\ell,\rm f}^{\mathcal{S}}=
	P_{\ell,\rm f}^{\mathcal{S}'}+{\rm O}(({\Lambda^{(\Nf)}/ M})^2)\,.
\end{equation}
In other words, at the non-perturbative level the function
$P_{\ell,\rm f}^{\mathcal{S}}$ intrinsically
comes with O($M^{-2}$) ambiguities. For this reason we shall 
often simply write it as $P_{\ell,\rm f}$, keeping the dependence
on $\mathcal{S}$ and the related ambiguities understood.

An interesting consequence of the above relations 
follows from multiplying eq.~(\ref{eq:MatchingNPT}) by
$\mathcal{S}^{(\Nf)}(0)/\Lambda^{(\Nf)}$, where
$\mathcal{S}^{(\Nf)}(0)$ stands for the low-energy
quantity $\mathcal{S}$ computed in the chiral limit
of the $\Nf$-flavor theory. Through this 
simple manipulation we find
that~\cite{Bruno:2014ufa,Athenodorou:2018wpk}
\begin{align}
	\nonumber
	{\mathcal{S}^{(\Nf)}(M)\over \mathcal{S}^{(\Nf)}(0)}
	&= Q^{\mathcal{S}}_{\ell,\rm f} \times 
	P^{\mathcal{S}}_{\ell,\rm f}(M/\Lambda^{(\Nf)})\\
	\label{eq:FactorizationQP}
	&= Q^{\mathcal{S}}_{\ell,\rm f} \times 
	P_{\ell,\rm f}(M/\Lambda^{(\Nf)})
	+{\rm O}\big(\big({\Lambda^{(\Nf)}/ M}\big)^2\big)\,,
\end{align}
where the factor
\begin{equation}
	Q^{\mathcal{S}}_{\ell,\rm f}\equiv
	{\mathcal{S}^{(\Nl)}/\Lambda^{(\Nl)}\over
	\mathcal{S}^{(\Nf)}(0)/\Lambda^{(\Nf)} }\,,
\end{equation}
is defined in terms of the massless $\Nf$- and $\Nl$-flavor 
theories. The interesting aspect of eq.~(\ref{eq:FactorizationQP}) 
is that the ratio on the l.h.s. can be computed within 
the fundamental theory, while the r.h.s. is a consequence 
of the decoupling of the heavy quarks. In particular, we see how
the mass dependence of the ratio
${\mathcal{S}^{(\Nf)}(M)/\mathcal{S}^{(\Nf)}(0)}$
is expressed in terms of the function 
$P^{\mathcal{S}}_{\ell, \rm f}$, while the factor
$Q^{\mathcal{S}}_{\ell, \rm f}$ is just an overall constant. 
In the limit of large mass $M$, the mass dependence of this 
ratio is therefore expected to be universal and described by 
perturbation theory. As a result, this relation allows us
to put at test the perturbative expansion of $P_{\ell, \rm f}$
and estimate the typical size of non-perturbative corrections.

To this end, it is convenient in practice to introduce the
mass-scaling function~\cite{Athenodorou:2018wpk} 
\begin{equation}
	\eta_{\ell,\rm f}^{M,\mathcal{S}}(M) \equiv 
	{M\over P^{\mathcal{S}}_{\ell,\rm f}} 
	{\partial P^{\mathcal{S}}_{\ell,\rm f}\over \partial M}\bigg|_{\Lambda^{(\Nf)}}\,.
\end{equation}
Considering eq.~(\ref{eq:FactorizationQP}), 
this can be computed from the mass dependence of  
any hadronic quantity as%
\footnote{Note that the number of light quarks on the
		  r.h.s. of this equation is implicitly given 
		  by the difference between $\Nf$ and the number 
		  $\Nh$ of heavy quarks of which the mass is varied.}
\begin{equation}
	\label{eq:etaM}
	\eta_{\ell,\rm f}^{M,\mathcal{S}}(M) = 
	{M\over \mathcal{S}^{(\Nf)}(M)} 
	{\partial {\mathcal{S}}^{(\Nf)}(M)\over \partial M}\bigg|_{\Lambda^{(\Nf)}}\,,
\end{equation}
with no need for determining $\mathcal{S}^{(\Nf)}(0)$ or
$Q_{\ell,\rm f}$.  As for the functions 
$P^{\mathcal{S}}_{\ell,\rm f}$,
the $\eta_{\ell,\rm f}^{M,\mathcal{S}}$ obtained
from different low-energy quantities $\mathcal{S}$ are 
expected to differ by O($(\Lambda^{(\Nf)}/M)^2$)
contributions. In the limit $M/\Lambda^{(\Nf)}\to\infty$, however,
the mass-scaling function becomes universal and can 
asymptotically be estimated in perturbation
theory~\cite{Athenodorou:2018wpk}. In the following we shall 
see how by studying the mass-scaling function we will be able 
to obtain valuable insight on the applicability of perturbation
theory in computing $P_{\ell,\rm f}$.

\subsubsection{Non-perturbative corrections to decoupling}
\label{subsubsec:NonperturbativeDec}

Non-perturbative effects in the decoupling of heavy quarks 
have been systematically investigated in a series of recent
papers~\cite{Bruno:2014ufa,Athenodorou:2018wpk,Knechtli:2017xgy,Cali:2017brl,Cali:2019enm,Hollwieser:2020qri}, which focus 
on the relevant case of the charm. The main question 
that these works contribute to answer is how good of an 
approximation $\Nf=2+1$ QCD is to the $\Nf=2+1+1$ flavor 
theory. From our perspective, we are particularly interested 
in understanding how precisely we can expect to obtain
$\Lambda_{\overline{\rm MS}}^{(4)}$ (and consequently
$\Lambda_{\overline{\rm MS}}^{(5)}$) from results in 
$\Nf=2+1$ QCD. As already mentioned, the first issue is to 
understand how accurately we can estimate 
$\Lambda_{\overline{\rm MS}}^{(4)}$ from 
$\Lambda_{\overline{\rm MS}}^{(3)}$ by relying on a 
perturbative approximation for 
$P_{3,4}(M_c/\Lambda_{\overline{\rm MS}}^{(4)})$.
Secondly, we must question how much we can rely on
setting the physical units of the theory using $\Nf=2+1$ QCD.

Studying the decoupling of the charm quark through
simulations of $\Nf=2+1+1$ and $\Nf=2+1$ QCD is very
challenging from the computational point of view. 
As a result, the physical effects one is after can  
easily end up being masked by the final uncertainties.
For this reason, the authors of refs.~\cite{Bruno:2014ufa,Athenodorou:2018wpk,Knechtli:2017xgy} 
have investigated non-perturbatively a model system 
given by QCD with $\Nf=2$ degenerate heavy quarks. 
When the mass of the doublet of quarks becomes 
large the heavy quarks eventually decouple, and the theory is
expected to be described at low energy by the pure Yang-Mills
theory, i.e. $\Nl=0$ flavor QCD. Studying this model rather 
than the realistic case, avoids the complications 
of simulating light quarks. This allows one 
to reach much finer lattice spacings than typically possible 
in large-volume simulations with light quarks, because the
volumes are similar to those in pure-gauge theory. Fine lattice
spacings are essential to have discretization errors under control
in the presence of quarks with masses close to that of the charm. 

By studying this model one expects to be able to reliably 
estimate the typical size of the effects induced by the 
charm in low-energy physics. In particular, the absence of 
the light quarks is not expected to change the picture very 
much and their effect is likely more than compensated by the
additional heavy quark present in the model. In fact, as we 
shall report below, some first results have been recently 
obtained for the more realistic situation where a single charm
quark decouples in the presence of 3 mass-degenerate lighter
quarks~\cite{Hollwieser:2020qri}. The results collected so far
confirm the findings of the model study.

\begin{figure*}[h]
	\centering
	%	\vspace*{5cm}       % Give the correct figure height in cm
	\resizebox{0.8\textwidth}{!}{%
		\includegraphics{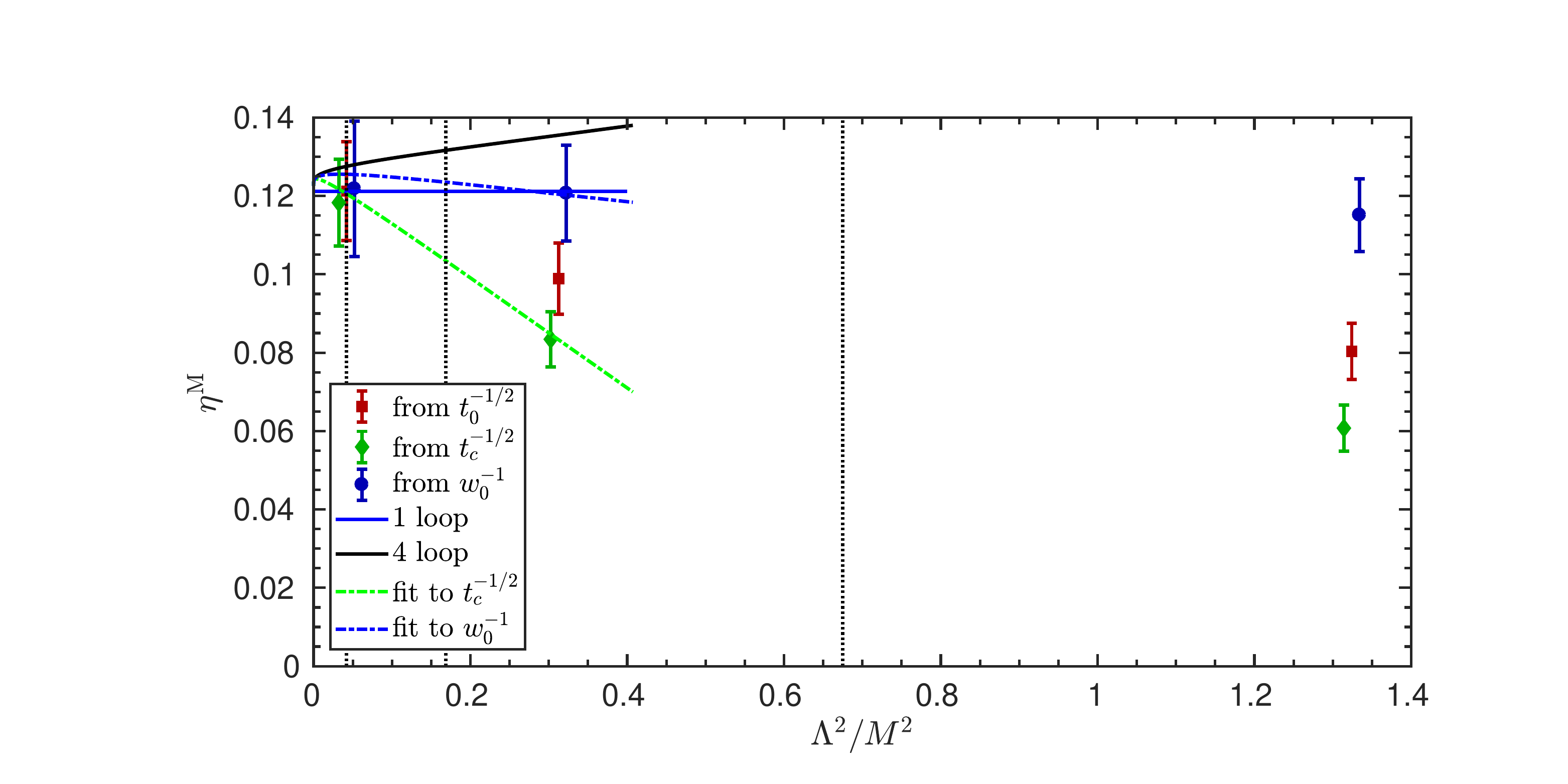}}
	\caption{Mass dependence of the mass-scaling functions
	$\eta^{M,\mathcal{S}}_{0,2}$ obtained from the hadronic 
	scales $\mathcal{S}=t_0^{-1/2},t_c^{-1/2},w_0^{-1}$
	\cite{Athenodorou:2018wpk}. The data for a given mass $M$ 
	are slightly displaced horizontally for better clarity. 
	The non-perturbative results are compared to the
	perturbative estimates at 1- and 4-loop order. The dash-dotted
	lines are fits to the data for $\mathcal{S}=t_c^{-1/2}$ and
	$w_0^{-1}$ used to estimate the size of the non-perturbative
	corrections to
	$\eta_{0,2}^{M,\mathcal{S}}$~\cite{Athenodorou:2018wpk}. 
	The vertical dotted lines mark the values of the quark mass
	$M_c,M_c/2$, and $M_c/4$.}
	\label{fig:etaM}     
\end{figure*}

\paragraph{Ratios of $\Lambda$-parameters}

\begin{figure*}[h]
	\centering
	%	\vspace*{5cm}       % Give the correct figure height in cm
	\resizebox{0.45\textwidth}{!}{%
	\includegraphics{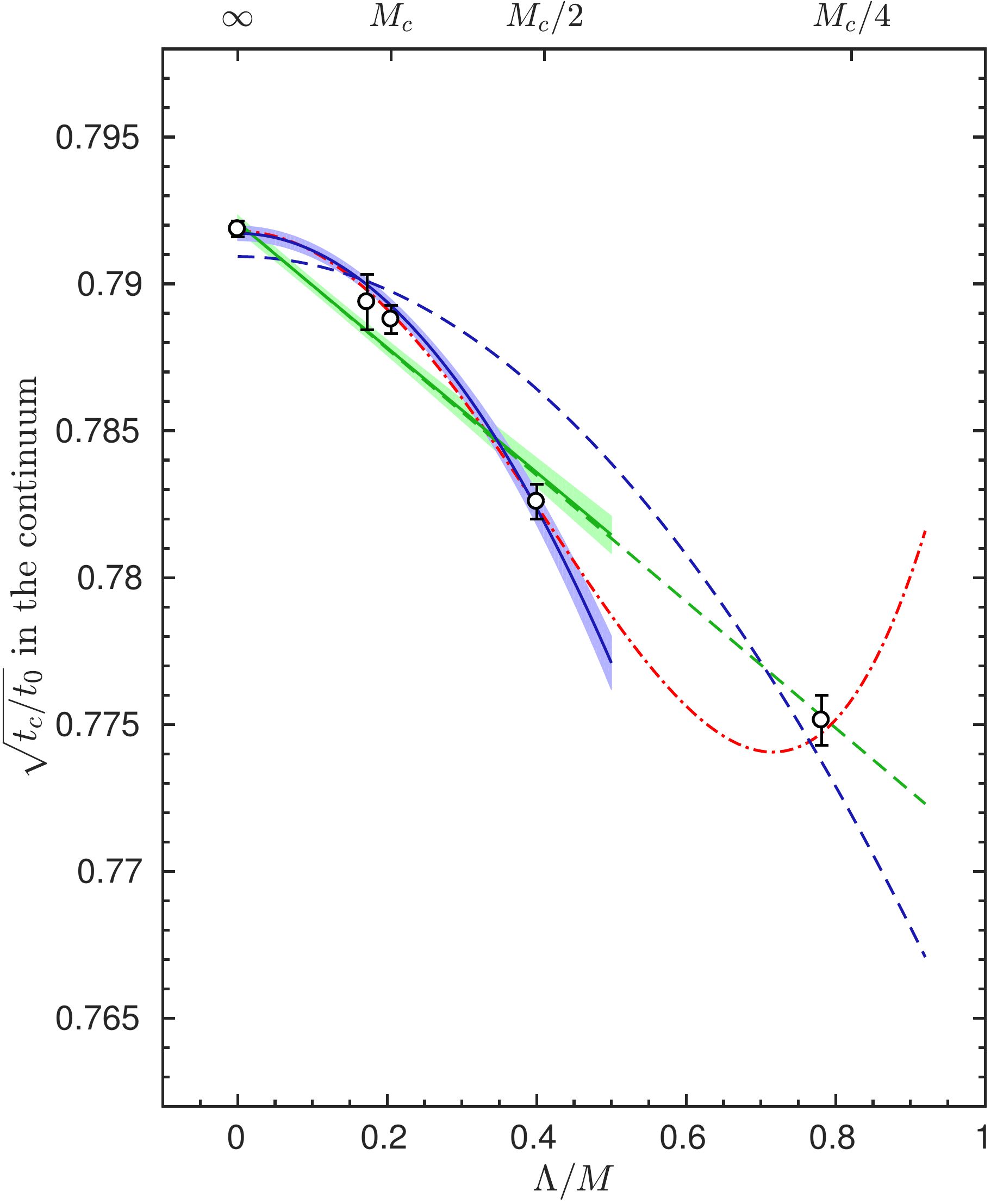}}
	\hspace*{10mm}
	\resizebox{0.45\textwidth}{!}{%
	\includegraphics{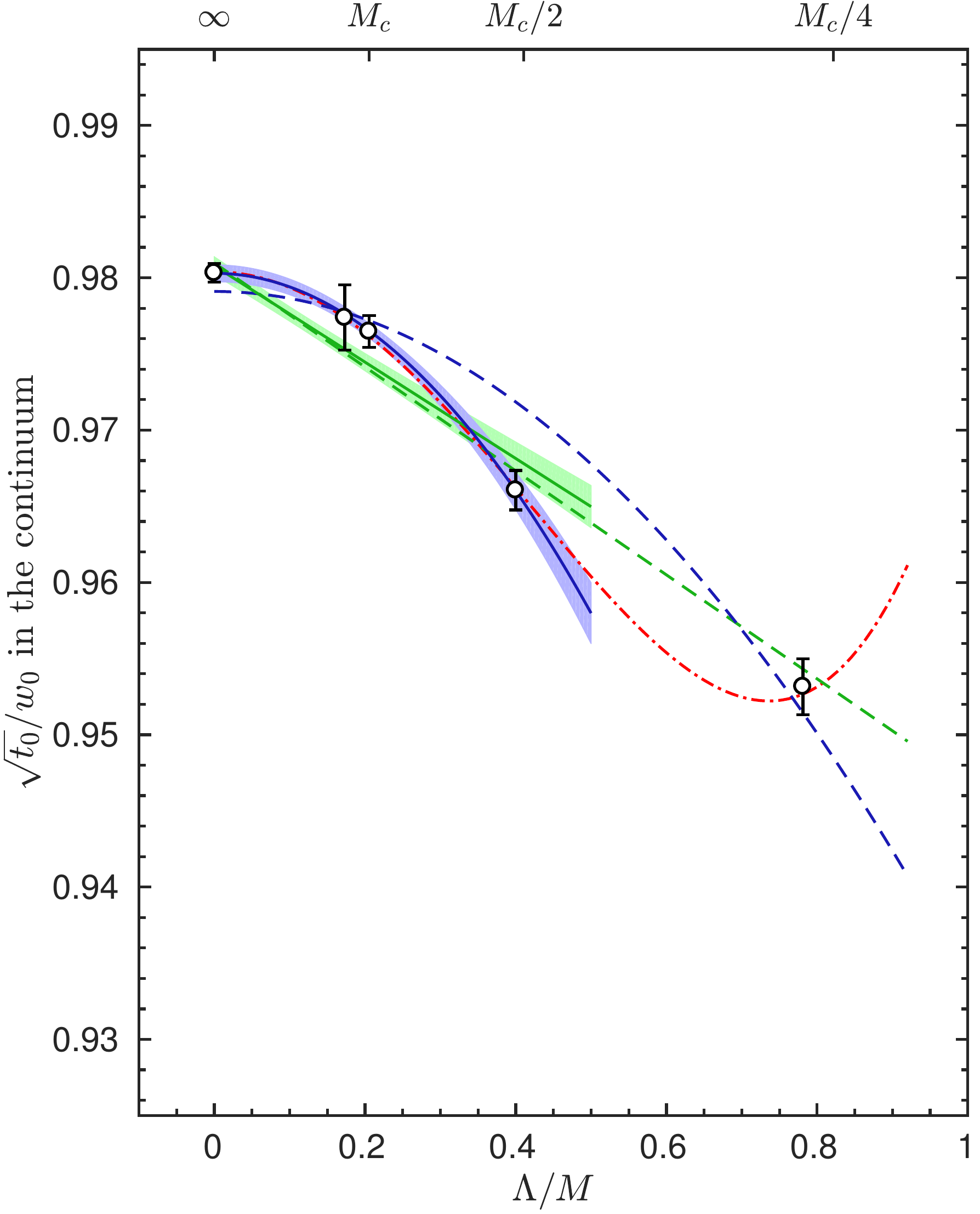}}
	\caption{Continuum extrapolated values for 
		$\sqrt{t_c/t_0}$ (left) and $\sqrt{t_0}/w_0$ (right) 
		as a function of $\Lambda/M$~\cite{Knechtli:2017xgy}. 
		The line in the blue band is the leading-order effective 
		theory  prediction: 
		$R(M) =R(\infty) +k_R\times\Lambda^2/M^2$,  $R=\sqrt{t_c/t_0},
		\sqrt{t_0}/w_0$, with $k_R$ a free parameter, fitted through points
		from $M=\infty$ down to $M/\Lambda=2.5$.  The  line in the 
		green band is instead a fit linear in $\Lambda/M$. For
		comparison the dashed lines represent the quadratic (blue) 
		and linear (green) fit through points from $M=\infty$ down to 
		$M/\Lambda = 1.28$. Also shown by the dashed-dotted red line 
		is a fit in this range adding to the leading-order prediction 
		a next-to-leading term proportional to $\Lambda^4/M^4$.}
	\label{fig:LowEnergyRatios}      
\end{figure*}

The first set of results that we want to discuss are 
from ref.~\cite{Athenodorou:2018wpk} and are shown 
in figure~\ref{fig:etaM}. They correspond to the 
determination of the mass-scaling functions
$\eta^{M,\mathcal{S}}_{0,2}$ based on the gluonic 
low-energy scales $\mathcal{S}=t_0^{-1/2},t_c^{-1/2},w_0^{-1}$
(cf.~eq.~(\ref{eq:etaM})). The precise definition of these
scales can be found in the given reference. The results refer
to the continuum limit of the model system introduced above, 
i.e.~$\Nf=2$ QCD with two heavy quarks. The RGI mass $M$ of 
the heavy quarks is varied from about $M_c/8$ to $1.2M_c$, 
where $M_c$ is the RGI mass of the charm quark.%
\footnote{The value of the charm-quark mass in simulations 
		  is set by targeting
		  $M_c/\Lambda_{\overline{\rm MS}}^{(2)}=4.87$, which
		  is obtained using 
		  $\Lambda_{\overline{\rm MS}}^{(2)}=310\,\MeV$
		  from ref.~\cite{Fritzsch:2012wq} and 
		  $M_c=1510\,\MeV$ from ref.~\cite{Heitger:2013oaa}
	  	  (cf.~ref.~\cite{Athenodorou:2018wpk} for the details).}
The plot also includes the results for $\eta^{M}_{0,2}$ at 
1- and 4-loop order in perturbation
theory~\cite{Athenodorou:2018wpk}. 

As one can see from the figure, the results for the mass-scaling
functions corresponding to different low-energy observables
significantly differ at the smaller values of $M$ in the range. 
As expected, however, they consistently approach each other as
$M\to\infty$. In particular, for values of the mass close to the
charm-quark mass all determinations well agree within errors,
indicating the smallness of the O($M_c^{-2}$) corrections. 

The results for $\eta^{M}_{0,2}$ are about $1/10$
for $M\approx M_c$, both in perturbation theory and
non-perturbatively. In fact, within the uncertainties
of roughly 10\% the perturbative and non-perturbative results
perfectly agree. Note that even though the relative precision 
on $\eta^{M}_{0,2}$ might not seem impressive, it corresponds 
to an absolute error on $\eta^{M}_{0,2}$ of about $0.01$.
This means, in particular, that it is reasonable to assume 
that the difference, $\Delta\eta^{M}_{0,2}$, between the
non-perturbative $\eta^{M}_{0,2}$ and its (4-loop) perturbative
approximation is bounded by this error for masses $M\gtrsim M_c$.
The scaling of the non-perturbative corrections 
$\Delta\eta^{M}_{0,2}$ to $\eta^{M}_{0,2}$ as a function of $M$ 
can then be assessed by studying the $M$-dependence seen in 
figure \ref{fig:etaM} and comparing the results for 
the different observables $\mathcal{S}$~\cite{Athenodorou:2018wpk}. 

Putting all this information together, the authors of
ref.~\cite{Athenodorou:2018wpk} obtain a conservative
estimate for the size of the non-perturbative contributions 
$\Delta P_{0,2}$ to $P_{0,2}$. We shall not report their 
detailed discussion here and refer the interested reader 
to ref.~\cite{Athenodorou:2018wpk}.%
\footnote{The deviation $\Delta\log(P_{0,2})$ of
	      the full function $\log(P_{0,2})$ from its 
	      perturbative approximation is obtained by 
	      integrating $\Delta\eta^M_{0,2}$ 
	      in $\log(M/\Lambda)$ from the value of $M/\Lambda$ 
	      of interest up to  $M/\Lambda=\infty$
	      (cf.~eq.~(\ref{eq:etaM}) and
	      ref.~\cite{Athenodorou:2018wpk}). 
	      In order to translate to $P_{0,2}$ the estimate 
	      for the non-perturbative corrections to 
	      $\eta^M_{0,2}$ made for $M\lesssim M_c$, one therefore
	      needs to make some assumptions on how these
	      approach the limit $M/\Lambda\to\infty$. Depending
          on the assumptions, more or less conservative 
          estimates are obtained.}
Summarizing  their conclusions, given the results for
$\Delta\eta^{M}_{0,2}$ obtained in the model, one can safely 
state that the non-perturbative contributions to 
$P_{0,2}(M_c/\Lambda_{\overline{\rm MS}}^{(2)})$ 
are \emph{at most} 2\% and quite likely at the level of 0.4\%. 
This translates into at least a 2\% precision of perturbation
theory in the conversion of $\Lambda$-parameters for the
investigated case.

What can we conclude from this about the phenomenologically
interesting case of $P_{3,4}(M_c/\Lambda_{\overline{\rm MS}}^{(4)})$? 
We first note that the dependence in perturbation theory of
$\eta^{M}_{\ell,\rm f}$ on $\Nl$ at fixed $\Nh$ is very mild. 
At leading order it amounts to about a 20\% effect in going 
from $\Nl=0$ to $\Nl=3$~\cite{Athenodorou:2018wpk}. For this
reason, the authors of ref.~\cite{Athenodorou:2018wpk} include 
an additional 50\% contribution to their estimate for the
non-perturbative corrections $\Delta\eta^{M}_{0,2}$ to account 
for the missing light-quark effects. Secondly, our intuition 
from both perturbative and non-perturbative considerations
suggests that most likely the effects of the decoupling of a 
single charm quark are about half of those of two quarks. From 
these observations, one concludes that one can safely neglect
non-perturbative effects in connecting 
$\Lambda_{\overline{\rm MS}}^{(3)}$ and 
$\Lambda_{\overline{\rm MS}}^{(4)}$ down to a precision of 1.5\%
or better~\cite{Athenodorou:2018wpk}.

\paragraph{Ratios of hadronic quantities}
The second important category of effects that we  
address are non-perturbative contributions from heavy quarks 
to dimensionless ratios of low-energy quantities. For 
us these are relevant in the context of setting the physical 
scale of the theory (see e.g.~ref.~\cite{Sommer:2014mea}). 
As well-known, in fact, besides the technical difficulties 
of computing the relevant ratios, an important issue that one 
faces in setting the scale in lattice QCD is the fact that one 
never really simulates the ``real world'', where experiments 
are conducted. Hence, when comparing the lattice results with
experiments in order to set the scale of the theory, one must
``correct'' the experimental quantities for effects that are 
not taken into account in the lattice simulations, or at least
verify that these are not relevant once compared with the rest 
of the uncertainties. The most common examples of effects to 
be considered are the difference in the $u$- and $d$-quark 
masses, electromagnetic effects, and the unphysical number of
quark flavors.

In the following we focus on the issue of estimating 
the effects of the charm quark in low-energy quantities, and
to which extent this can be omitted in lattice simulations. 
From the point of view of determining 
$\Lambda_{\overline{\rm MS}}^{(4)}$ from the results 
of $\Lambda_{\overline{\rm MS}}^{(3)}$, thus, the relevant 
question is how accurate the determination of the physical scale 
is from simulations in the 3-flavor theory. This amounts to 
quantify how accurate it is to compute in $\Nf=2+1$ rather than
$\Nf=2+1+1$ QCD, the ratios of low-energy quantities that are 
used to set the physical scale.

As presented in eq.~(\ref{eq:RatioS'S}), the effects of
heavy quarks in dimensionless ratios of low-energy
quantities are expected to be of O($M^{-2}$), provided 
that the mass of the heavy quarks is large enough compared 
to the energy scales of the observables considered. 
In ref.~\cite{Knechtli:2017xgy} 
a systematic study was conducted in order to assess the range 
of heavy-quark masses for which the O($M^{-2}$) scaling of 
the heavy-quark effects predicted by the effective theory 
actually sets in. In addition, the authors estimated the 
typical size of these contributions when $M\approx M_c$. For 
their computations, they considered the very same model of 
QCD with two degenerate heavy quarks previously introduced, 
and the same range of masses, $M\approx M_c/8-1.2M_c$. The 
results for several different ratios of low-energy quantities 
obtained in the fundamental $\Nf=2$ QCD theory and in the
effective pure-gauge theory were compared. Note that since 
the effective theory is purely gluonic only gluonic quantities 
were considered. These, however, are all relevant observables 
entering realistic scale-setting determinations. 

In figure \ref{fig:LowEnergyRatios} we show two examples 
given by the ratios $\sqrt{t_c/t_0}$ (left panel) and
$\sqrt{t_0}/w_0$ (right panel), evaluated in the continuum 
limit. The results from $\Nf=2$ QCD for different values of 
the heavy-quark masses $M\gtrsim M_c/4$ are shown, as well as 
those from the effective pure-gauge theory which correspond 
to $M=\infty$. Several fits to the data are proposed. Let us 
focus first on the quadratic (blue) and linear (green) fits in
$\Lambda/M$. Among these four fit types, the data seem to favor 
the leading-order effective theory predictions: $R(M) =R(\infty)
+k_R\times\Lambda^2/M^2$, $R=\sqrt{t_c/t_0},\sqrt{t_0}/w_0$,
with $k_R$ free parameters, restricted to masses 
$M/\Lambda \gtrsim 2.5$, i.e. $M\gtrsim M_c/2$ (blue bands). 
Both linear fits in $\Lambda/M$, either excluding (green bands) 
or including (green dashed lines) the point at $M\approx M_c/4$
have significantly larger $\chi^2$ per degree of freedom than 
the previous fits (cf.~Table 3 in ref.~\cite{Knechtli:2017xgy}).
The pure $(\Lambda/M)^2$ fits which include the $M\approx M_c/4$
results (blue dashed lines), instead, are clearly off. 

Although the results do not completely exclude a linear $\Lambda/M$
dependence, the fits obtained by adding to the leading-order
prediction a next-to-leading term $\propto(\Lambda/M)^{4}$ and
including data down to $M\approx M_c/4$ (red dashed lines), 
further support the findings from the previous fits. Indeed, 
the close agreement for $M\gtrsim M_c/2$ of these fits and the
leading-order predictions restricted to this range, reinforce 
the conclusion that the O($M^{-2}$) scaling sets in for masses
$M\gtrsim M_c/2$, while for smaller masses higher-order
contributions become relevant, or the expansion has broken
down entirely.

\begin{figure}[h]
	\centering
	%	\vspace*{5cm}       % Give the correct figure height in cm
	\resizebox{0.515\textwidth}{!}{%
	\includegraphics{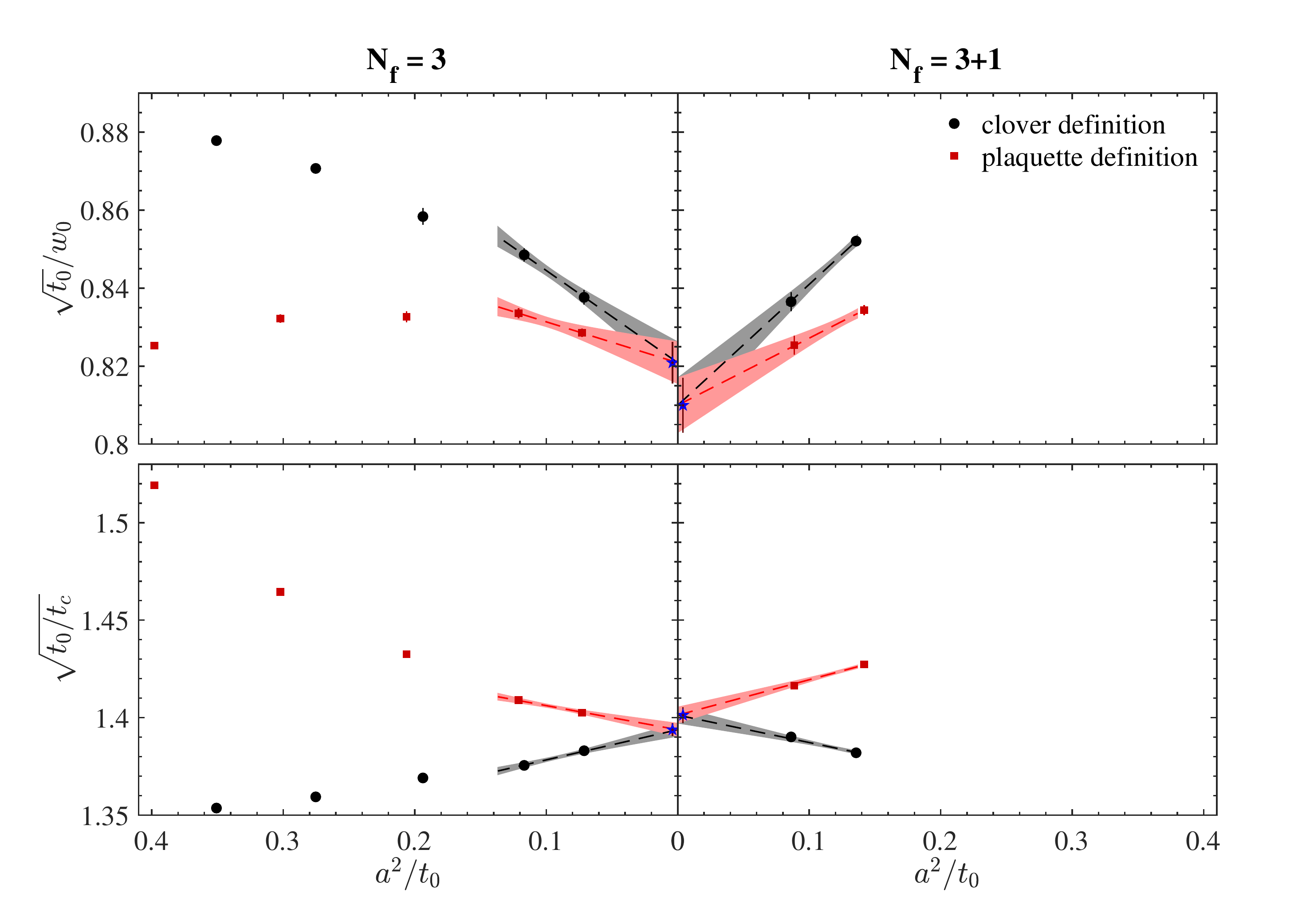}}
	\caption{Comparison of continuum limit extrapolations
	of $\sqrt{t_0/t_c}$ (bottom) and $\sqrt{t_0}/w_0$ (top)
	of the $\Nf= 3 + 1$ data from ref.~\cite{Hollwieser:2020qri}
	(right) with corresponding $\Nf= 3$ CLS results (left) 
	from refs.~\cite{Bruno:2014jqa,Bruno:2016plf} including the
	finest ensemble J500, cf.~ref.~\cite{Bali:2019yiy}.}
	\label{fig:ScaleNf3p1}    
\end{figure}

In general the corrections induced by the heavy quarks are 
small. They amount to $2-3\%$ at the smallest masses in the plot, 
around $M_c/4$, but they are reduced significantly below the
percent level, to about $0.4\%$, once $M\approx M_c$. 
As discussed above, we expect that this result provides a reliable
estimate for the magnitude of these effects in the realistic case 
of the decoupling of the charm quark in $\Nf=2+1+1$ QCD. 
The simultaneous decoupling of two charm-like quarks rather than
just one, likely compensates the missing effects from the
light quarks; this at least in the purely gluonic quantities under
consideration.

These expectations are confirmed by the recent results 
of ref.~\cite{Hollwieser:2020qri}. In figure \ref{fig:ScaleNf3p1} 
we show their continuum limit extrapolations for different
discretizations of the ratios $\sqrt{t_0/t_c}$ and $\sqrt{t_0}/w_0$.
In this case, the results from $\Nf=3$ and $\Nf=3+1$ QCD are
compared. Both simulations include 3 mass-degenerate quarks with 
a mass around the physical average mass of the $u$-, $d$- and
$s$-quarks. The $\Nf=3+1$ simulations include in addition a 
fourth quark with a mass set to the physical value of the
charm. By comparing the results of the two set of simulations 
one can directly test decoupling in a close-to-real situation. 
As one can see from the figure, the differences in these very
precise ratios of low-energy gluonic quantities are far below 
the percent level and of the order of magnitude found in the 
model. Interestingly, the ensembles generated in
ref.~\cite{Hollwieser:2020qri} open the possibility to study
systematically the effects of the charm quark also in fermionic
low-energy observables. 

\paragraph{Conclusions}
Let us summarize what we have learned from the studies above. 
1) The ratio $\Lambda_{\overline{\rm MS}}^{(4)}/
\Lambda_{\overline{\rm MS}}^{(3)}$ can be safely computed 
in perturbation theory with a precision of \emph{at least}
1.5\%, and realistically much better. In fact, it is important 
to stress that the estimate for non-perturbative corrections 
to the function $P_{3,4}(M_c/\Lambda_{\overline{\rm MS}}^{(4)})$ 
is \emph{very} conservative and the actual size of these effects 
is much likely quite smaller, i.e. at the level of $0.3\%$ or 
so~\cite{Athenodorou:2018wpk}. 2) Power corrections of 
O($M^{-2}_c$) in low-energy observables are also found to 
be very small, i.e. well-below the percent level
\cite{Knechtli:2017xgy}. All in all, this means that
$\Lambda_{\overline{\rm MS}}^{(5)}$ can be accurately predicted 
at the 1-2\% level from $\Lambda_{\overline{\rm MS}}^{(3)}$.
In this respect, we note that the competitive precision of about
0.7\% on the $\alpha_s(M_Z)$ determination of
ref.~\cite{Bruno:2017gxd} (to be discussed below), corresponds to
an uncertainty of 3.5\% on $\Lambda_{\overline{\rm MS}}^{(3)}$.
This uncertainty is also conservative. In conclusions, there is
plenty of room for improvement within $\Nf=3$ QCD.

% end

%% file: sect4.tex
% sect4.tex

\section{Renormalization by decoupling}
\label{sec:RenByDec}

\subsection{The QCD coupling from $\Nf=3$ QCD}
\label{subsec:LambdaNf3}

In the previous section we established that by relying 
on perturbative decoupling relations for the charm and 
bottom quarks, precise determinations of 
$\Lambda_{\overline{\rm MS}}^{(5)}$ are possible from 
results in the $\Nf=3$ flavor theory. In order to be 
able to make this statement the systematic studies on 
the non-perturbative effects induced by the charm quark 
in low-energy quantities have been instrumental. 

Having this settled, as discussed in detail in
Sect.~\ref{sec:PrecisionDeterminations}, the main challenges 
for an accurate determination of $\alphas$ on the lattice 
are: 1) controlling discretization errors in continuum limit
extrapolations of the chosen non-perturbative coupling(s), 
and 2) estimating the uncertainties associated with the use 
of perturbation theory in extracting the $\Lambda$-parameter. 
In this respect, the combined application of finite-volume
renormalization schemes and finite-size scaling techniques 
has proven to be extremely effective in dealing with these
difficulties, paving the way for robust and precise
lattice determinations of the QCD coupling. 

The determination of ref.~\cite{Bruno:2017gxd}, in particular,
relies on these techniques to compute 
$\Lambda_{\overline{\rm MS}}^{(3)}$. The calculation reaches a
final precision on $\Lambda_{\overline{\rm MS}}^{(3)}$ of 
about $3.5\%$, which translates into a $0.7\%$ uncertainty on
$\alphas(M_Z)$. The strength of this result lies in the fact 
that all systematic uncertainties are carefully kept under control
at this competitive level of accuracy. The calculation is therefore
a prominent example of the current state-of-the-art determinations
of $\alphas$ from the lattice~\cite{Aoki:2019cca}. Below we want 
to briefly recall the main steps that led to this result in order 
to understand what the challenges are in improving on it. For a
more detailed presentation we refer the interested reader to the
original
references~\cite{Brida:2016flw,DallaBrida:2016kgh,Bruno:2017gxd,DallaBrida:2018rfy} and 
reviews~\cite{Korzec:2017ypb,DallaBrida:2018cmc,Ramos:2019,Sint:2019bmq,DallaBrida:2019zmm}.

\begin{figure}[h]
	\centering
	%	\vspace*{5cm}       % Give the correct figure height in cm
	\resizebox{0.475\textwidth}{!}{%
		\includegraphics{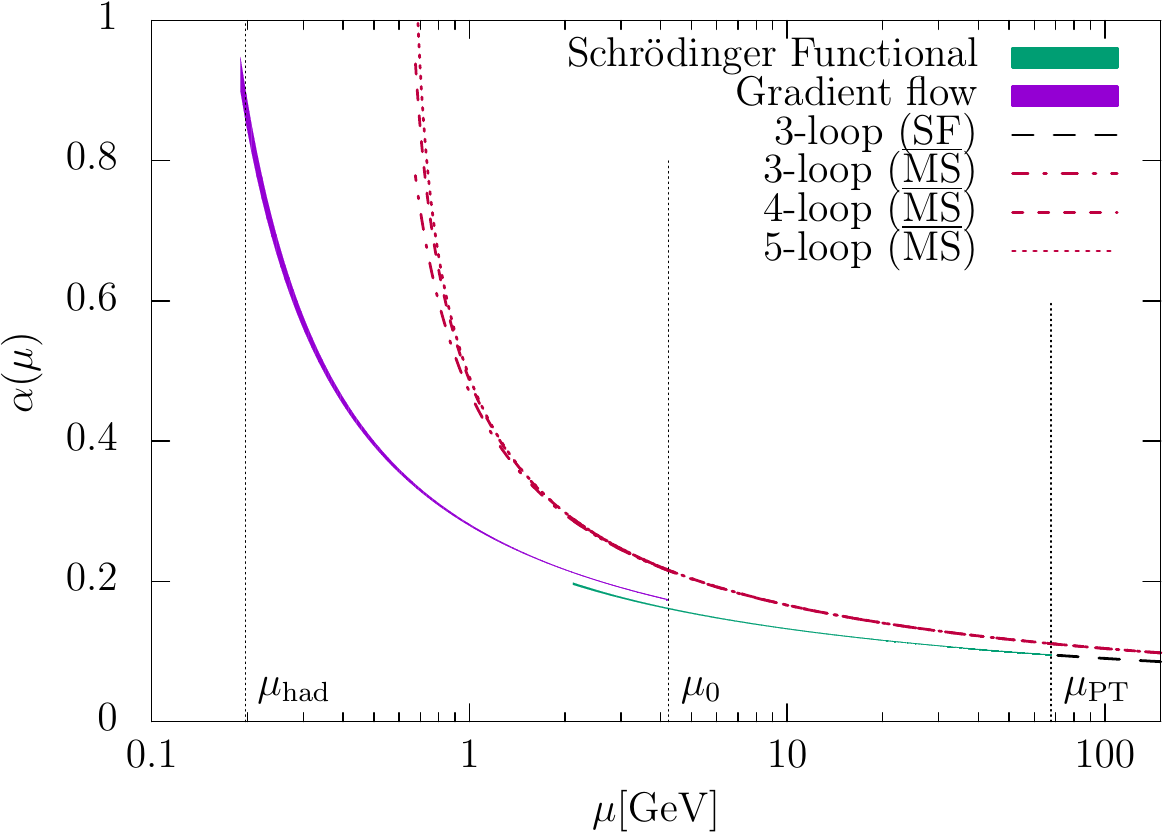}}
	\caption{Running couplings of $\Nf= 3$ QCD from 
		integrating the non-perturbative $\beta$-functions in the 
		SF- and GF-schemes~\cite{Brida:2016flw,DallaBrida:2016kgh}.
		They  are  matched non-perturbatively at the scale 
		$\mu_0$ defined by  $\bar{g}^2_{\rm SF_{\nu=0}}(\mu_0)  =  2.012$ by computing $\bar{g}^2_{\rm GF}(\mu_0/2)=2.6723(64)$~\cite{DallaBrida:2016kgh}.
		The scales $\mu_{\rm PT}=16\mu_0$ and $\mu_{\rm had}$
		defined by $\bar{g}^2_{\rm GF}(\mu_{\rm had})=11.31$ are
		also shown, as well as the perturbative prediction
		for the ${\rm SF}_{\nu=0}$-coupling for $\mu>\mu_{\rm PT}$
		using the 3-loop $\beta$-function. 
		The red curves correspond to the results for
		$\alpha^{(3)}_{\overline{\rm MS}}(\mu)$	obtained from
		$\Lambda_{\overline{\rm MS}}^{(3)}
		=341(12)\,\MeV$~\cite{Bruno:2017gxd}, considering different
		perturbative orders for the $\beta$-function in the 
		$\overline{\rm MS}$-scheme.}
	\label{fig:AlphaRunning}     
\end{figure}

The $\Lambda_{\overline{\rm MS}}^{(3)}$ determination of
ref.~\cite{Bruno:2017gxd} was obtained from the study of the
non-perturbative running of some convenient finite-volume 
schemes from a scale of about $0.2\,\GeV$ to roughly 
$70\,\GeV$. The very high energies reached non-perturbatively
allowed for a systematic and robust assessment of the uncertainties 
related to the application of perturbation theory. This study 
has been presented in Sect.~\ref{subsubsec:AccuracyNf3}, where 
the high-energy end of the running in the SF-schemes and the 
result for $\Lambda_{\overline{\rm MS}}^{(3)}/\mu_{0}$ with 
$\mu_{0}\approx 4.3\,\GeV$, have been discussed in detail. 
The rest of the determination is built on the following steps.

Firstly, we have the computation of the lower end of the
running and corresponding determination of the ratio of
finite-volume scales $\mu_{0}/\mu_{\rm had}$ 
with $\mu_{\rm had}\approx 200\,\MeV$~\cite{DallaBrida:2016kgh}. 
For this step, a novel finite-volume coupling defined in terms 
of the Yang-Mills gradient flow was employed. This allowed for
reaching much greater precision than otherwise possible using
the schemes considered at high energy~\cite{Brida:2014joa}. 
Note that a non-perturbative matching between the finite-volume
schemes at $\mu_0$ was performed in order to continue the running 
at lower energies. For illustration, we show in figure
\ref{fig:AlphaRunning} the non-perturbative running of the
finite-volume couplings over the energy range covered,
together with the results for the coupling in the 
$\overline{\rm MS}$-scheme obtained from the corresponding
determination of $\Lambda_{\overline{\rm MS}}^{(3)}$ (see below).

In a second step, the relation $\mu_{\rm had}/f_{\pi K}$
was established passing through the intermediate technical 
scale $\mu_{\rm ref}^*=1/\sqrt{8t^*_0}$~\cite{Bruno:2017gxd}. 
Here, $f_{\pi K}\equiv\frac{1}{3}(2f_K+f_\pi)$ is a 
convenient combination of the pion and kaon decay constants, 
while the scale $\mu_{\rm ref}^*$ is given in terms of the 
flow time $t^*_0$ defined in the SU(3) flavor-symmetric
limit~\cite{Bruno:2016plf}. This step involved a combination of
small-volume and large-volume simulations in the hadronic
regime~\cite{Bruno:2017gxd}. From the experimental value of 
$f_{\pi K}$ the precise physical units for $\mu_{\rm had}$ could 
be inferred and hence those of $\Lambda_{\overline{\rm MS}}^{(3)}$.
Finally, perturbation theory was used for the functions 
$P_{3,4}(M_c/\Lambda_{\overline{\rm MS}}^{(4)})$ 
and $P_{4,5}(M_b/\Lambda_{\overline{\rm MS}}^{(5)})$  
to obtain $\Lambda_{\overline{\rm MS}}^{(5)}$ and from this
$\alphas(M_Z)$. Splitting the determination of 
$\Lambda_{\overline{\rm MS}}^{(5)}$ over the above
steps was the key to keep all systematic errors under control.
With the proper choice of observables and techniques 
the hard multi-scale problem of relating the low- and 
high-energy sectors of QCD could be solved in full confidence.

\begin{figure}[h]
	\centering
	%	\vspace*{5cm}       % Give the correct figure height in cm
	\resizebox{0.475\textwidth}{!}{%
	\includegraphics{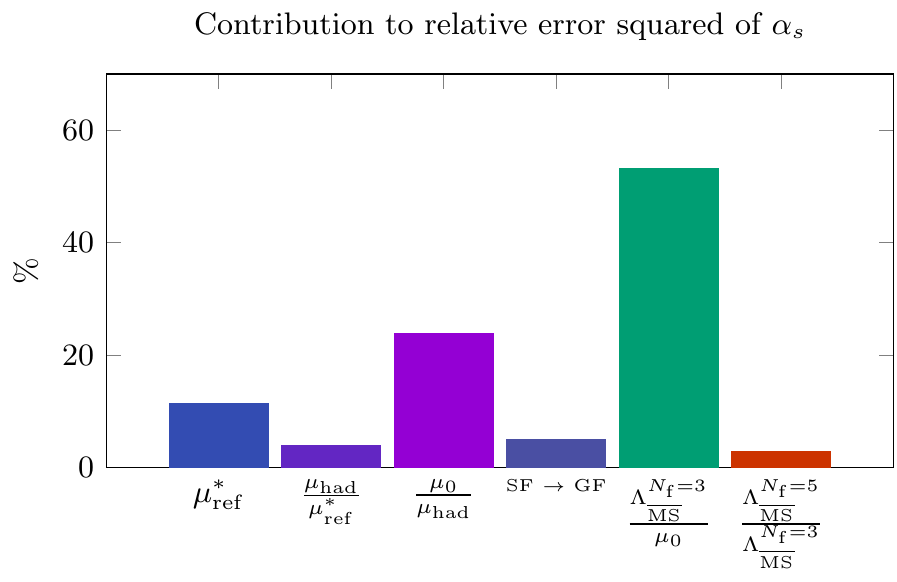}}
	\caption{Contribution in percentage to the relative error
	squared of $\alphas(M_Z)$ from the different steps of the 
	determination of ref.~\cite{Bruno:2017gxd} 
	(cf. text for more details). As the reader can see, the 
	dominant source of uncertainty is the non-perturbative 
	running at high energy $\mu\approx4-70\,\GeV$
	($\Lambda_{\overline{\rm MS}}^{(\Nf=3)}/\mu_0$)~\cite{Brida:2016flw,DallaBrida:2018rfy},
	followed by the running at low energy $\mu\approx 0.2-4\,\GeV$
	($\mu_0/\mu_{\rm had}$)~\cite{DallaBrida:2016kgh}, and 
	by scale setting 
	($\mu_{\rm ref}^*$)~\cite{Bruno:2016plf}.
	Note that the error from decoupling 
	($\Lambda_{\overline{\rm MS}}^{(5)}/
	\Lambda_{\overline{\rm MS}}^{(3)}$) is only	perturbative. 
	However, even adding the very conservative uncertainty 
	estimated in Sect.~\ref{subsubsec:NonperturbativeDec} for 
	the non-perturbative contributions to the decoupling of 
	the charm quark, the total error is still dominated by 
	the one from the running.}
	\label{fig:ErrorBudget}      
\end{figure}

It is now instructive to look at the error budget of 
this $\alphas$ determination. This is given in figure
\ref{fig:ErrorBudget} which shows the contribution in 
percentage to the relative error squared on $\alphas(M_Z)$ from 
the different steps described above~\cite{Korzec:2017ypb}.
As it is clear from the figure, the main source of uncertainty 
comes from the determination of the non-perturbative running 
from $\mu_{\rm had}\approx 0.2\,\GeV$ up to 
$\mu_{\rm PT}\approx 70\,\GeV$, where perturbation theory 
is applied to extract $\Lambda_{\overline{\rm MS}}^{(3)}/\mu_{\rm PT}$ (cf.~Sect.~\ref{subsubsec:AccuracyNf3}). In particular, 
the error accumulated by running from $\mu_{0}$ to $\mu_{\rm PT}$
(labeled as $\Lambda_{\overline{\rm MS}}^{(\Nf=3)}/\mu_0$ in 
the plot) contributes roughly 60\% of the total budget. 

It is important to recall at this point that the error 
coming from the running is completely dominated by 
statistical uncertainties. In particular, thanks to 
the fact that $\mu_{\rm PT}\approx 70\,\GeV$ was reached
non-perturbatively, the uncertainties due to the use of
perturbation theory are well below the statistical errors 
(cf.~Sect.~\ref{subsubsec:AccuracyNf3}). In this respect, 
we want to stress the important difference between this 
and the majority of other lattice determination of $\Lambda_{\overline{\rm MS}}^{(3)}$, 
where perturbation theory is applied at scales 
$\mu_{\rm PT}\lesssim 2-3\,\GeV$. In these cases, a large 
fraction of the final error comes from the \emph{systematic}
uncertainties related to the truncation of the perturbative
series and possible remnants of non-perturbative contributions
(cf.~ref.~\cite{Aoki:2019cca}). As we have seen, estimating 
these systematics reliably is very challenging, particularly so 
when precision is desired but the accessible range of 
scales is limited to low energy. In this situation, 
a reduction of the final uncertainties is highly non-trivial, 
and can eventually come only from reaching significantly higher
energy scales. Without a step-scaling approach this is in practice
extremely demanding in QCD given the present computational and
algorithmic capabilities.%
\footnote{We recall that with the current state-of-the-art 
		  algorithms, the cost of QCD simulations for reducing 
		  the lattice spacing $a$ at fixed physical volume 
	  	  and number of independent configurations scales
	  	  $\propto a^{-7}$
	  	  (cf.~e.g.~ref.~\cite{Schaefer:2012tq}). Hence, 
	  	  reducing the lattice spacing by a factor $2$ or so
	  	  requires a factor of O($100$) in the cost.}

In the case of the step-scaling method, on the other hand, 
reducing the uncertainties on the current $\alphas(M_Z)$
determination is a question of reducing the statistical
uncertainties coming from the running of the coupling(s) 
in $\Nf=3$ QCD. This is in principle a straightforward task.
However, reducing the total error by an \emph{important} 
factor, say a factor 2 or so, is yet a non-trivial 
challenge from the computational point of view.

Rather than following a brute force approach for the reduction
of the error in the computation of the running in $\Nf=3$ QCD,
in the following we shall discuss a novel strategy 
which promises the desired error reduction in a
significantly cheaper way~\cite{DallaBrida:2019mqg}.  
It is based on the ideas presented in the previous section 
on heavy-quark decoupling. The distinct feature of the approach 
is that one can replace the non-perturbative computation 
of the running in $\Nf=3$ QCD needed to determine 
$\Lambda_{\overline{\rm MS}}^{(3)}$ with the  
running in the pure Yang-Mills theory.%
\footnote{We take the opportunity to point out the study of 
		  ref.~\cite{Brower:2014dfa} where ideas based on
		  heavy-quark decoupling are considered in order to 
		  address other RG-related problems in the context 
		  of Beyond the Standard Model Physics.}

\subsection{The coupling from decoupling}

\subsubsection{General strategy and master formula}

Let us begin by considering QCD with $\Nf$ flavors of 
heavy quarks of RGI mass $M$. In this theory all quarks 
are massive and there are no light quarks. As we have 
seen in Sect.~\ref{subsec:HQET}, as the mass $M$ becomes 
larger and larger this theory is expected to be approximated
better and better by an effective theory given 
by the pure Yang-Mills theory. 
In particular, once the $\Lambda$-parameters of the fundamental 
and effective theory are properly matched, dimensionless 
low-energy observables can be computed in the effective theory 
up to corrections of O($M^{-n}$), where, in general, $n=2$
if the theories are matched at leading order. 

The decoupling of heavy quarks is also valid for 
the interesting case of couplings defined in massive
renormalization
schemes~\cite{Appelquist:1974tg,Bernreuther:1981sg}. This 
is a direct consequence of the fact that such couplings 
are defined in terms of dimensionless observables in the 
massive theory. Following the notation of Sect.~\ref{subsubsec:LambdaParams}, we 
indicate the generic renormalized  massive coupling in 
the $\Nf$-flavor theory as $\bar{g}^{(\Nf)}_{\Obs}(\mu,M)$, 
where $\Obs$ denotes the short-distance observable 
used to define the coupling (cf.~eq.~(\ref{eq:alphaO})), 
while $\mu$ is the renormalization scale. From the decoupling 
of heavy quarks applied to the observable $\Obs$ it follows 
that,
\begin{equation}
	\label{eq:CouplingDecouplingNP}
	\bar{g}^{(\Nf)}_\Obs(\mu,M)=
	\bar{g}^{(0)}_\Obs(\mu)+{\rm O}(M^{-2})\,,
\end{equation}
where $\bar{g}^{(0)}_\Obs(\mu)$ refers to the corresponding
coupling in the pure Yang-Mills theory evaluated at the 
\emph{same} physical scale $\mu$. Note that here and below 
we shall  loosely denote as ${\rm O}(M^{-k})$ terms, 
contributions that contain terms of 
${\rm O}((\Lambda/M)^{k})$ as well as ${\rm O}((\mu/M)^{k})$.

The decoupling relation, eq.~(\ref{eq:CouplingDecouplingNP}),
can be equivalently recast in terms of the renormalization scales
$\mu$ implicitly defined by the couplings. Specifically, given a
numerical value for the coupling $g_M$, we define the scales
$\mu_{\rm dec}^{(\Nf)}$ and $\mu_{\rm dec}^{(0)}$ through
\begin{equation}
	\label{eq:mudec}
	\bar{g}^{(\Nf)}_\Obs(\mu_{\rm dec}^{(\Nf)},M)=
	g_M
	=\bar{g}^{(0)}_\Obs(\mu_{\rm dec}^{(0)})\,.
\end{equation}
From the theory of decoupling it then follows that
\begin{equation}
	\mu_{\rm dec}^{(0)}=
	\mu_{\rm dec}^{(\Nf)}+{\rm O}(M^{-2})\,.
\end{equation}
We stress again that we assume that
the $\Lambda$-parameters in the two theories are 
properly matched.

From this basic observation the master formula proposed in
ref.~\cite{DallaBrida:2019mqg} follows. We start by
considering the  relation in eq.~(\ref{eq:MatchingNPT}), 
and take for the low-energy scale $\mathcal{S}$ the 
renormalization scale $\mu_{\rm dec}$ defined above 
in terms of the given couplings. In formulas,
\begin{equation}
	\label{eq:PreMasterEquation0}
	{\Lambda^{(0)}_{\overline{\rm MS}}\over \mu_{\rm dec}^{(0)}}= 
	P_{0,\rm f}(M/\Lambda^{(\Nf)}_{\overline{\rm MS}})
	{\Lambda^{(\Nf)}_{\overline{\rm MS}}\over \mu_{\rm dec}^{(\Nf)}}+{\rm O}(M^{-2})\,,
\end{equation}
where for later convenience we took the $\Lambda$-parameters
in the $\overline{\rm MS}$-scheme.%
\footnote{Note that the relation is non-perturbatively valid	
		  at this point as $\Lambda_{\overline{\rm MS}}$ can
	  	  be expressed exactly in terms of any non-perturbative
  	  	  scheme (cf.~Sect.~\ref{subsubsec:LambdaParams}).}
Now, rather than interpreting the above equation as a matching 
relation for the $\Lambda$-parameters that defines the function
$P_{0,\rm f}$, we shall turn tables and use it to \emph{predict} 
the ratio ${\Lambda^{(\Nf)}_{\overline{\rm MS}}/\mu_{\rm dec}^{(\Nf)}}$ in terms of ${\Lambda^{(0)}_{\overline{\rm MS}}/ \mu_{\rm dec}^{(0)}}$. 

To this end, we first replace the function $P_{0,\rm f}$ with 
its perturbative expansion $P_{0,\rm f}^{\rm PT}$ in the 
${\overline{\rm MS}}$-scheme to some order 
$n$ (cf.~eqs.~(\ref{eq:Plf}),(\ref{eq:LambdaMuPT})
and Sects.~\ref{subsubsec:DecouplingPT},
\ref{subsubsec:NonperturbativeDecDef}),
\begin{equation}
	\label{eq:PreMasterEquation1}
	P_{0,\rm f}(M/\Lambda^{(\Nf)}_{\overline{\rm MS}})=
	P^{\rm PT}_{0,\rm f}(M/\Lambda^{(\Nf)}_{\overline{\rm MS}})+
	{\rm O}(g_*^{2n-2},M^{-2})\,.
\end{equation}
We therefore have that,
\begin{equation}
	\label{eq:PreMasterEquation2}
	{\Lambda^{(\Nf)}_{\overline{\rm MS}}\over \mu_{\rm dec}^{(\Nf)}}
	P^{\rm PT}_{0,\rm f}(M/\Lambda^{(\Nf)}_{\overline{\rm MS}})=
	{\Lambda^{(0)}_{\overline{\rm MS}}\over \mu_{\rm dec}^{(0)}}
	+{\rm O}(g_*^{2n-2},M^{-2}).
\end{equation}
As a second step, using eq.~(\ref{eq:LambdaParam}) and
the definition in eq.~(\ref{eq:mudec}) we write,
\begin{equation}
	\label{eq:PureGaugeLambda}
	{\Lambda^{(0)}_{\overline{\rm MS}}\over \mu_{\rm dec}^{(0)}}
	= {\Lambda^{(0)}_{\overline{\rm MS}}\over
	   \Lambda^{(0)}_\Obs}\,\varphi^{(0)}_{\rm g,\Obs}(g_M)\,,
\end{equation}
which only involves quantities in the pure-gauge theory.
We recall that $\mathcal{O}$ appearing here is the observable 
used to calculate the coupling in
eq.~(\ref{eq:mudec}). Moreover, as discussed in 
Sect.~\ref{subsubsec:LambdaParams}, the change of scheme given 
by the ratio ${\Lambda^{(0)}_{\overline{\rm MS}}/
\Lambda^{(0)}_\Obs}$ can be computed exactly through a
1-loop calculation (cf.~eq.~(\ref{eq:MStoOLambda})). 

Finally, introducing the dimensionless variables,
\begin{equation}
	\rho\equiv
	{\Lambda^{(\Nf)}_{\overline{\rm MS}}/ 
	\mu_{\rm dec}^{(\Nf)}}\,,
	\qquad
	z\equiv {M/\mu_{\rm dec}^{(\Nf)}}\,,
\end{equation}
using eqs.~(\ref{eq:PreMasterEquation2}) and
(\ref{eq:PureGaugeLambda}) we arrive at the master
equation,
\begin{equation}
	\label{eq:MasterEquation}
	\rho\,P^{\rm PT}_{0,\rm f}(z/\rho) =
	{\Lambda^{(0)}_{\overline{\rm MS}}\over
	\Lambda^{(0)}_\Obs}\,\varphi^{(0)}_{\rm g,\Obs}(g_M)	
	+{\rm O}(g_*^{2n-2},M^{-2}),
\end{equation}
which can be solved for $\rho$ once the pure-gauge function
$\varphi^{(0)}_{\rm g,\Obs}(g_M)$ is known. As promised, 
the master formula allows us to replace the non-perturbative
computation of the running of the coupling from the low-energy
scale $\mu_{\rm dec}$ up to infinite energy in $\Nf$-flavor
QCD with the corresponding running in the pure-gauge theory. 
A few remarks are in order at this point. 

First of all, it is important to stress the fact that 
eq.~(\ref{eq:MasterEquation}) is \emph{exact} in the limit 
where $M\to\infty$. In this limit both the perturbative
O($g_*^{2n-2}$) corrections and the non-perturbative O($M^{-2}$)
contributions vanish. The basic idea that is applied in this
strategy is in fact similar to when we extract 
$\Lambda_{\overline{\rm MS}}^{(4)}$ from 
$\Lambda_{\overline{\rm MS}}^{(3)}$
by replacing the non-perturbative function
$P_{3,4}(M_c/\Lambda^{(4)}_{\overline{\rm MS}})$ with its
perturbative approximation to some order, neglecting both
higher-order terms and non-perturbative corrections. The 
crucial difference in the present case is that the approximation 
can be made systematically better by considering larger values 
of $M$, which, at least in principle, is a free parameter of 
the strategy. In this respect, note that the perturbative
corrections to $P_{\rm 0, f}(M/\Lambda^{(\Nf)}_{\overline{\rm MS}})$
only depend on the value of $M/\Lambda^{(\Nf)}_{\overline{\rm MS}}$,
while the choice of scale $\mu_{\rm dec}^{(\Nf)}$ does not
matter. From the results presented in
Sect.~\ref{subsubsec:DecouplingPT}, we infer that
the perturbative errors due to the truncation of 
the perturbative series for 
$P_{\rm 0, f}(M/\Lambda^{(\Nf)}_{\overline{\rm MS}})$
are already small for
$M/\Lambda_{\overline{\rm MS}}^{(\Nf)}\approx 5$,
for the relevant values of $\Nf$. On the 
other hand, controlling the non-perturbative O($M^{-2}$) 
terms requires to have both 
O($(\Lambda^{(\Nf)}_{\overline{\rm MS}}/M)^2$) and
O($(\mu_{\rm dec}^{(\Nf)}/M)^2$) terms under control.
Whether this is possible in practice must be assessed
by carefully studying the limit $M\to\infty$ with
the accessible values of $M$.

In comparing  eqs.~(\ref{eq:MatchingNPT}) and
(\ref{eq:PreMasterEquation0}), the attentive reader might 
have noticed that we omitted the $M$-dependence 
on the scale $\mu^{(\Nf)}_{\rm dec}$. This was intentional. 
As we shall see in the later subsection, in practice 
it is convenient in fact to define a single scale,
$\mu^{(\Nf)}_{\rm dec}$, common to all the $\Nf$-flavor 
theories defined by the different values of $M$.%
\footnote{The fact that we can set a unique scale for all 
	theories with different $M$ values is a direct 
	consequence of the fact that they all share the very 
	same $\Lambda$-parameter and this is $M$ independent.}
This means that, as anticipated by our notation in
eq.~(\ref{eq:mudec}), we will have different values for $g_M$
depending on the value of $M$ considered. The value of $g_M$ 
is found by computing 
$\bar{g}^{(\Nf)}_\Obs(\mu^{(\Nf)}_{\rm dec},M)$
for the given $M$ at the common scale $\mu^{(\Nf)}_{\rm dec}$.
In lattice QCD, setting a common scale among the theories with
different $M$ values can be achieved via the bare
parameters. This amounts to consider the very same lattice
discretization and establish a line of constant physics
along which $\mu_{\rm dec}^{(\Nf)}$ is  kept fixed. The 
massive couplings are then evaluated at matching values 
of the bare coupling along this line of constant physics
for the different bare quark masses corresponding to the 
target RGI masses. From their continuum limit extrapolations
we find the values $g_M$.

Finally, in order to extract $\Lambda_{\overline{\rm MS}}^{(\Nf)}$ 
from the results for $\rho$ determined from
eq.~(\ref{eq:MasterEquation}) it is necessary to know the 
value of $\mu^{(\Nf)}_{\rm dec}$ in physical units. This is 
obtained by establishing the relation 
$\mu^{(\Nf)}_{\rm dec}/\mu^{(\Nf)}_{\rm phys}$,
where $\mu^{(\Nf)}_{\rm phys}$ denotes a convenient low-energy
scale computed in $\Nf$-flavor QCD at \emph{physical} values of the
quark masses. The scale $\mu^{(\Nf)}_{\rm phys}$ can thus
be related to its experimental counterpart. Of course, it goes
without saying that in order to be able to set the scale 
accurately in terms of experimentally measurable quantities,
as well as to perturbatively match 
$\Lambda_{\overline{\rm MS}}^{(\Nf)}$ and 
$\Lambda_{\overline{\rm MS}}^{(5)}$, we must consider $\Nf=3$
or $4$.

\subsubsection{Another hard multi-scale problem?}
\label{subsubsec:MultiScaleProblem2}

The general strategy presented above is certainly very compelling.
As any other strategy, though, in practical implementations
it comes with its challenges. In particular, in 
order to have all systematic effects under control, it is necessary
to carefully address how to accommodate in lattice simulations 
the different scales that enter the problem. As we shall see, a 
naive approach can easily end up facing severe limitations.

In the general situation, first of all, the space-time volume 
has to be large enough for finite-volume effects to be under 
control in all relevant observables. This means that the
infrared cutoff set by the linear extent $L$ of the lattice must 
be much smaller than all other scales. Secondly, in order to have
small decoupling corrections in eq.~(\ref{eq:MasterEquation}) we 
must have that the heavy-quark mass $M$ is larger than all 
other physical scales, as in particular  $\mu^{(\Nf)}_{\rm dec}$ 
and $\Lambda_{\overline{\rm MS}}^{(\Nf)}$. Note that although
$\mu^{(\Nf)}_{\rm dec}$ is in principle arbitrary, in practice
it is not convenient to take this scale to be much larger than
$\Lambda_{\overline{\rm MS}}^{(\Nf)}\sim \mu^{(\Nf)}_{\rm phys}$. 
Last but not least, all scales have to be
well below the ultraviolet cutoff set by the lattice spacing. 
Putting all these constraints together we find 
(cf.~eq.~(\ref{eq:Window})),
\begin{equation}
	\label{eq:WindowProblem2}
	L^{-1}\ll  
	\mu^{(\Nf)}_{\rm phys} \sim
	\Lambda_{\overline{\rm MS}}^{(\Nf)}\sim
	\mu^{(\Nf)}_{\rm dec} 
	\ll
	M
	\ll
	a^{-1}\,.
\end{equation}
It is clear from this series of inequalities that
having all these scales comfortably resolved on a 
\emph{single} lattice is challenging and requires
a very large $L/a$. Just to give an 
example, if one attempts the calculations using 
a state-of-the-art large-volume ensemble with, say,
$L/a=100$, $m_\pi L=4$, $m_\pi=140\,\MeV$,
which results in $a\approx 0.056\,\fm$, 
eq.~(\ref{eq:WindowProblem2}) 
translates into $M\ll3.5\,\GeV$. 

\subsubsection{Finite-volume couplings rescue us again}
\label{subsec:FiniteVolumeDecoupling}

Some of the constraints encoded in eq.~(\ref{eq:WindowProblem2})
can be lifted by considering for the coupling in 
eq.~(\ref{eq:CouplingDecouplingNP}) a finite-volume scheme,
i.e. $\bar{g}_\Obs(\mu,M)\equiv\bar{g}_\Obs(L^{-1},M)$
(cf.~Sect.~\ref{subsubsec:FV}). With this choice, the 
determination of $\bar{g}_\Obs(\mu^{(\Nf)}_{\rm dec},M)$ does 
not require the physical volume to be large. Furthermore, if 
the decoupling scale $\mu^{(\Nf)}_{\rm dec}$ is also 
defined in terms of a finite-volume coupling, i.e.
$\mu^{(\Nf)}_{\rm dec}\equiv L^{-1}_{\rm dec}$, one has some
additional freedom in the choice of the value of 
$\mu^{(\Nf)}_{\rm dec}$. Any sizable scale separation
between $\mu^{(\Nf)}_{\rm dec}$ and the hadronic scale 
$\mu^{(\Nf)}_{\rm phys}$ can in fact be bridged through
step-scaling within the $\Nf$-flavor theory
(cf.~Sect.~\ref{subsubsec:FV}). Taking $\mu^{(\Nf)}_{\rm dec}$ 
larger at fixed $z=M/\mu^{(\Nf)}_{\rm dec}$, allows us to reach
larger values of $M/\Lambda_{\overline{\rm MS}}^{(\Nf)}$,
as well as to profit from simulating at smaller values of 
$a$. On the other hand, the larger $\mu^{(\Nf)}_{\rm dec}$ 
is, the smaller is the range of energy scales for which, 
through eq.~(\ref{eq:MasterEquation}), the running of 
the coupling in the $\Nf$-flavor theory is replaced 
by that in pure-gauge. In practice, choosing 
$\mu^{(\Nf)}_{\rm dec}={\rm O}(1\,\GeV)$ is  
a good compromise.

By employing finite-volume couplings and scales
the only constraints that we have to meet are to 
have small decoupling corrections in eq.~(\ref{eq:MasterEquation}), 
and to keep discretization errors under control.
The first condition requires $z=ML_{\rm dec}\gg1$,
while discretization effects are small once  
$aM\ll1$ and $a/L_{\rm dec}\ll1$. Putting these 
inequalities together we find the conditions
\begin{equation}
	L_{\rm dec}/a\gg z\gg1.	
\end{equation}
If we take, say,
$\mu^{(\Nf)}_{\rm dec}=1\,\GeV$ and 
$L_{\rm dec}/a=50$, then $M\ll50\,\GeV$.
In summary, using finite-volume schemes, for a given set of
lattice sizes $L/a$ we can reach much finer lattice spacings
$a$, as the extent of the lattice $L$ does not have to be large
in physical units. Smaller $a$ values allow us to consider
larger $M$ values, while having $aM$ reasonably small.
Larger values of $M$ make for a more precise approximation 
$P_{\rm 0,f}^{\rm PT}(M/\Lambda)$. Still, large $z$ values 
have to be reached in order to control non-perturbative 
decoupling corrections.

\paragraph{Heavy-quark decoupling in a finite volume}

As pointed out for the case of computing the running
through a step-scaling procedure (cf.~Sect.~\ref{subsubsec:FV}), 
the choice of finite-volume coupling is dictated by several 
technical aspects, as for instance, statistical precision and
discretization errors. For the strategy based on decoupling 
an additional factor becomes relevant which is the size of
non-perturbative contributions in the decoupling of heavy quarks
(cf.~eq.~(\ref{eq:CouplingDecouplingNP})). In a finite volume, 
the situation can be quite different from one coupling
definition to another, as even the leading power in $M^{-1}$
may be different.

Most finite-volume couplings that are used in practice are 
based on the QCD Schr\"odinger functional (SF) 
\cite{Luscher:1992an,Sint:1993un,Sint:1995rb}. In the 
SF the quark and gluon fields satisfy Dirichlet boundary
conditions at the space-time boundaries located at $x_0=0$ and 
$T$, where $T$ is the temporal extent of the space-time volume
(cf.~eqs.~(\ref{eq:SFbc})-(\ref{eq:SFbc2})). These boundary
conditions guarantee many compelling
features~\cite{Luscher:1992an,Sint:1993un}. However, they 
come with the price of having, for instance, additional 
discretization effects of
O($a$)~\cite{Luscher:1992an,Luscher:1996sc}. 
Using Symanzik effective theory these can be understood as
dimension 4 counterterms localized at the space-time
boundaries~\cite{Luscher:1992an,Luscher:1996sc}.
In close analogy with Symanzik effective theory an analysis of 
the effective Lagrangian for heavy quarks in the presence of SF
boundary conditions shows that the same boundary fields 
appear as O($M^{-1}$) counterterms. More precisely, considering
the relevant case $\Nl=0$, the Lagrangian $\mathcal{L}_1$ in
eq.~(\ref{eq:EffectiveLagrangian}) has the form
(cf.~eq.~(\ref{eq:CountertermsLagrangians}))
\begin{gather}
	\nonumber
	\mathcal{L}_1= \omega_b(g)
	\big[
	\mathcal{B}(0)+\mathcal{B}(T)\big]\,,\\[1ex]
	\label{eq:OMTermsSF}
	\mathcal{B}(x_0)
	=-{1\over g^2}\int \rmd \boldsymbol{x}\,
	  \tr\{F_{0k}(x)F_{0k}(x)\}|_{x_0}\,,
\end{gather}
where $F_{\mu\nu}(x)$ is the gluon-field strength tensor,
while $\omega_b$ is a coefficient function that can be fixed 
by matching with the fundamental $\Nf$-flavor theory.
Note that for simplicity we listed the only gluonic
operator that is relevant for the class of the SF boundary
conditions normally employed (cf.~eqs.~(\ref{eq:SFbc})). 
As a result, if couplings based on the SF are considered, the
decoupling relations
eqs.~(\ref{eq:CouplingDecouplingNP}),(\ref{eq:MasterEquation})
must be corrected to have leading O($M^{-1}$) corrections rather
than O($M^{-2}$)~\cite{Sint:1995ch,Sint:2007ug,DallaBrida:2019mqg}.%
\footnote{The list of O($M^{-1}$) counterterms for the more
	   	  general case, including the situation where the 
		  effective theory also contains light quarks,
		  can be inferred from the O($a$) counterterms
		  of the Symanzik effective theory for the
		  SF~\cite{Luscher:1996sc}.
		  It should be clear that the analogy only refers to 
		  the fields entering the counterterms and, in general,
		  it does not extend to their coefficient functions
		  $\omega_b$. In particular, one expects different
		  bases of counterterm fields to appear at higher orders 
		  in the effective theories, as the two theories have
		  different defining symmetries. From these observations,
		  one concludes that also finite-volume couplings based 
		  on open-SF~\cite{Luscher:2014kea} or
		  open~\cite{Luscher:2011kk} boundary conditions are
		  affected by O($M^{-1}$) corrections.}
On the other hand, finite-volume couplings defined through some 
variant of periodic boundary conditions, as for instance
regular periodic~\cite{Fodor:2012td} or
twisted~\cite{deDivitiis:1994yz,Ramos:2014kla} boundary
conditions, have leading decoupling corrections of O($M^{-2}$), 
as observables in infinite space-time.
	  
There are several possibilities to deal with the issue 
of O($M^{-1}$) corrections in SF-based couplings. A 
straightforward one is to consider a sufficiently large time 
extent $T$, and take for the observable $\Obs$ that defines 
the couplings fields which are localized in the middle of the
space-time manifold, i.e. at $x_0=T/2$. This maximizes the 
distance from the boundaries and therefore the correlation 
between $\Obs$ and the fields responsible for O($M^{-1}$) effects.
In fact, at low energy (i.e. relatively large physical $L$ and $T$)
the O($M^{-1}$) contaminations  are expected to be exponentially
suppressed with the distance of $\Obs$ from the boundaries.

More elegant solutions have been proposed which eliminate
entirely the issue. For example, one could consider for the 
heavy quarks a twisted rather than a standard mass~\cite{Frezzotti:2000nk,Frezzotti:2001ea,DellaMorte:2001ys,Pena:2004gb}. 
In this case, one can show that $\mathcal{L}_1=0$ and the 
decoupling in the SF is realized with O($M_{\rm tw}^{-2}$)
corrections, where $M_{\rm tw}$ is the heavy twisted mass of 
the quarks~\cite{Sint:2007ug}. Equivalent in the 
continuum is  the situation where the heavy quarks have a 
standard mass but the SF boundary conditions are chirally
rotated~\cite{Sint:2005qz,Sint:2007ug,Sint:2010eh}
(see also
refs.~\cite{Sint:2007zz,Sint:2010xy,Lopez:2012as,Lopez:2012mc,Brida:2016rmy,DallaBrida:2018tpn}). The issue with these solutions is that they 
require an even number of flavors $\Nf$. They are therefore a
promising approach for the case of $\Nf=4$ QCD. For $\Nf=3$, 
one may consider having a doublet of twisted-mass quarks
and a regular massive quark (or equivalently a doublet of 
chirally rotated quarks and a regular SF quark 
(see e.g.~ref.~\cite{DallaBrida:2018tpn})). 
This would reduce the O($M^{-1}$) contributions 
to those of only a single flavor. 

Another possibility that we want to mention, which is valid
for any choice of $\Nf$, is to match the effective and 
fundamental theory at O($M^{-1}$)~\cite{Sommer:2020}. In other
words, by equating the results for some convenient observable 
in the effective and fundamental theory one can determine the
coefficient $\omega_b$ appearing in eq.~(\ref{eq:OMTermsSF}). 
Once this is determined, the O($M^{-1}$) terms can be taken 
into account in the effective theory by computing the insertion 
of the counterterm in eq.~(\ref{eq:OMTermsSF}) in the relevant
observables. This guarantees that the decoupling corrections 
are of O($M^{-2}$).

\subsection{$\Lambda_{\overline{\rm MS}}^{(3)}$ from 
			the decoupling of heavy quarks}
\label{subsec:CouplingFromDecoupling}

\subsubsection{Definitions}

In this subsection we present the results of ref.~\cite{DallaBrida:2019mqg} 
where the master formula, eq.~(\ref{eq:MasterEquation}), 
was first applied for the computation of 
$\Lambda_{\overline{\rm MS}}^{(3)}$. The study considers 
$\Nf=3$ QCD which is set on the lattice in terms of
non-perturbatively O($a$)-improved Wilson quarks and the
tree-level Symanzik O($a^2$)-improved gauge
action~\cite{Bulava:2013cta}.%
\footnote{For ease of presentation in the following we  
		  use continuum notation to introduce the relevant
		  definitions while referring to the literature 
		  for the corresponding lattice expressions.}
	  
The theory is defined in a finite volume with time extent 
$T$ and spatial size $L$. The quark fields $\psi,\psibar$ 
and the gauge field $A_\mu$ satisfy Dirichlet boundary 
conditions in the time direction, specifically
($P_\pm\equiv\frac{1}{2}(1\pm\gamma_0)$)~\cite{Luscher:1992an,Sint:1993un,Luscher:1996sc},
\begin{gather}
	\nonumber
	P_+\psi(x)|_{x_0=0}=0
	=
	P_-\psi(x)|_{x_0=T}\,,\\[1ex]
	\nonumber
	\psibar(x)P_-|_{x_0=0}=0=
	\psibar(x)P_+|_{x_0=T}\,,\\[1ex]
	\label{eq:SFbc}
	A_{k}(x)|_{x_0=0,T}=0\,,	
\end{gather}
$k=1,2,3$, while in the spatial directions we have 
\begin{gather}
	\nonumber
	\psi(x+\hat{k}L)=e^{\frac{i}{2}}\psi(x)\,,
	\quad
	\psibar(x+\hat{k}L)=\psibar(x)e^{-\frac{i}{2}}\,,\\
	\label{eq:SFbc2}
	A_\mu(x+\hat{k}L)=A_\mu(x)\,.
\end{gather}
The finite-volume couplings that we consider in the following 
are constructed in terms of the gradient flow field 
$B_\mu(t,x)$ which is defined by the 
equations~\cite{Narayanan:2006rf,Luscher:2010iy},
\begin{equation}
	\label{eq:GradienFlowEq}
	\partial_t B_\mu(t,x)=D_\nu G_{\nu\mu}(t,x)\,,
	\quad
	B_\mu(0,x)=A_\mu(x)\,,
\end{equation}
where $t>0$ is the flow time and  
\begin{equation}
	G_{\mu\nu}=\partial_\mu B_\nu - \partial_\nu B_\mu
	+[B_\mu,B_\nu]\,,
\end{equation}
is the flow-field strength tensor. On the lattice several 
discretizations of the GF equations have been considered,
here we employ the Symanzik O($a^2$)-improved definition
proposed in ref.~\cite{Ramos:2015baa}, also known as 
Zeuthen flow.

Gauge-invariant composite fields made out of the flow field
$B_\mu(t,x)$ are automatically finite~\cite{Luscher:2011bx}, 
and are thus ideal quantities to define renormalized 
couplings~\cite{Luscher:2010iy,Fodor:2012td,Fritzsch:2013je}. 
The definition to which we apply decoupling
(cf.~eq.~(\ref{eq:CouplingDecouplingNP})) is the massive
finite-volume scheme given by
\begin{equation}
	\label{eq:GFTCoupling}
	[\bar{g}^{(3)}_{\rm GFT}(\mu,M)]^2\equiv 
	{t^2\over\mathcal{N}} 
	\langle E_{\rm mag}(t,x)\rangle_{{\rm SF},Q=0}\big|_{T=2L}^{x_0=L,\mu=L^{-1},\sqrt{8t}=cL}\,,
\end{equation}
where $\mathcal{N}$ is a normalization
constant~\cite{Fritzsch:2013je} and 
\begin{equation}
	\label{eq:Emag}
	E_{\rm mag}(t,x)=-\frac{1}{2}\tr\{G_{kl}(t,x)G_{kl}(t,x)\}
\end{equation}
is the magnetic component of the energy density of the 
flow field $B_\mu(t,x)$. On the lattice, we define $E_{\rm mag}$ 
in terms of the O($a^2$)-improved definition given in
ref.~\cite{Ramos:2015baa}. 

Note how in eq.~(\ref{eq:GFTCoupling}) the renormalization 
scale $\mu$ is set in terms of the finite spatial extent 
$L$, as appropriate for a finite-volume coupling. In order 
for the coupling to depend on a single scale (apart from $M$), 
the flow time $t$ is also linked to $L$ through the constant
$c$~\cite{Fodor:2012td,Fritzsch:2013je}.
The value of this constant is in principle arbitrary, but
experience suggests that $c=0.3$ is a good compromise 
between statistical precision and discretization
errors~\cite{Fritzsch:2013je}. From the point of view of 
decoupling, we note that the larger the value of $c$ is, 
the more sensitive the coupling is to the O($M^{-1}$) 
counterterms located at $x_0=0,T$. This is so because for 
larger values of $t$ the footprint of the flow field 
$B_\mu(t,x)$ extends closer to the boundaries.

In order to attenuate the sensitivity to the O($M^{-1}$) 
terms, in eq.~(\ref{eq:GFTCoupling}) we consider a 
space-time volume with $T=2L$, and place the energy 
density $E_{\rm mag}(t,x)$  at $x_0=T/2$, in order to 
maximize the distance from the boundaries
(cf.~Sect.~\ref{subsec:FiniteVolumeDecoupling}). 
Taking only the magnetic part of the flow energy 
density also helps in reducing the O($M^{-1}$) contaminations.%
\footnote{These choices made for reducing 
		  O($M^{-1}$) effects are also effective in 
		  reducing the sensitivity to the O($a$) 
		  effects stemming from the space-time
		  boundaries 
		  (cf.~Sect.~\ref{subsec:FiniteVolumeDecoupling}).}
Lastly, we note that the expectation value  
$\langle\cdots\rangle_{\rm SF,Q=0}$ in eq.~(\ref{eq:GFTCoupling}) 
is meant to be considered in the presence of the SF boundary 
conditions, eqs.~(\ref{eq:SFbc})-(\ref{eq:SFbc2}),
and restricted to gauge fields in the trivial
topological sector~\cite{Fritzsch:2013yxa,DallaBrida:2016kgh}. 
The latter constraint is imposed in order to circumvent 
issues related to topology freezing at small lattice 
spacings~\cite{DelDebbio:2002xa,Schaefer:2010hu}.

Having introduced the massive scheme of choice, we now
move to the definition of the decoupling scale 
$\mu^{(3)}_{\rm dec}$. We define this in terms of a 
\emph{massless} finite-volume coupling. Its definition slightly 
differs from that of eq.~(\ref{eq:GFTCoupling}). Specifically, 
we take~\cite{Fritzsch:2013je,DallaBrida:2016kgh},
\begin{equation}
	\label{eq:GFcoupling}
	[\bar{g}^{(3)}_{\rm GF}(\mu)]^2\equiv 
	{t^2\over\mathcal{N}'} 
	\langle E_{\rm mag}(t,x)\rangle_{{\rm SF},Q=0}\big|_{T=L,M=0}^{x_0=L/2,\mu=L^{-1},\sqrt{8t}=cL}\,,
\end{equation}
where as before we set $c=0.3$. The main differences
with respect to the definition in eq.~(\ref{eq:GFTCoupling}) 
are that the coupling is evaluated at vanishing (renormalized)
quark masses and the temporal extent is shorter, i.e. $T=L$. 
The reason for considering this specific definition is because 
its non-perturbative running in the $\Nf=3$ theory is known 
very precisely in the range of scales $\mu\approx0.2-4\,\GeV$
(see ref.~\cite{DallaBrida:2016kgh} and
Sect.~\ref{subsec:LambdaNf3}). This gives us the freedom to 
choose for $\mu^{(3)}_{\rm dec}$ the most convenient value 
in this range. Its physical units can in fact be inferred
from combining the knowledge of the non-perturbative 
$\beta$-function of 
$\bar{g}^{(3)}_{\rm GF}(\mu)$~\cite{DallaBrida:2016kgh} and 
the physical scales $\mu_{\rm phys}^{(3)}$ determined in 
large-volume hadronic
simulations~\cite{Bruno:2016plf,Bruno:2017gxd}.%
\footnote{The latter are given by a combination of pion and 
		  kaon decay constants
		  (cf.~refs.~\cite{Bruno:2016plf,Bruno:2017gxd} 
		  and Sect.~\ref{subsec:LambdaNf3} for more 
		  details).}
 
In this study, the decoupling scale is specified
by the condition
\begin{equation}
	\label{eq:LCP}
	[\bar{g}^{(3)}_{\rm GF}(\mu^{(3)}_{\rm dec})]^2=
	3.95\equiv u_0\,,
\end{equation}
which using the information mentioned above is found
to correspond to
\begin{equation}
	\label{eq:mudecPhysUnits}
	\mu_{\rm dec}^{(3)}=789(15)\,\MeV\equiv L_{\rm dec}^{-1}.
\end{equation}
This scale is convenient in practice as, given our choice 
of lattice resolutions (see below), it allows us to simulate
at values of the lattice spacing sensibly smaller than 
those typically accessible to large-volume simulations. As we 
shall see, this enables us to simulate quark masses up to 
$M\approx 4M_c$, while having $aM$ effects under good 
control. Additionally, we can profit from some perturbative
information in the treatment of O($aM$) effects, which is 
expected to be accurate enough at the values of the bare 
coupling corresponding to the relevant lattice spacings. 
Lastly, the value of $\mu_{\rm dec}^{(3)}$ is low enough in 
energy that only a very limited part of the running in $\Nf=3$
QCD is needed in order to connect it to the scales 
$\mu^{(3)}_{\rm phys}$ and set its physical units.

\subsubsection{Determinations of the massive coupling}

The next step in the strategy is to determine the value of the
coupling $\bar{g}^{(3)}_{\rm GFT}(\mu^{(3)}_{\rm dec},M)$ 
for some large quark masses. To this end, we must evaluate the
coupling $\bar{g}^{(3)}_{\rm GFT}(\mu^{(3)}_{\rm dec},M)$ for
several values of the lattice spacing at fixed $\mu^{(3)}_{\rm dec}$
and given $M$, and extrapolate it to the continuum limit.

The condition in eq.~(\ref{eq:LCP}) defines the line of 
constant physics along which $\mu_{\rm dec}^{(3)}$ is 
constant. Given a set of lattice sizes $L/a$, by 
tuning the bare coupling $g_0$ so that the massless GF-coupling 
has the prescribed value $u_0$, we can identify the values 
of $a$ for which $L=L_{\rm dec}$ is fixed in physical units 
as $a/L\to0$. In this respect, note that in order to compute 
the massive coupling $\bar{g}^{(3)}_{\rm GFT}(\mu^{(3)}_{\rm dec},M)$ and the massless coupling 
$\bar{g}^{(3)}_{\rm GF}(\mu^{(3)}_{\rm dec})$ at matching values 
of the lattice spacing up to O($(aM)^2$) corrections, the two 
must be evaluated at the same value of the O($a$)-improved bare
coupling $\tilde{g}_0$~\cite{Luscher:1996sc}. 
The latter, we recall, is defined as~\cite{Luscher:1996sc},
\begin{equation}
	\label{eq:g0tilde}
	\tilde{g}_0^2\equiv g_0^2(1+b_{\rm g}(\tilde{g}_0)am_{\rm q})\,,
	\quad
	m_{\rm q}=m_0-m_{\rm cr}(\tilde{g}_0)\,,
\end{equation}
where $m_0$ is the bare quark mass, $m_{\rm cr}$ is its critical 
value at which the (renormalized) quark masses vanish, and 
$b_{\rm g}(\tilde{g}_0)$ is a function of the bare coupling 
to be determined. In the massless theory $m_{\rm q}=0$, and the
improved bare coupling coincides with $g_0$. The values of 
$g_0$ determined from the condition (\ref{eq:LCP}) in terms of
the massless coupling therefore specify the values of $\tilde{g}_0$
at which the massive couplings should be evaluated. According to
the Symanzik improvement programme, the coefficient 
$b_{\rm g}(\tilde{g}_0)$ can be tuned in order to remove 
O($am_{\rm q}$) effects in the matching between the massless and
massive renormalization schemes. At present, however, this is only
known to 1-loop order in lattice perturbation theory, where 
$b_{\rm g}(g_0)=0.012\,g_0^2\times \Nf+{\rm O}(g_0^4)$~\cite{Sint:1995ch,Luscher:1996sc}. 

\begin{figure*}[h]
	\centering
	%	\vspace*{5cm}       % Give the correct figure height in cm
	\resizebox{0.875\textwidth}{!}{%
		\includegraphics{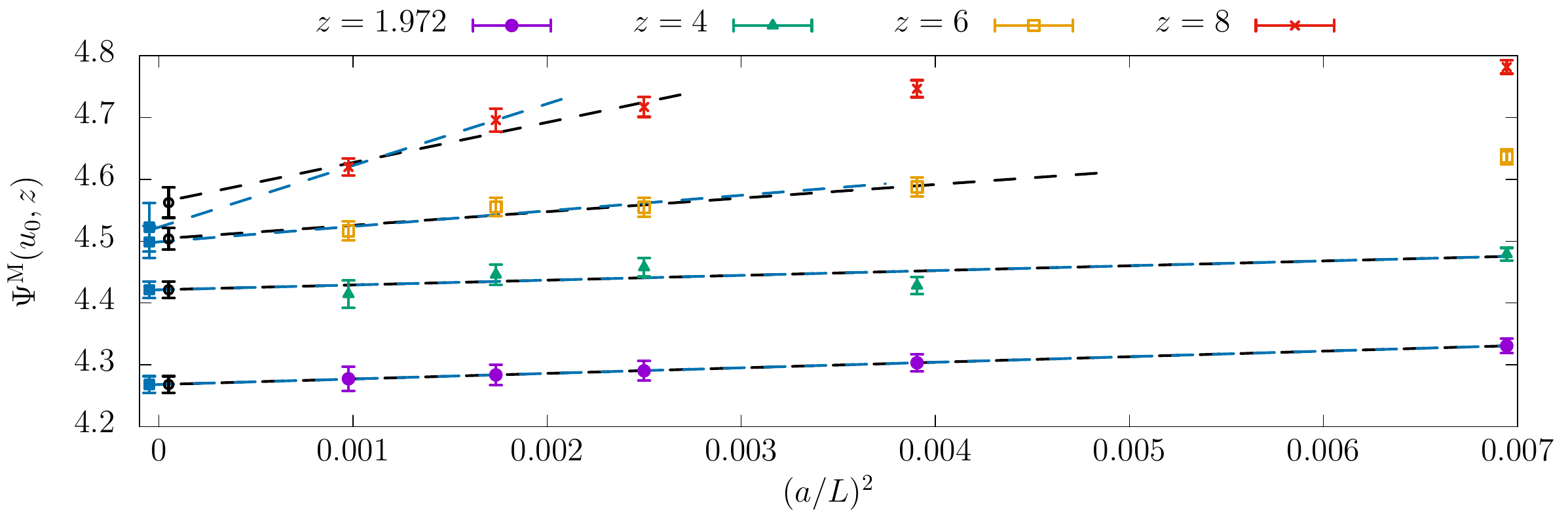}}
	\caption{Continuum limit extrapolations of the massive 
	couplings $\Psi^M(u_0,z)$ for $z=1.972,4.0,6.0,8.0$
	(cf.~eq~(\ref{eq:PsiM}))~\cite{DallaBrida:2019mqg}. 
	Two cuts $(aM)^2 < 1/8,1/4$ are applied in order to estimate 
	the systematic uncertainties in the extrapolations.}
	\label{fig:MassiveCouplings}       
\end{figure*}

Together with the bare coupling, the bare quark masses 
must be set in order to guarantee a given value for the RGI mass 
$M$ in the continuum limit. This is achieved by considering 
a value of $z=ML_{\rm dec}$, and by solving for a given
lattice size $L_{\rm dec}/a$ the following
equation for the bare quark masses,
\begin{equation}
	\label{eq:m0}
	z={L_{\rm dec}\over a} {M\over 
	\overline{m}^{(3)}_{\rm SF}(\mu^{(3)}_{\rm dec})}
	Z_{\rm m}(\tilde{g}_0,a/L_{\rm dec})a\tilde{m}_{\rm q}\,.
\end{equation}
In this equation, 
\begin{equation}
	\tilde{m}_{\rm q}=m_{\rm q}
	(1+b_{\rm m}(\tilde{g}_0)am_{\rm q})\,,
	\quad
	m_{\rm q}=m_0-m_{\rm cr}(\tilde{g}_0)\,,
\end{equation}
is the O($a$)-improved definition for the bare quark mass,
which replaces the regular bare mass $m_{\rm q}$ in order to 
eliminate O($am_{\rm q}$) effects in massive
schemes~\cite{Luscher:1996sc}. To this end, the function 
$b_{\rm m}(\tilde{g}_0)$ must be properly chosen. The function
$Z_m(\tilde{g}_0,a/L)$, instead, refers to the renormalization
constant that relates the bare quark mass to the 
renormalized quark mass $\overline{m}_{\rm SF}(\mu)$ in 
the SF-scheme of
refs.~\cite{Sint:1998iq,Capitani:1998mq,DellaMorte:2005kg,Campos:2018ahf}. This, together with the improvement coefficient $b_{\rm m}$,
and the critical mass $m_{\rm cr}$, are known non-perturbatively
for the relevant parameters (see ref.~\cite{DallaBrida:2019mqg}).
The matching factor 
$M/\overline{m}^{(3)}_{\rm SF}(\mu^{(3)}_{\rm dec})$ then
allows us to convert the renormalized quark mass
in the SF-scheme at the scale $\mu_{\rm dec}^{(3)}$ to the RGI 
mass $M$. It can be obtained from the results of
ref.~\cite{Campos:2018ahf}. Once $am_{\rm q}$ for the given $z$ 
is known from eq.~(\ref{eq:m0}) at the values of $\tilde{g}_0$ 
given by the condition eq.~(\ref{eq:LCP}), using the 1-loop 
results for $b_{\rm g}$ we can infer from eq.~(\ref{eq:g0tilde}) 
the values of $g_0$ at which the massive couplings must be 
computed in simulations~\cite{DallaBrida:2019mqg}.

Having set the bare parameters we can finally evaluate the
functions,
\begin{equation}
	\label{eq:PsiM}
	\Psi^M(u_0,z)=\lim_{a/L_{\rm dec}\to0}
	\big[\bar{g}^{(3)}_{\rm GFT}(\mu^{(3)}_{\rm dec},M)\big]^2\big|_{[\bar{g}^{(3)}_{\rm GF}(\mu^{(3)}_{\rm dec})]^2=u_0}\,.
\end{equation}
In figure \ref{fig:MassiveCouplings} the results for 
the extrapolations in eq.~(\ref{eq:PsiM}) are shown.
Several values of $z$ have been considered, ranging
from $z\approx2-8$, which correspond to RGI masses
$M\approx 1.6-6.3\,\GeV$. The different values of $z$ 
will allow us to assess the size of the non-perturbative 
corrections to decoupling in eq.~(\ref{eq:MasterEquation}). 
The range of lattice sizes considered is $L/a=12-32$. 

As one can see from the figure, as expected, the continuum limit
extrapolations become more challenging as $z$ becomes larger.
However, at the smaller values of $a/L$, the data seem to 
be well described by O($a^2$) discretization errors.
In order to assess systematic effects in the extrapolations,
fits with different cuts in $aM$ have been considered,
specifically $(aM)^2<1/8,1/4$. The results from the different
fits are compatible, with the results for $(aM)^2 < 1/8$ 
having significantly larger errors at large values of $z$, 
where fewer points are left after the cut is imposed. 
We take as final results those with cut $(aM)^2<1/8$. 

\begin{figure*}[h]
	\centering
	%	\vspace*{5cm}       % Give the correct figure height in cm
	\resizebox{0.875\textwidth}{!}{%
	\includegraphics{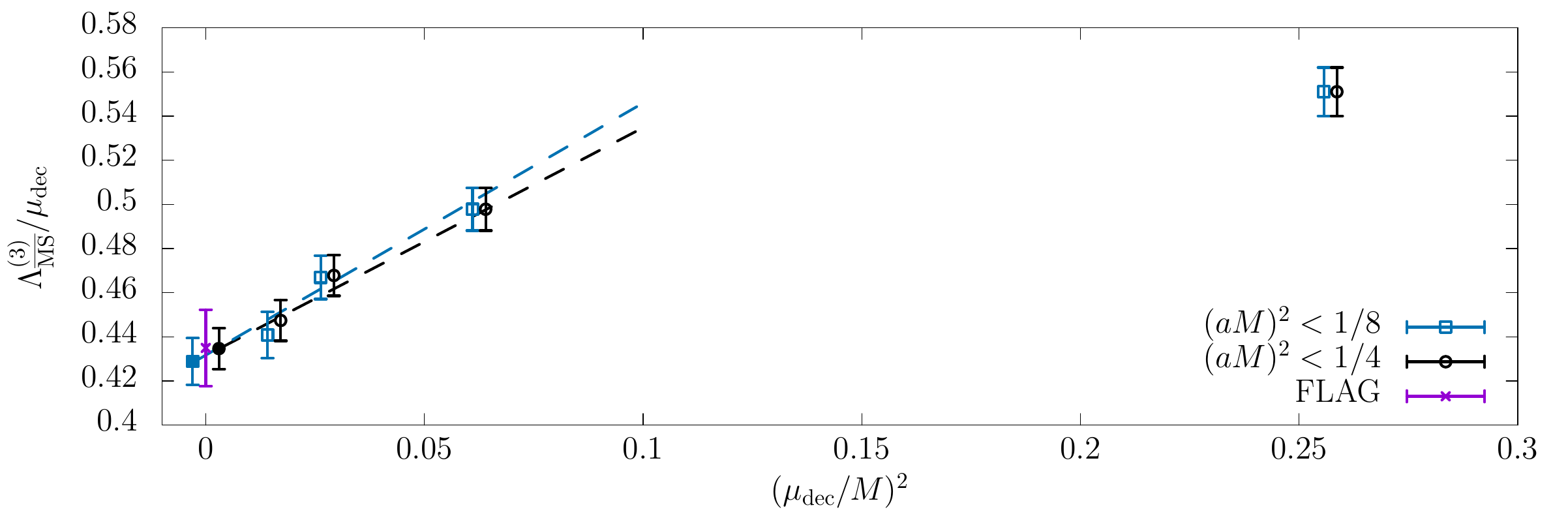}}
	\caption{Values for $\rho=\Lambda^{(3)}_{\overline{\rm MS}}/
	\mu_{\rm dec}^{(3)}$ determined from the decoupling relation,
	eq.~(\ref{eq:MasterEquation})~\cite{DallaBrida:2019mqg}.
	As $z=M/\mu^{(3)}_{\rm dec}$ gets larger, the 
	results for $\rho$ approach the FLAG value 
	$\Lambda^{(3)}_{\overline{\rm MS}}=343(12)\,\MeV$~\cite{Aoki:2019cca}
	in units of $\mu^{(3)}_{\rm dec}=789(15)\,\MeV$
	(cf.~eq.~(\ref{eq:mudecPhysUnits})). The filled symbols illustrate
	possible extrapolations for $z\to\infty$. The results of the
	extrapolations show significantly smaller statistical 
	uncertainties than the FLAG result. (Note that 
	$\mu_{\rm dec}^{(3)}$ gives a negligible contribution to 
	the uncertainty of $\Lambda_{\overline{\rm MS}}^{(3)}/
	\mu_{\rm dec}^{(3)}$ from FLAG.) The uncertainties on 
	$\rho$ may be further reduced with modest computational 
	effort by improving the determination of 
	$\Lambda^{(0)}_{\overline{\rm MS}}/\mu_{\rm dec}^{(0)}$ 
	from ref.~\cite{DallaBrida:2019wur}.
	The data both at finite $z$ and the extrapolations for
	$z\to\infty$ have been slightly shifted horizontally for 
	better clarity.}
	\label{fig:LambdaFromDecoupling}       
\end{figure*}

\subsubsection{Results for $\Lambda_{\overline{\rm MS}}^{(3)}$}

A last step separates us from applying eq.~(\ref{eq:MasterEquation})
to compute $\Lambda_{\overline{\rm MS}}^{(3)}$. In order 
to use eq.~(\ref{eq:MasterEquation}) directly in the 
GFT-scheme of eq.~(\ref{eq:GFTCoupling}) the non-perturbative
running of the corresponding coupling in the pure-gauge 
theory, $\bar{g}^{(0)}_{\rm GFT}(\mu)$, should be known. This, 
however, has never been computed. On the other hand, the 
running of the pure-gauge coupling in the GF-scheme, 
$\bar{g}^{(0)}_{\rm GF}(\mu)$, is known very
precisely~\cite{DallaBrida:2019wur}. In other to resolve the 
issue, all we have to do is to match non-perturbatively the 
GFT- and GF-schemes in the pure Yang-Mills theory. More precisely,
we need to determine the values of the coupling 
$g_M=\bar{g}^{(0)}_{\rm GF}(\mu^{(0)}_{\rm dec})$
corresponding to $\bar{g}^{(0)}_{\rm GFT}(\mu^{(0)}_{\rm dec})=
\sqrt{\Psi^M(u_0,z)}$ for the relevant values of $z$. 
Given this relation we can compute,
\begin{equation}
	{\Lambda^{(0)}_{\rm GF}\over\mu_{\rm dec}^{(0)}}
	=\varphi_{\rm g,\rm GF}^{(0)}(g_M)
	=\varphi_{\rm g,\rm GFT}^{(0)}(\sqrt{\Psi^M})\,,
	\quad
	g_M=\sqrt{\chi(\Psi^{M})}\,,
\end{equation}
where 
$\chi([\bar{g}^{(0)}_{\rm GFT}(\mu)]^2)=
[\bar{g}^{(0)}_{\rm GF}(\mu)]^2$. 
The function $\chi$ can easily be obtained in the relevant 
range of couplings $[\bar{g}^{(0)}_{\rm GFT}]^2=\Psi^M(u_0,z)$,
by computing $[\bar{g}^{(0)}_{\rm GF}]^2$ and 
$[\bar{g}^{(0)}_{\rm GFT}]^2$ at several matching 
values of $L/a$ and $g_0$ in this range, and extrapolating 
their relation to the continuum limit~\cite{DallaBrida:2019mqg}.

The results for $\rho=\Lambda_{\overline{\rm MS}}^{(3)}/\mu^{(3)}_{\rm dec}$
as obtained from eq.~(\ref{eq:MasterEquation}) for different
values of $z$ are shown in figure
\ref{fig:LambdaFromDecoupling}. For the estimates the function 
$P^{\rm PT}_{0,3}(M/\Lambda_{\overline{\rm MS}}^{(3)})$ 
is evaluated at 5-loop order. In this respect we note that, 
as expected from the discussion in
Sect.~\ref{subsubsec:DecouplingPT}, the perturbative
uncertainties in $\rho$  estimated from the effect of the last
known terms of $P^{\rm PT}_{0,3}$ are completely negligible 
compared to the other sources of uncertainties (cf. Table 2
of ref.~\cite{DallaBrida:2019mqg}). As one can see from the 
figure, excluding the point at $z\approx2$, the non-perturbative
corrections to decoupling are  small. At larger $z$ values they 
are compatible with O($z^{-2}$) scaling, indicating that the
O($z^{-1}$) corrections due to the SF boundary conditions are
subdominant. For values of $M\approx 6.3\,\GeV$ (i.e. $z=8$) 
the estimated $\rho$ agrees well with the fully $\Nf=3$ flavor
theory results for $\Lambda_{\overline{\rm MS}}^{(3)}/
\mu^{(3)}_{\rm dec}$, where $\Lambda_{\overline{\rm MS}}^{(3)}$ 
is given by the FLAG average value~\cite{Aoki:2019cca} and
$\mu_{\rm dec}^{(3)}$ is taken from eq.~(\ref{eq:mudecPhysUnits}).
If one attempts a $z\to\infty$ extrapolation of the data the
agreement becomes even better.

\subsection{Summary and miscellaneous remarks}
\label{subsec:Summary}

The results of ref.~\cite{DallaBrida:2019mqg} put on solid 
grounds the application of the decoupling relation
eq.~(\ref{eq:MasterEquation}) as a novel strategy to determine 
the QCD coupling from lattice QCD. The remarkable feature of 
this approach is that the non-perturbative running of the 
coupling from the low-energy scale $\mu_{\rm dec}$ up to 
high energy is done entirely in the pure-gauge theory. This 
opens up the possibility to significantly reduce the current 
error on $\alphas$ (cf.~Sects.~\ref{subsec:LambdaNf3} and 
\ref{sec:Conclusions}).

In order to translate the results for 
$\Lambda_{\overline{\rm MS}}^{(0)}/\mu^{(0)}_{\rm dec}$
into $\Lambda_{\overline{\rm MS}}^{(\Nf)}/\mu^{(\Nf)}_{\rm dec}$,
the strategy relies on two crucial ingredients. The first 
ingredient is the computation of a massive coupling
$\bar{g}^{(\Nf)}_{\Obs}(\mu,M)$ at the low-energy
scale $\mu^{(\Nf)}_{\rm dec}$ in an unphysical set-up with
$\Nf$-flavors of degenerate massive quarks of mass 
$M\gg\mu^{(\Nf)}_{\rm dec}$. Exploiting the decoupling of the
massive quarks, the scales $\mu^{(\Nf)}_{\rm dec}$ and
$\mu^{(0)}_{\rm dec}$ can be connected through the massive 
coupling and so the fundamental $\Nf$-flavor and effective
pure-gauge theories. The second ingredient is the use of 
high-order perturbation theory for estimating the ratio of 
the $\Lambda$-parameters of the two theories, 
$\Lambda_{\overline{\rm MS}}^{(0)}/\Lambda_{\overline{\rm MS}}^{(\Nf)}$, given by the function 
$P_{\rm 0,f}(M/\Lambda_{\overline{\rm MS}}^{(\Nf)})$.
Control on the determination of the massive coupling can 
be achieved by employing a suitable finite-volume scheme. 
Thanks to the fact that the physical volume does not need 
to be large, small lattice spacings can be simulated, and 
safe continuum limit extrapolations of 
$\bar{g}^{(\Nf)}_{\Obs}(\mu_{\rm dec}^{(\Nf)},M)$
with $\mu^{(\Nf)}_{\rm dec}={\rm O}(1\,\GeV)$ can be taken
for quark masses up to a few GeV. At these large masses,
perturbation theory for the function 
$P_{\rm 0,f}(M/\Lambda_{\overline{\rm MS}}^{(\Nf)})$ 
works extremely well and non-perturbative O($M^{-2}$) corrections
to the decoupling relations are found to be small.

From the results of figure \ref{fig:LambdaFromDecoupling}
we can appreciate how the strategy based on decoupling promises
great accuracy. The $z\to\infty$ extrapolations give in fact
results for $\rho$ which are about a factor 2 more precise 
than those obtained using the current FLAG estimate for 
$\Lambda_{\overline{\rm MS}}^{(3)}$ and $\mu_{\rm dec}^{(3)}$
from eq.~(\ref{eq:mudecPhysUnits}). In order to set this result 
on firmer grounds, however, a robust $z\to\infty$ extrapolation 
must be performed. To this end, it is important that the 
continuum limit extrapolations for the massive couplings
$\Psi^M(u_0,z)$ are made more solid at the largest (most relevant)
$z$ values by investing some additional computational effort.
The lattices used for the computations entering figure
\ref{fig:MassiveCouplings} are in fact rather modest. The 
largest simulated lattices have $L/a=32$, $T/a=64$. In addition,
the large quark masses make these simulations significantly 
cheaper than the more common massless SF simulations. A factor 
two finer lattices are hence  within reach with affordable
computational resources. These lattices will allow us to consider
larger quark masses too, and thus improve even further 
the control on the $z\to\infty$ extrapolations. All in all, we 
can expect that after these improvements the final determination 
for $z\to\infty$ will be at least as precise as the results in
figure \ref{fig:LambdaFromDecoupling} promise, but will include
conservative estimates for all systematics.

It is important to note at this point that a significant
fraction of the error on $\rho$ at finite $z$ comes from 
the uncertainty on $\Lambda_{\overline{\rm MS}}^{(0)}/
\mu_{\rm dec}^{(0)}$ from ref.~\cite{DallaBrida:2019wur},
which is about $1.5\%$ (cf.~figure \ref{fig:LambdaFromDecoupling}). 
A reduction of this error down to $0.5\%$ or so is desirable
and in principle possible. Given the importance of the result 
for the determination of $\Lambda_{\overline{\rm MS}}^{(3)}$,
however, it is crucial that this error reduction is achieved
robustly. As discussed in Sect.~\ref{subsubsec:YangMillsLambda},
even though the determination of $\Lambda$ in the pure Yang-Mills 
theory is very much simplified from the computational point 
of view compared to QCD, the problem is yet non-trivial and care
must be taken, especially if such a high precision is desired. 
For this reason, it is mandatory that
the results for $\Lambda_{\overline{\rm MS}}^{(0)}$ are
corroborated by investigating different strategies where the
estimates of systematic uncertainties are put to a stringent 
test. As we have seen, there is currently tension among different
determinations of $\Lambda_{\overline{\rm MS}}^{(0)}$,
some of which quote the desired sub-percent precision
(cf.~Sect.~\ref{subsubsec:YangMillsLambda} and
ref.~\cite{DallaBrida:2019wur}). Studies as the ones of
refs.~\cite{Husung:2020pxg,Nada:2020jay} are hence 
encouraged in order to set the actual accuracy at which we 
currently know $\Lambda_{\overline{\rm MS}}^{(0)}$.

This corroboration goes hand in hand with the exploration
of new strategies for the determination of 
$\Lambda^{(0)}_{\overline{\rm MS}}$. In this respect we point 
out the results of ref.~\cite{Nada:2020jay}, where an alternative
way to do step-scaling for GF-based couplings was proposed 
and tested. In short, the change of renormalization scale in the
coupling is first achieved by changing the flow time at fixed
physical volume and in a second step the physical volume is 
changed at fixed flow time. This has to be compared with the
traditional situation where both flow time and spatial size are
changed at once. One of the interesting features of the approach 
is that it amounts to a reanalysis of data gathered from a
traditional step-scaling study. However, the systematics to deal 
with are quite independent given the different continuum limit
extrapolations involved. By comparing the two analysis one
can stringently test the assumptions made in one or the other
approach.

Furthermore, it would be interesting to employ 
other definitions of finite-volume schemes based on either
different observables and/or set-ups. For instance, the 
GF-coupling with twisted boundary conditions explored in
refs.~\cite{Ramos:2014kla,Lin:2015zpa,Ishikawa:2017xam,Bribian:2020xfs} is promising. Differently from the SF case it enjoys 
full translational invariance and yet its perturbative
expansion in finite volume appears feasible~\cite{Bribian:2019ybc}.
Of course, standard periodic boundary conditions are also an 
option~\cite{Fodor:2012td} despite the difficulties with 
perturbation theory in this set-up. In fact, in cases
where the perturbative information is limited or the relevant
perturbative expansion is poorly convergent, a viable option
is to non-perturbatively match the given finite-volume scheme 
to some other scheme for which the perturbative $\beta$-function 
is known to high-loop order and it is well behaved 
(see e.g.~refs.~\cite{DallaBrida:2016kgh,DallaBrida:2019wur}). 
This may allow for a significantly more precise determination 
of $\Lambda_{\overline{\rm MS}}^{(0)}$
(cf.~Sect.~\ref{subsubsec:YangMillsLambda}). In this respect, 
we note that a powerful framework for automated
numerical high-loop calculations in finite volume has been 
recently developed and successfully
applied~\cite{DallaBrida:2017tru,DallaBrida:2017pex}.

Another idea that may be interesting to consider is the
determination of $\Lambda^{(0)}_{\overline{\rm MS}}$ based
on the infinite-volume $\beta$-function of GF-based couplings,
following the strategy of
refs.~\cite{Fodor:2017die,Hasenfratz:2019hpg,Fodor:2019ypi}
(see also ref.~\cite{Luscher:2014kea}). In this approach, the 
infinite-volume results are obtained by extrapolations from 
small-volume simulations, which might already be at hand from 
a conventional step-scaling study. If the (non-trivial) 
infinite-volume extrapolations can be performed in a controlled 
way and the convergence to the perturbative regime of the chosen
scheme is fast enough, this strategy may allow for interesting
crosschecks of the results from step-scaling. In this case, the
framework developed in ref.~\cite{Artz:2019bpr} can be used to
obtain the necessary infinite-volume perturbative information 
to high-loop order. 

Besides applying different strategies for the calculation of
$\Lambda_{\overline{\rm MS}}^{(0)}$, different techniques
can be considered for the QCD part of the decoupling strategy
as well. A simple extension is to consider different schemes 
for the massive finite-volume coupling. In the case of $\Nf=3$
QCD, periodic and twisted boundary conditions would avoid 
entirely the issue with O($M^{-1}$) contaminations. 
For $\Nf=4$ QCD twisted boundary conditions cannot be 
implemented~\cite{Parisi:1984cy}, but twisted-mass fermions 
with SF boundary conditions or regular massive quarks with 
chirally rotated boundary conditions are available options
(cf.~Sect.~\ref{subsec:FiniteVolumeDecoupling}).

A substantially different approach is to avoid entirely 
finite-volume couplings and rely on heavy-quark decoupling  
in hadronic quantities. Particularly interesting observables 
to consider are the popular gluonic scales 
$\mathcal{S}=t_0^{-1/2}, t_c^{-1/2},w_0^{-1},
r_0$~\cite{Luscher:2010iy,Athenodorou:2018wpk,Borsanyi:2012zs,Sommer:1993ce}. 
In this case, the decoupling relation is applied more directly in 
the form of eq.~(\ref{eq:MatchingNPT}), specifically,
\begin{equation}
	{\Lambda^{(\Nf)}_{\overline{\rm MS}}\over 
	\mathcal{S}^{(\Nf)}(M)}
	P^{\rm PT}_{0,\rm f}\bigg({M\over
	{\Lambda^{(\Nf)}_{\overline{\rm MS}}}} \bigg)
	=
	{\Lambda^{(0)}_{\overline{\rm MS}}\over \mathcal{S}^{(0)}}\,
	+{\rm O}(g_*^{2n-2},M^{-2}).
\end{equation} 
Once the right hand side is known from computing the running 
in pure-gauge theory, the perturbative approximation 
$P^{\rm PT}_{0,\rm f}({M/\Lambda^{(\Nf)}_{\overline{\rm MS}}})$
to the ratio of $\Lambda$-parameters can be used to solve
the above equation for 
$\Lambda_{\overline{\rm MS}}^{(\Nf)}/\mathcal{S}^{(\Nf)}(M)$.
All that is left to do to determine $\Lambda_{\overline{\rm MS}}^{(\Nf)}$ is then to fix the physical units of the 
low-energy quantity $\mathcal{S}^{(\Nf)}(M)$ computed in a 
theory with $\Nf$ heavy quarks of RGI mass $M$. This can be 
obtained by relating $\mathcal{S}^{(\Nf)}(M)$ to some 
convenient physical scale $\mu_{\rm phys}^{(\Nf)}$ evaluated
at physical quark masses via the bare parameters. As discussed 
in Sect.~\ref{subsubsec:MultiScaleProblem2}, it might be 
difficult to reach large masses $M$ with this approach, while 
having discretization errors and finite-volume effects under
control. Some compromises are likely necessary in order to 
reach high enough $M$ values to be able to control decoupling
corrections. On the other hand,  the studies of
refs.~\cite{Knechtli:2017xgy,Athenodorou:2018wpk} 
show that interesting results may be obtained if masses
close to that of the charm can be reliably reached. This 
makes the strategy worth being explored.

% end

%% file: sect5.tex
% sect5.tex

\section{Conclusions}
\label{sec:Conclusions}

In this contribution we presented a novel strategy for 
the determination of the QCD coupling using lattice QCD. 
It exploits the decoupling of heavy quarks at low energy 
to connect the pure Yang-Mills theory and 
QCD with $\Nf$ flavors of quarks. The main result is that 
the computation of the running of the coupling from a 
known low-energy scale $\mu_{\rm dec}={\rm O}(1\,\GeV)$ 
up to high energies can be done entirely in the pure-gauge
theory instead of $\Nf$-flavor QCD. Considering $\Nf=3$ or 4, 
this paves the way for unprecedented precision determinations 
of $\Lambda_{\overline{\rm MS}}^{(\Nf)}$ from which 
$\Lambda_{\overline{\rm MS}}^{(5)}$ and $\alphas$ 
can be obtained. In ref.~\cite{DallaBrida:2019mqg} the 
potential of these methods was successfully established 
in the determination of $\Lambda_{\overline{\rm MS}}^{(3)}$. 
We now want to put this result into context of a future  
precision $\alphas$ extraction.

As presented in Sect.~\ref{subsec:CouplingFromDecoupling},
the results for $\Lambda_{\overline{\rm MS}}^{(3)}/
\mu^{(3)}_{\rm dec}$ from decoupling have an uncertainty 
which is about half the one obtained using the FLAG 
average $\Lambda_{\overline{\rm MS}}^{(3)}=343(12)\,\MeV$~\cite{Aoki:2019cca} 
and $\mu_{\rm dec}^{(3)}=789(15)\,\MeV$ from
eq.~(\ref{eq:mudecPhysUnits})
(cf.~figure	\ref{fig:LambdaFromDecoupling}). 
As discussed in Sect.~\ref{subsec:Summary}, by investing some
modest computational effort, this result can be 
set on very solid grounds by improving the continuum limits of 
the massive couplings (cf.~figure \ref{fig:MassiveCouplings}) 
and performing a robust $z\to\infty$ extrapolation
(cf.~figure \ref{fig:LambdaFromDecoupling}). Considering lattices
twice as large as the ones simulated in
ref.~\cite{DallaBrida:2019mqg} is in fact affordable. 
With such lattices we can expect that the continuum limit
extrapolations of $\Psi^M(u_0,z)$ in eq.~(\ref{eq:PsiM})
can be performed with confidence also at the largest 
masses investigated so far ($M\approx6\,\GeV$). In addition, 
we will be able to consider some larger $z$ values, too.

The precision on 
$\Lambda_{\overline{\rm MS}}^{(3)}/\mu^{(3)}_{\rm dec}$
can be further improved significantly by reducing the 
uncertainties coming from the pure-gauge determination of 
$\Lambda_{\overline{\rm MS}}^{(0)}/\mu^{(0)}_{\rm dec}$, 
which has the (conservative) error of
1.5\%~\cite{DallaBrida:2019wur}. As noticed in 
Sect.~\ref{subsec:Summary}, this error can in principle be 
reduced by a substantial factor, e.g. down to 0.5\%. However, 
while it is certainly possible to reach such a high precision 
in a given computation, it is crucial that the 
results of different analysis and groups corroborate it. 
At present, there is in fact  tension among 
determinations of $\Lambda_{\overline{\rm MS}}^{(0)}$ 
involving results with sub-percent accuracy
(cf.~Sect.~\ref{subsubsec:YangMillsLambda}). It is 
important to understand the origin of these
differences. We hope that the renovated interest in this 
quantity brought by this new strategy motivates the community 
to resolve the issue and contribute to a high-precision
determination.

Once the above steps are achieved the precision on
$\Lambda_{\overline{\rm MS}}^{(3)}$ will be limited 
by the present error on $\mu^{(3)}_{\rm dec}$, which is
about 2\%. A reduction of this error down to $1\%$ is 
however foreseeable. It requires, first of all, to improve
the results for the running of the GF-coupling 
$\bar{g}^{(3)}_{\rm GF}(\mu)$ at energies 
$\mu<\mu_{\rm dec}$~\cite{DallaBrida:2016kgh}. We recall
that this is needed in order to connect 
$\mu^{(3)}_{\rm dec}$ with the hadronic scales
$\mu^{(3)}_{\rm phys}$ used to set the physical units
of the theory 
(cf.~Sect.~\ref{subsec:CouplingFromDecoupling}).
Work in this direction has already started as part of the 
HQET efforts by the ALPHA Collaboration
(cf.~ref.~\cite{Fritzsch:2018yag}). Secondly, the scale setting
in terms of the physical scales $\mu_{\rm phys}^{(3)}$ must be
improved as well. In practice, this means to obtain a more precise
determination for a convenient low-energy reference scale in 
$\Nf=3$ QCD in physical units, 
like for example, $\mu_{\rm ref}^*=1/\sqrt{8t_0^*}$
(cf.~ref.~\cite{Bruno:2016plf}). A precision of $1\%$ or better
on this or similar scales is desirable. This is expected
to be possible by exploiting the new CLS ensembles close to 
the physical
point~\cite{Bruno:2014jqa,Bali:2019yiy,Gerardin:2019rua}.
Also in this case, however, corroboration from different 
strategies and groups is important.

Through all these steps a determination of 
$\Lambda_{\overline{\rm MS}}^{(3)}$ with a final  
uncertainty of 1-2\% appears feasible. As discussed in
Sect.~\ref{subsubsec:NonperturbativeDec}, at 
this level of precision $\Lambda_{\overline{\rm MS}}^{(5)}$
can yet be obtained from $\Lambda_{\overline{\rm MS}}^{(3)}$
by relying on the perturbative decoupling of the charm quark,
eventually including some conservative estimate for the 
unaccounted non-perturbative corrections. A determination of
$\alphas(M_Z)$ at the level of 0.4\% is therefore within reach
thanks to the novel techniques. To further halve the error
on $\alphas(M_Z)$, on the other hand, requires several issues to
be reconsidered. Non-perturbative decoupling effects might 
become relevant, and one might need to include electromagnetic
and $m_u\neq m_d$ effects in the lattice computations in order 
to set the physical scale of the theory to a greater level of
accuracy (cf.~Sect.~\ref{subsubsec:NonperturbativeDec} and see
also discussion in ref.~\cite{DelDebbio:2021ryq}).

Before concluding we want to note that even though our emphasis 
was on the determination of $\Lambda_{\overline{\rm MS}}^{(\Nf)}$,
the ideas presented can be extended to solve other 
RG problems. A clear case is that of the quark masses,
where one can replace their running in $\Nf$-flavor QCD
with the one in the quenched approximation. In 
ref.~\cite{Cali:2019enm} a similar application was in fact 
explored in order to study the non-perturbative charm-quark 
effects in the determination of the charm-quark mass itself. 
More complicated composite operators, like for instance 
four-quark operators, require more thought. First of all, 
a study of the quality of their perturbative decoupling relations 
is necessary in order to establish whether the strategy has any
chance to be applied in the first place. Then, an investigation 
of the non-perturbative decoupling corrections must follow. 

In conclusion, we can affirm that the decoupling of
heavy quarks enters at full right in the renormalization 
toolkit of the lattice field theorist. Many more 
applications of these powerful ideas in lattice QCD and 
lattice field theory in general are likely to come.

\section*{Acknowledgments}

I would like to thank the organizers of the Special Issue:
``Lattice Field Theory during the COVID-19 pandemic''
for their kind invitation and the opportunity to share
these results. I am really grateful to all my colleagues for 
the rewarding collaboration over the past years.
A special thanks goes to:  Patrick Fritzsch, Roman H\"ollwieser,
Francesco Knechtli, Tomasz Korzec, Alberto Ramos, Stefan Sint, 
and Rainer Sommer. In this occasion, I particularly thank
F. Knechtli and R. Sommer for their valuable comments 
on this manuscript. Finally, it is with gratitude that I thank 
the University of Cyprus, the Cyprus Institute, and the members 
of their lattice QCD groups, for their kind hospitality during 
the preparation of this work.

%end

%% file: biblio.tex
% biblio.tex

\bibliographystyle{JHEP}
\bibliography{bibfile}

% end